\documentclass[11pt,a4paper]{article}
\usepackage{jheppub}

\usepackage{lmodern}

\usepackage[utf8]{inputenc}
\usepackage{color}
\usepackage{float}
\usepackage{amsmath,bm}
\usepackage{amssymb}
\usepackage{graphicx}
\usepackage{setspace}
\usepackage{subfigure}
\usepackage{amsthm}
\usepackage{amsfonts}
\usepackage{braket}
\usepackage{multirow}
\usepackage{ulem}
\usepackage{slashed}
\usepackage{babel}
 \hypersetup{
	colorlinks=true,
	linkcolor=cyan,
	urlcolor=blue,
	citecolor=red,
}

\title{Black holes Entangled by Radiation} 

\author[a]{Yuxuan Liu,}

\author[b]{Zhuo-Yu Xian,}

\author[a]{Cheng Peng,}

\author[c,d]{Yi Ling}

\affiliation[a]{Kavli Institute for Theoretical Sciences (KITS), University of Chinese Academy of Sciences, Beijing 100190, China}

\affiliation[b]{Institute for Theoretical Physics and Astrophysics and W\"urzburg-Dresden Cluster of Excellence ct.qmat, Julius-Maximilians-Universit\"at W\"urzburg, 97074 W\"urzburg, Germany}

\affiliation[c]{Institute of High Energy Physics, Chinese Academy of Sciences, Beijing 100049, China}

\affiliation[d]{School of Physics, University of Chinese Academy of Sciences,
Beijing 100049, China 
}

\emailAdd{liuyuxuan@ucas.ac.cn}

\emailAdd{zhuo-yu.xian@physik.uni-wuerzburg.de}

\emailAdd{pengcheng@ucas.ac.cn}

\emailAdd{lingy@ihep.ac.cn}

\abstract{We construct three models to describe the scenario where two eternal black holes are separated by a flat space, and can eventually be entangled by exchanging radiation. 
In the doubly holographic setup, we compute the entanglement entropy and mutual information among subsystems and obtain the dynamic phase structure of the entanglement.
The formation of entanglement between these two black holes is delayed by the space which the radiation must travel through.
If the black holes exchange sufficient Hawking modes, the final state is characterized by a connected entanglement wedge; otherwise, the final entanglement wedge contains two separate islands. 
In the former case, the entanglement wedge of the black holes forms at the time scale proportional to the size of the flat space between them. While in both cases, the unitarity of the evolution is preserved.
When the sizes of the black holes are not equal, we observe a loss of entanglement between the smaller black hole and the radiation at late times. 
On the field theory side, we consider two Sachdev-Ye-Kitaev (SYK) clusters coupled to a Majorana chain, which resemble two black holes connected by a radiation region. 
We numerically compute the same entanglement measures and obtain similar phase structures as the bulk results. In general, a time delay of the entanglement between the SYK clusters is found in cases with a long Majorana chain. In particular, when the SYK clusters are different in size, similar entanglement loss between the smaller SYK cluster and the Majorana chain is observed.
Finally, we investigate a chain model composed of EPR clusters with particles exchanging between neighboring clusters and reproduce the features of entanglement observed in the previous models.}

\arxivnumber{}

\begin{document}

\maketitle

\section{Introduction}
Understanding black hole physics in the context of quantum mechanics is a notoriously difficult task. 
It leads to the famous problem of whether a black hole evaporates in a unitary fashion, which is also called the black hole information paradox \cite{hawking1974black,hawking1975particle,hawking1976breakdown}. 
On one side, the nearly thermal spectrum of Hawking radiation requires the entropy to keep growing, according to Hawking's earlier calculation. 
On the other side, however, the quantum nature of black holes demands that the entropy of radiation should decrease at late times and its dynamical evolution should obey the rules described by the Page curve \cite{Page:1993wv,Page:2004xp,Page:2013dx}. 
See \cite{susskind1993stretched,harlow2013quantum,Almheiri:2012rt,Maldacena:2013xja} for further debates.

Recently, a breakthrough was made to understand the unitary evolution in the context of Gauge/Gravity duality \cite{penington2020entanglement,almheiri2019entropy}. 
By extremizing the generalized gravitational entropy \cite{Lewkowycz:2013nqa,Engelhardt:2014gca}, the entanglement contributed by Hawking radiation can be captured by the quantum extremal surface (QES), through the island formula \cite{Almheiri:2019hni} \begin{equation}\label{eq_QESinRad}
S[\mathcal{R}]=\min_{\mathcal I} \left\{ \mathop{\text{ext}}\limits_{\mathcal I} \left[\frac{\operatorname{Area}[\partial \mathcal{I}]}{4 G_{N}} + S[\mathcal{R} \cup \mathcal{I}]\right]\right\}.
\end{equation}
Here $\mathcal{R}$ and $\mathcal{I}$ represent the radiation and the islands, respectively.
At the beginning of evolution, the entropy increases due to the accumulation of Hawking modes.
While at late times, the growth of entropy is pinched off by the emergence of the island in the gravitational region.
As a result, the entropy of radiation during evolution is in line with the Page curve \cite{Almheiri:2020cfm}. 
See \cite{Chen:2019uhq,Alishahiha:2020qza,Hashimoto:2020cas,anegawa2020notes,hartman2020islands,chen2021evaporating,bhattacharya2021topological,deng2021defect,wang2021islands,He:2021mst,gautason2020page,krishnan2020page,sybesma2021pure,chou2021page,Hollowood:2021lsw,Suzuki:2022xwv,Suzuki:2022yru,Bhattacharya:2021nqj,Bhattacharya:2021dnd,Caceres:2021fuw,Bhattacharya:2021jrn,Caceres:2020jcn} for more recent discussions on islands. 
The island formula (\ref{eq_QESinRad}) was firstly proposed in the model of two-dimensional gravity coupled to two-dimensional CFT matter sectors \cite{Almheiri:2019hni}. 
In addition to this time-dependent case, the island also emerges in the two-dimensional static geometry, where the eternal black hole is in equilibrium with two flat baths \cite{Almheiri:2019yqk}.
The full-time behavior of entropy for this model can be obtained by considering the holographic dual of this two-dimensional theory, known as double holography, where the QES becomes a standard RT/HRT surface in the three-dimensional geometry \cite{Almheiri:2019hni,Chen:2019uhq}. 

The existence of islands in higher dimensions was first demonstrated in \cite{Almheiri:2019psy}, where the lower-dimensional black hole is replaced by a brane with tension where Neumann boundary conditions are applied
\cite{Takayanagi:2011zk,Chu:2018ntx,Miao:2018qkc}. 
In this context, however, the dynamical Page curve can be obtained only under the condition that there are sufficient degrees of freedom (DOF) on the brane \cite{Ling:2020laa,Chen:2020uac,Geng:2020fxl,Geng:2020qvw},  where the prescriptions of inputting enough DOF have been proposed as well. 
See \cite{Krishnan:2020fer,Geng:2020fxl,Geng:2020qvw,Chen:2020uac,Chen:2020hmv,Hernandez:2020nem,grimaldi2022quantum,Miao:2020oey,Akal:2020wfl,akal2021entanglement,Omidi:2021opl} for more brane-world construction in higher dimensions.

From the aspect of path integral formalism, the emergence of islands corresponds to the dominance of another saddle point, known as the wormhole saddle point, where Euclidean wormholes connect the different replicas \cite{Penington:2019kki,Almheiri:2019qdq} (See also \cite{rozali2020information,karlsson2020replica}).
Inspired by this thought-provoking viewpoint, the notion of ``wormhole'' was extensively discussed as the bridge between disjoint universes \cite{balasubramanian2021islands,balasubramanian2021entanglement1,balasubramanian2021entanglement,miyata2022evaporation,miyata2021entanglement}, where the wormhole behind horizons is lengthened, as well as in baby universes \cite{marolf2021observations,marolf2020transcending,balasubramanian2020spin,peng2021baby}, which connect to parent universes through a wormhole. 

To better understand the black hole information paradox, measuring alone the entanglement entropy of radiation is not sufficient. 
On one side, other measures were introduced to investigate the entanglement properties of radiation in \cite{renner2021black,li2020reflected,basak2022islands,Kawabata:2021hac,kawabata2021replica,vardhan2021mixed,akal2021page}. 
On the other side, the entanglement properties inside the gravity system were studied in \cite{Ling:2021vxe}, where the emergence of the island trimmed off the growth of reflected entropy. 

Further, an intriguing topic is to investigate two black holes entangled by exchanging Hawking radiation, based on the setup of thermofield double (TFD) states \cite{Maldacena:2001kr,Almheiri:2019yqk,Gu:2017njx,Hartman:2013qma}.
We consider two black holes separated by a finite space. Intuitively, the Hawking radiation emitted by one black hole can eventually be absorbed by the other, if they travel through this space. By this process, two black holes will develop more and more entanglement. 
Due to unitarity, the entanglement will eventually reach an upper bound determined by the DOF of subsystems.
In this paper, we strive to model these phenomena by analyzing the entanglement properties between the separate black holes and envisaging the formation of entanglement, from both the bulk gravity and boundary perspective. (See \cite{Balasubramanian:2020hfs,anderson2021islands,engelhardt2022canonical,Balasubramanian:2021xcm} for other discussions on the formation of wormholes.) 
For these purposes, we propose the following three kinds of models to simulate the entanglement between black holes and radiation:

First, we construct a doubly holographic model with two Planck branes in $(3+1)$ dimensions.
The tension and Dvali-Gabadadze-Porrati (DGP) terms \cite{Dvali:2000hr} are imposed on both branes to enhance the DOF. 
Second, inspired by the work in \cite{Chen:2020wiq}, we propose a quantum mechanical model with two separate SYK clusters coupled to a Majorana chain connecting them. 
Third, we will further simplify the model to be one composed of several EPR clusters which form a chain structure, with the outermost clusters being black holes and the inner clusters being radiation. 
See also \cite{geng2021holographic,geng2021entanglement,Afrasiar:2022ebi} for other models in BCFT, BTZ black hole and black string with branes.

The paper is organized as follows: in section \ref{sec_dhs}, we generalize the doubly holographic setup in \cite{Ling:2020laa} by introducing two Planck branes with Neumann boundary conditions on them. Then we apply the Einstein-DeTurck formalism to solve this gravitational system. 

In section \ref{sec_ee&mi}, we elaborate on the description of the holographic entanglement entropy and mutual information in different phases. More importantly, we find a novel phase concerning the appearance of a wormhole. 

In section \ref{sec_pc&eps}, we illustrate the dynamical phase structures and transitions of entanglement during the evolution in the probe limit. Subsequently, we take the backreaction into account and explore the effect of DOF in black holes on the phase structure. 

In section \ref{sec_SYK}, we propose a quantum mechanics model that connects SYK models with Majorana chains. The eigensystems are calculated by exact diagonalization and the related measures in quantum information theory are computed. 

In section \ref{sec_tm}, we construct a toy model which consists of several EPR clusters. By allowing them to exchange particles with the nearby clusters, we qualitatively simulate the process of black holes entangled by radiation. 

Finally, the conclusions and discussions are given in section \ref{sec_c&d}.

\section{The doubly holographic setup}\label{sec_dhs}
\begin{figure} 
  \centering
 \subfigure[]{\label{fig_DDH_l1}
  \includegraphics[scale=0.63]{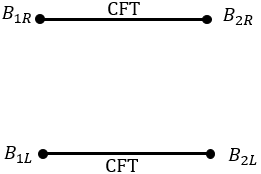}}
  \hspace{0pt}
  \subfigure[]{\label{fig_DDH_l2}
  \includegraphics[scale=0.63]{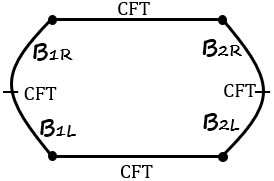}}
  \hspace{0pt}
  \subfigure[]{\label{fig_DDH_l3}
  \includegraphics[scale=0.63]{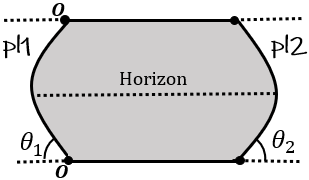}}
\caption{(a): Boundary perspective. The $(d-1)$-dimensional quantum systems $B_{1L}$, $B_{1R}$ and $B_{2L}$, $B_{2R}$ are in equilibrium with  two $d$-dimensional finite-sized baths. (b): Brane perspective. Two $d$-dimensional eternal black holes are in equilibrium with two $d$-dimensional finite-sized baths. (c): Bulk gravity perspective. The Planck branes are back-reacted to the ambient geometry.}
\end{figure}

In this section, we will present a general setup for two charged eternal black holes exchanging Hawking modes. 
Consider two $d$-dimensional eternal black holes which are asymptotic $AdS_{d}$, coupled to two finite-sized flat baths, where Hawking modes living in this combined system are described by the $d$-dimensional CFT matter sector, as shown in Fig.~\ref{fig_DDH_l2}. 
Here the matter sectors in baths are served as radiation $\mathcal{R}$, while $\mathcal{B}_1$ and $\mathcal{B}_2$ denote two different eternal black holes, with the subscripts $L, R$ denoting the left- and right-sided black holes in each eternal black hole $\mathcal{B}_i$.
Specifically, the asymptotic boundaries between black holes and finite-sized baths are glued together, where we impose the transparent boundary condition on the matter sector. 
This setup is called the brane perspective in literature \cite{Chen:2020hmv}, and there are two other equivalent perspectives. 
The first is the boundary perspective, where two $d$-dimensional eternal black holes together with the matter sectors are replaced by $(d-1)$-dimensional quantum systems $B_{1L}$, $B_{1R}$ and $B_{2L}$, $B_{2R}$ -- Fig.~\ref{fig_DDH_l1}. 
The second is the bulk gravity perspective, where the matter sector is dual to a $(d+1)$-dimensional bulk and $\mathcal{B}_1$ and $\mathcal{B}_2$ are described by Planck branes $\bm{pl_1}$ and $\bm{pl_2}$ in the bulk, as shown in Fig.~\ref{fig_DDH_l3}. 
In the first model, the setup will be constructed from the bulk gravity perspective, while the dynamical processes will be described from both the bulk gravity perspective and the brane perspective.

From the bulk gravity perspective, the action of the $(d+1)$-dimensional gravity theory is specified as
\begin{align}\label{eq_Action}
I=&\frac{1}{16\pi G_N^{(d+1)}} \Bigg[ \int d^{d+1}x
\sqrt{-g}\left(R+\frac{d(d-1)}{L^2}\right)+2\int_{\bm{\partial}}d^{d}x\sqrt{-h_{\bm{\partial}}}K_{\bm{\partial}}\nonumber\\
&-\int d^{d+1}x \sqrt{-g}\frac{1}{2}F^2+ 2\sum_{i=1}^2 \left(\int_{\bm{pl_i}}d^{d}x\sqrt{-h_i}\left(K_i-\alpha_i\right)-\int_{\bm{pl_i}\cap \bm{\partial}}
d^{d-1}x \sqrt{-\Sigma_i}\, \theta_i\right)\Bigg]\nonumber \\
&+ \sum_{i=1}^2\left[\frac{1}{16 \pi G_{b,i}^{(d)}}\int d^d x
\sqrt{-h_i}R_{h_i}+\frac{1}{8\pi G_{b,i}^{(d)}}\int_{\bm{pl_i}\cap \bm{\partial}} d^{d-1}x \sqrt{-\Sigma_i} \, k_i\right].
\end{align}
Here $K_{\bm{\partial}}$ is the extrinsic curvature on the
conformal boundary $\bm{\partial}$. 
The electromagnetic curvature is $F=\text{d}A$. 
$K_i$ is the extrinsic curvature and the parameter $\alpha_i$ is proportional to the tension on the brane $\bm{pl_i}$, which will be fixed later. 
The last term in the second line is the junction term at the intersection of the brane $\bm{pl_i}$ and the conformal boundary $\bm{\partial}$, where $\theta_i$ is the angle between the brane $\bm{pl_i}$ and the boundary, while $\Sigma_i$ is the metric on $\bm{pl_i}\cap \bm{\partial}$. 
The first term in the last line is the DGP term \cite{Chen:2020uac}, where $G_{b,i}^{(d)}$ is the additional Newton constant on each brane and $R_{h_i}$ is the intrinsic curvature on each brane. 
The second term in the last line is the junction term at the intersection $\bm{pl_i}\cap \bm{\partial}$ of the brane and the conformal boundary, where $k_i$ is the extrinsic curvature on $\bm{pl_i}\cap \bm{\partial}$. 
Taking the variation of the action, we obtain the equations of motion as
\begin{align}
  R_{\mu\nu}+\frac{d}{L^2} g_{\mu\nu}&=\left(T_{\mu\nu}-\frac{T}{d-1}g_{\mu\nu}\right), \quad\text{with}\quad T_{\mu\nu}=F_{\mu a}F_{\nu}{}^{a}-\frac{1}{4}F^2g_{\mu\nu}\label{eq_eineq},\\
  \nabla_\mu F^{\mu\nu}&=0,
\end{align}
where $T$ is the trace of the energy-stress tensor $T_{\mu\nu}$.

\begin{figure}\label{fig_pb2}
  \center{
  \includegraphics[scale=0.5]{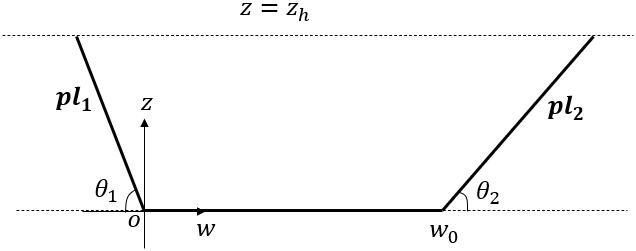}
  \caption{A simple setup of Planck branes. Here Planck branes are anchored on the conformal boundary at $(z, w) = (0, 0)$, $(z, w) = (0, w_0)$ respectively and penetrate into the bulk with angles $\theta_1$ and $\theta_2$.}}
\end{figure}

\subsection{Boundary conditions on the Planck branes}
In AdS/CFT setup with infinite volume, the $(d+1)$-dimensional
bulk is asymptotic to $AdS_{d+1}$ which in Poincaré coordinates is described by
\begin{equation}\label{eq_ads}
ds^2=\frac{L^2}{z^2}\left(-dt^2+dz^2+dw^2+\sum_{i=1}^{d-2} dw_j^2\right),
\end{equation}
with the conformal boundary at $z=0$.
Let $\theta_i$ be the angle between the $i$-th Planck brane and the conformal boundary as shown in Fig.~\ref{fig_pb2}. Then Planck branes $\bm{pl_1}$ and $\bm{pl_2}$ are described by hypersurfaces
\begin{align}\label{eq_cons} 
   w + \cot \theta_1 z =0 \quad \text{and} \quad   (w-w_0) - \cot \theta_2 z =0 
\end{align}
respectively near the boundary. One should cut the bulk ending on these two branes and restrict the region between them
\cite{Randall:1999vf,Dvali:2000hr,Karch:2000ct,Takayanagi:2011zk}.
We will also impose these constraints (\ref{eq_cons}) deep into
the bulk and find its backreaction to the geometry.

For the boundary terms in (\ref{eq_Action}), we impose
Neumann boundary conditions on each Planck brane, which are
\begin{equation}\label{eq_BCSonBrane}
  (K_i)_{AB}-K_i (h_i)_{AB}+\alpha_i (h_i)_{AB}=\lambda_i L \left[\frac{1}{2}R_{h_i} (h_i)_{AB}-(R_{h_i})_{AB}\right],
\end{equation}
where $h_i$ is the induced metric on the brane $\bm{pl_i}$,  and $\lambda_i \equiv \frac{G_N^{(d+1)}}{G_{b,i}^{(d)}L}$  can be regarded as the effective coupling of the DGP term on this brane. The parameter $\alpha_i$ in action (\ref{eq_Action}) is fixed to be a constant by solving (\ref{eq_BCSonBrane}) near the conformal boundary to concrete the tension term on the brane.

In addition to (\ref{eq_BCSonBrane}), we also impose Neumann boundary conditions for the gauge field $A_\mu$ on the brane $\bm{pl_i}$ \cite{Takayanagi:2011zk,Nozaki:2012qd,Chu:2018ntx,Miao:2018qkc}, which is
\begin{align}\label{eq_BCSonBrane1}
  n^\mu F_{\mu \nu} (h_i)^\nu{}_B=0,
\end{align}
where $n_\mu$ is the normal vector to that brane and the subscript $B$ denotes the coordinates along the brane.

\subsection{The Einstein-DeTurck formalism}
The line element of the standard RN-AdS$_{d+1} (d\geq 3)$ geometry is
\begin{align}\label{eq_RNBH}
  ds^2&=\frac{L^2}{z^2}\left[-f(z)dt^2+\frac{dz^2}{f(z)}+dw^2+\sum_{j=1}^{d-2}dw_j^2\right],\\
  A=&\mu\left(1-z^{d-2}\right)dt, \\
  f(z)=&1-
  \left(1+\frac{d-2}{d-1} \frac{\mu^2}{L^2}\right) z^d + \frac{d-2}{d-1}\frac{\mu^2}{L^2}  z^{2d-2},
\end{align}
where $\mu$ is the chemical potential of the boundary theory, the outer horizon has been scaled to be $z=1$ and one can recover it by transforming coordinates
\begin{align}\label{eq_dim}
\{t,z,w,w_j\}\to \{t,z,w,w_j\}z_h^{-1},\quad j=1,2,...,d-2.
\end{align}
In order to restrict the region between the Planck branes $(0 \leq x \leq 1)$, and outside the event horizon, we transform the coordinates to
\begin{equation}\label{eq_CoordinateTransform}
  w= w_0 x + \left[(\cot \theta_2 +\cot \theta_1) x - \cot \theta_1 \right] z, \qquad z=1-y^2.
\end{equation}
That is, for $x=0$ we have $w+\cot \theta_1 z =0$, while for $x=1$ we have $(w-w_0) -\cot\theta_2 z=0$ (see Appendix~\ref{app_bhewpb} for more discussions). 
With the presence of Planck branes, the geometry is not precisely RN-AdS, and the deformation of the ambient geometry due to the backreaction of Planck branes can be described by introducing the Deturck method \cite{Dias:2015nua}.

Instead of solving (\ref{eq_eineq}) directly, we solve the so-called Einstein-DeTurck equation, which is
\begin{align}\label{eq_EDE}
  R_{\mu\nu}+3 g_{\mu\nu}&=\left(T_{\mu\nu}-\frac{T}{2}g_{\mu\nu}\right)+\nabla_{(\mu}\xi_{\nu)},
\end{align}
where $$\xi^{\mu}:=\left[\Gamma_{\nu \sigma}^{\mu}(g)-\Gamma_{\nu \sigma}^{\mu}(\bar{g})\right] g^{\nu \sigma}$$ is the DeTurck vector and $\bar{g}$ is the reference metric. 
Here $\bar{g}$ is required to satisfy the same boundary conditions as $g$ only on Dirichlet boundaries, but not on Neumann boundaries
\cite{Almheiri:2019psy}.

Now we introduce the metric ansatz in the four-dimensional case and derive the boundary conditions in the doubly holographic setup. 
Since the translational symmetry along the $x$ direction is broken, the most general ansatz of the ambient geometry is 
\begin{align}\label{eq_Background}
  ds^2=&\frac{L^2}{(1-y^2)^2}\left[-y^2 P(y) Q_1 dt^2 +\frac{4 Q_2}{P(y)}dy^2+
  Q_4\left(dx - 2 y Q_3 dy \right)^2+Q_5dw_1^2\right]\\
  A=&y^2\,Q_6\,dt, \label{eq_GaugeField}\\
  P(y)=&2-y^2+(1-y^2)^2-\frac{1}{2}(1-y^2)^3\frac{\mu^2}{L^2},
\end{align}
with $\{Q_1,Q_2,Q_3,Q_4,Q_5,Q_6\}$ being the functions of
$(x,y)$. All the boundary conditions are listed in
Tab.~\ref{tab_BCS}, with the tension on the brane
\begin{equation}\label{eq_tension}
  \alpha_i=\frac{2 \cos\theta_i-\lambda_i \sin^2 \theta_i}{L}.
\end{equation}
(see Appendix~\ref{app_fixalpha} for derivation).
Moreover, the boundary conditions at the
horizon $y=0$ also imply that $Q_1(x,0)=Q_2(x,0)$, which
fixes the temperature of the black hole as
\begin{equation}\label{eq_HawkingTemp4}
  T_h=\frac{6-\mu^2/L^2}{8\pi}. 
\end{equation}
The reference metric $\bar{g}$ is given by $Q_1=Q_2=Q_5=1$,
$$ Q_3=\frac{x \left(\cot \theta_1+\cot \theta_2\right)-\cot\theta_1}{w_0-\left(y^2-1\right)
\left(\cot \theta_1+\cot \theta_2\right)} \quad \text{and} \quad Q_4 = \left(w_0-\left(y^2-1\right)\left(\cot \theta_1+\cot \theta_2\right)\right){}^2. $$

\begin{table}
  \centering
  \begin{tabular}{|c| c| c |c |c| c |c|}
    \hline
      & 1 & 2 & 3 & 4 & 5 & 6 \\
    \hline
    $\mathbf{y=1}$ & $Q_1=1$ & $Q_2=1$ & $Q_3=Z_1(x,1)$ & $Q_4=Z_2(x,1)$ & $Q_5=1$ & $Q_6=\mu$\\
    \hline
    $\mathbf{y=0}$ & $\partial_y Q_1=0$ & $\partial_y Q_2=0$ & $\partial_y Q_3=0$ & $\partial_y Q_4=0$ & $\partial_y Q_5=0$ & $\partial_y Q_6=0$\\
    \hline
    $\mathbf{x=1}$ & $n^\mu F_{\mu \nu} h^\nu{}_i=0$ & $n_\mu \xi^{\mu}=0$ & $Q_3=Z_1(1,y)$ & \multicolumn{3}{c|}{Equation (\ref{eq_BCSonBrane}) at $x=1$}\\
    \hline
    $\mathbf{x=0}$  & $n^\mu F_{\mu \nu} h^\nu{}_i=0$ & $n_\mu \xi^{\mu}=0$ & $Q_3=Z_1(0,y)$ & \multicolumn{3}{c|}{Equation (\ref{eq_BCSonBrane}) at $x=0$}\\
    \hline
  \end{tabular}
  \caption{Boundary conditions, where $Z_1(x,y):=\frac{-\cot \theta_1 + x (\cot \theta_1+\cot \theta_2)}{w_0-\left(y^2-1\right) (\cot \theta_1+\cot \theta_2)}$ and $Z_2(x,y)=\left[w_0-(y^2-1)(\cot \theta_1+\cot \theta_2)\right]^2$.} \label{tab_BCS}
\end{table}

With the metric ansatz (\ref{eq_Background}), the ambient geometry is numerically solved by the Newton-Raphson method. 
Specifically, we discretize the Einstein-DeTurck equations (\ref{eq_EDE}) both on the $x$ and $y$ directions with the Chebyshev Pseudo-spectral method.
 
Distinguishing the DeTurck soliton and the solution of Einstein equations is always important. 
Note that the possible solution of (\ref{eq_EDE}) with $\xi \neq 0 $ is called DeTurck soliton, which is not the solution of the Einstein equations (\ref{eq_eineq}). 
In literature, it is shown that the DeTurck soliton does not exist for static geometries with Dirichlet boundary conditions, while for Neumann boundary conditions, this conclusion has never been proved \cite{Almheiri:2019psy,Figueras:2011va,Figueras:2016nmo}.
Although there is no rigorous proof, the problem we address is well-posed elliptic, where the local uniqueness of solutions is guaranteed.
That is to say, a soliton must be distinguishable from any genuine solution. 
Therefore, in practice, we just need to monitor the gauge vector $\xi$, and make sure it will converge with the increase of the grids -- see Appendix~\ref{app_converge} for details.

\begin{figure}
  \centering
 \subfigure[]{\label{fig_Q1}
  \includegraphics[scale=0.24]{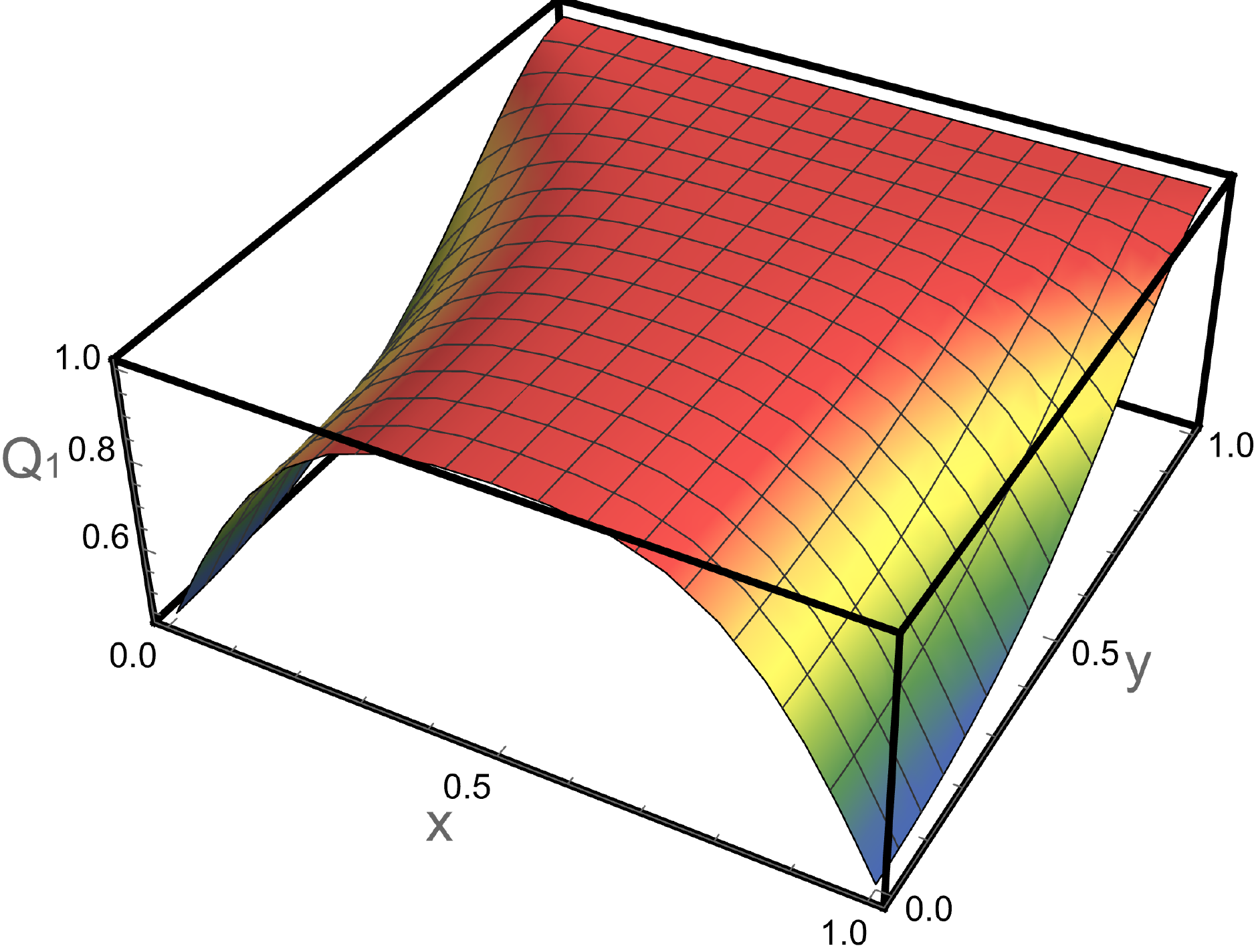}}
  \hspace{0pt}
  \subfigure[]{\label{fig_Q2}
  \includegraphics[scale=0.24]{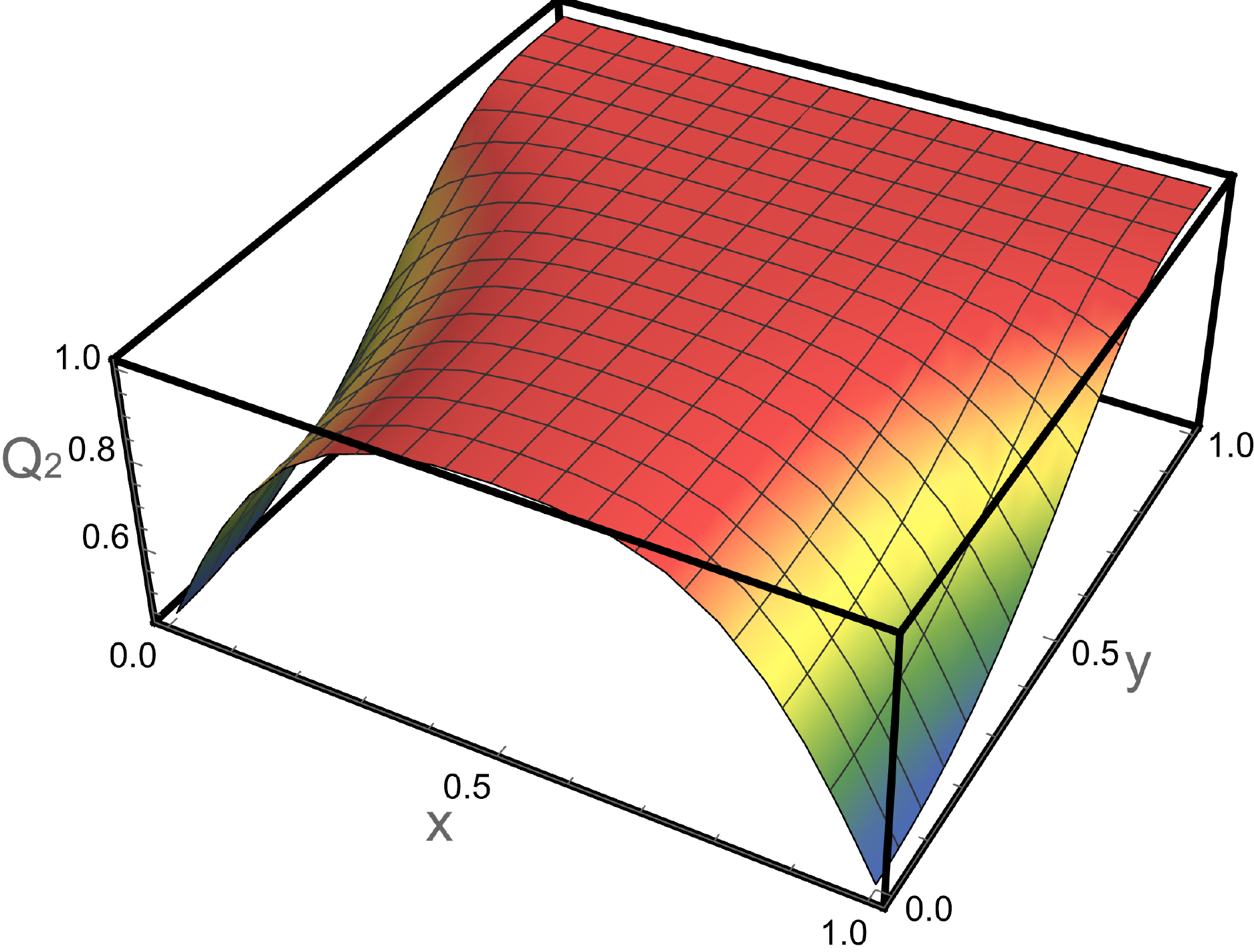}}
  \hspace{0pt}
  \subfigure[]{\label{fig_Q3}
  \includegraphics[scale=0.24]{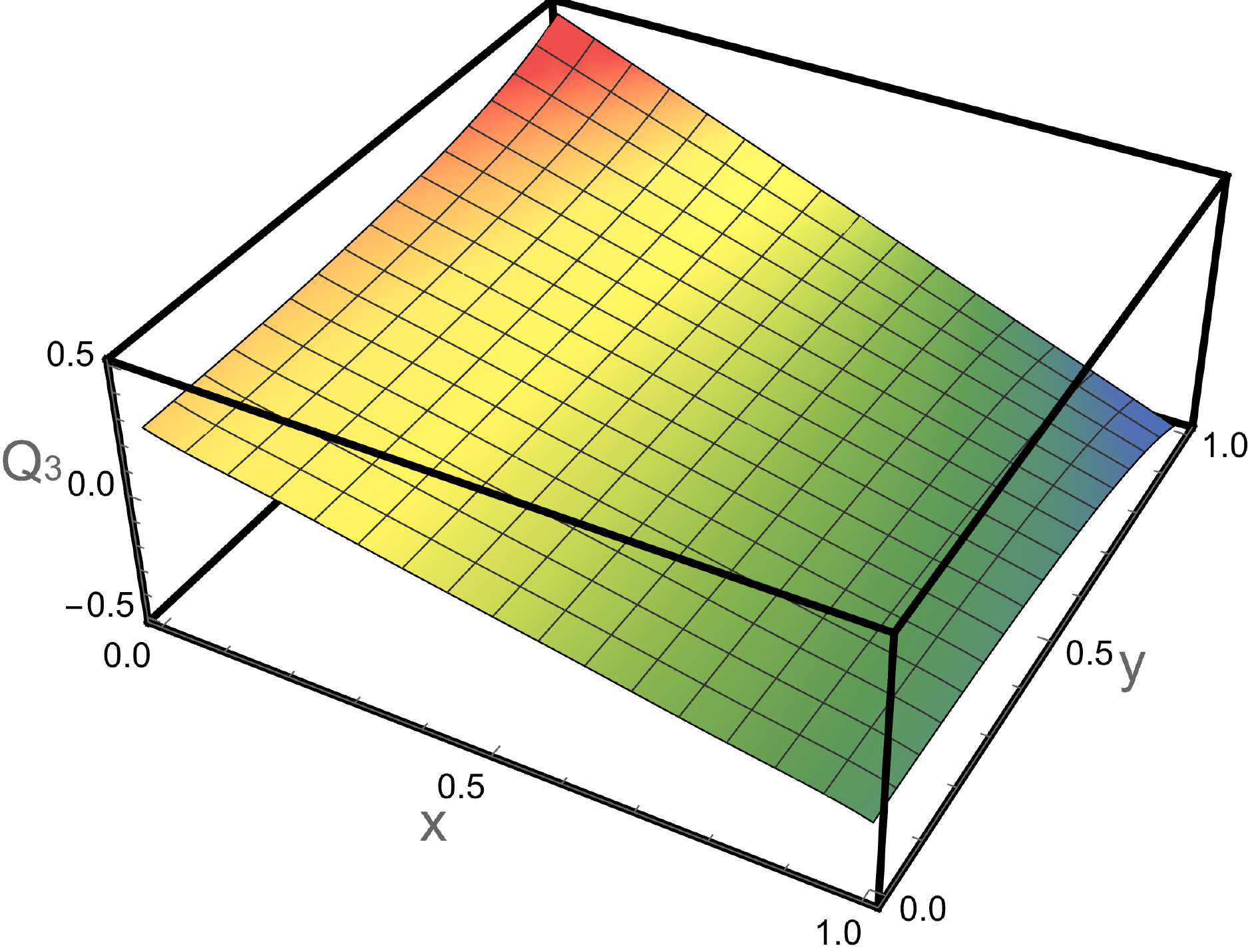}}
  \hspace{0pt}
  \subfigure[]{\label{fig_Q4}
  \includegraphics[scale=0.24]{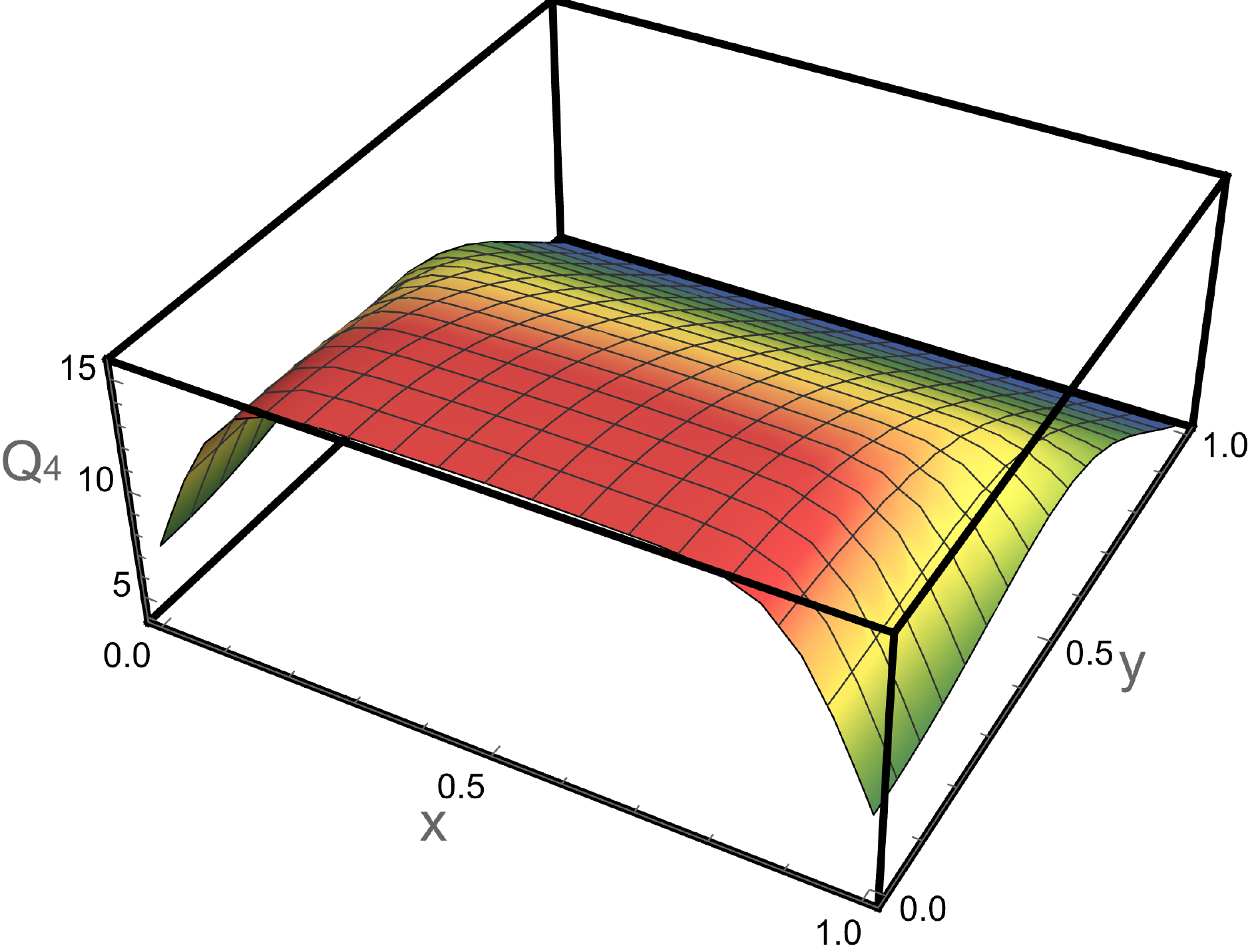}}
  \hspace{0pt}
  \subfigure[]{\label{fig_Q5}
  \includegraphics[scale=0.24]{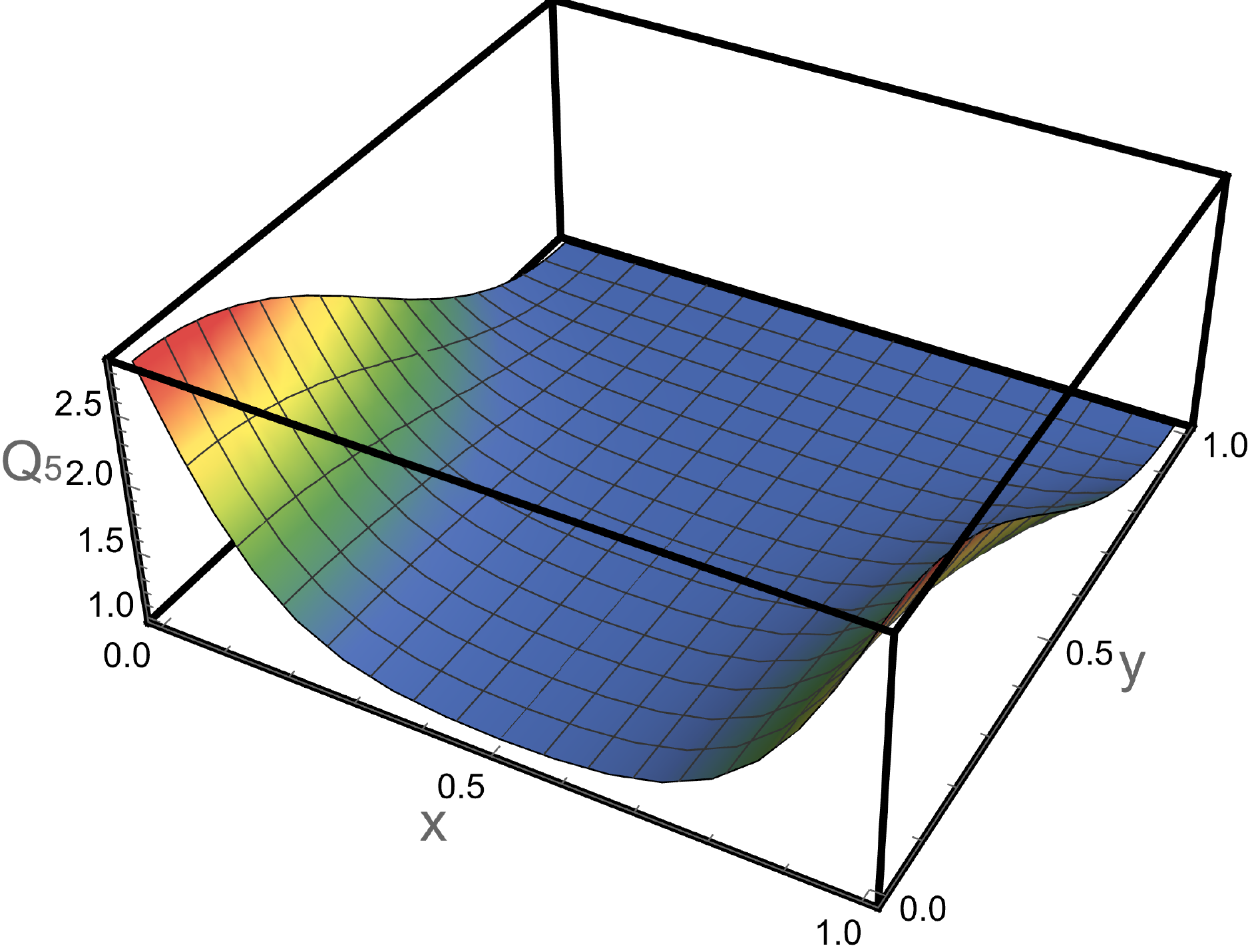}}
  \hspace{0pt}
  \subfigure[]{\label{fig_psi}
  \includegraphics[scale=0.24]{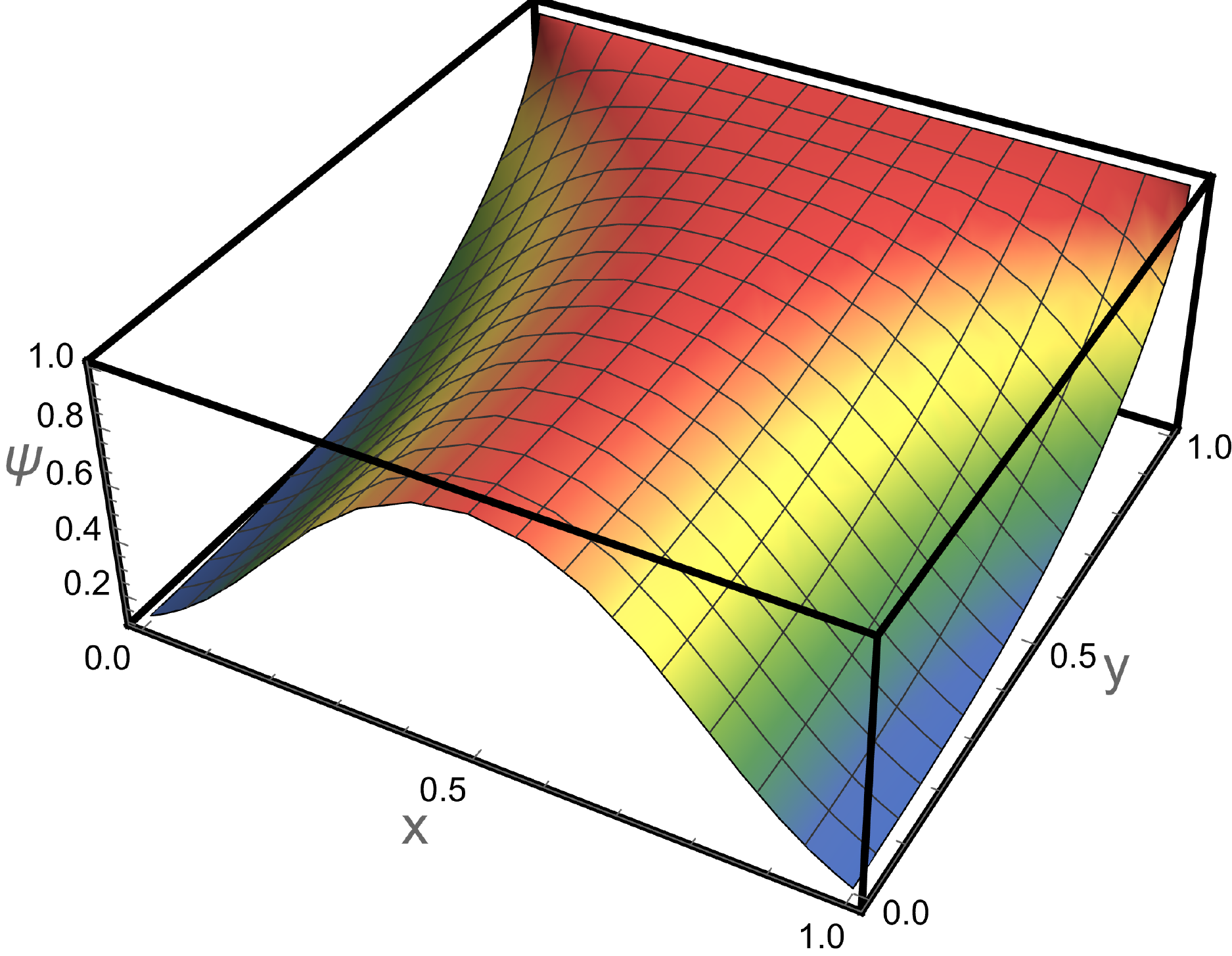}}
\caption{Free parameters are fixed to be $\{\theta_1, \theta_2, w_0, \mu, L\}=\{\pi/4, \pi/4, 2, 1, 1\}$, while $\{Q_1,Q_2,Q_3,Q_4,Q_5,Q_6\}$ as functions of $(x,y)$ are shown from (a) to (f).}\label{fig_DTF}
\end{figure}

Owing to the Neumann boundary conditions on the brane, there is no analytical expression of the metric. See Fig.~\ref{fig_DTF} for a concrete example, where the values of $\{Q_1,Q_2,Q_3,Q_4,Q_5,Q_6\}$ are obviously deviated from those of $\bar{g}$.

\section{The entanglement entropy and mutual information}\label{sec_ee&mi}

\begin{figure}
  \centering
 \subfigure[]{\label{fig_SDH_tv}
  \includegraphics[scale=0.6]{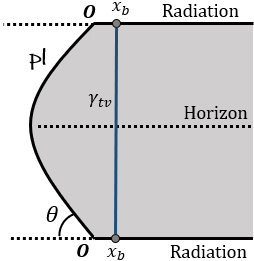}}
  \hspace{40pt}
  \subfigure[]{\label{fig_SDH_pl}
  \includegraphics[scale=0.6]{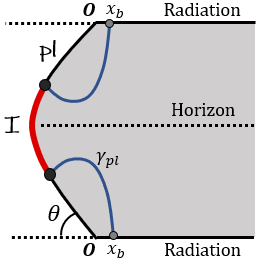}}
\caption{In the ordinary doubly holographic model, (a): no island emerges at early times; (b): an island emerges after Page time.}\label{fig_SDH}
\end{figure}

\subsection{Quantum extremal surfaces and phase structures}
From the brane perspective, the entanglement entropy of the radiation $\mathcal{R}$ is measured by a quantum extremal surface.
Further from the bulk gravity perspective, the QES is equivalently described by a $(d-1)$-dimensional HRT surface and a lower-dimensional area term on the brane $\bm{pl}$ \cite{Almheiri:2019psy,Chen:2020uac,Chen:2020hmv,Hernandez:2020nem}, namely
\begin{align}\label{eq_SR}
S[\mathcal R]=\frac{1}{4G_N^{(d+1)}} \min_{\mathcal I} \left\{\mathop{\text{ext}}\limits_{\mathcal I} \left[\textbf{Area}(\gamma_{\mathcal I\cup\mathcal R}) + \lambda L \,\textbf{Area}(\partial\mathcal I)\right]\right\},
\end{align}
where $\gamma_{\mathcal I\cup\mathcal R}$ is the corresponding HRT surface sharing the boundary with $\mathcal I\cup\mathcal R$.
In the ordinary doubly holographic setup with one Planck brane, there are two possible phases characterized by distinct configurations of HRT surfaces -- Fig.~\ref{fig_SDH}.
The first phase $\gamma_{tv}$ occurs at early times with the absence of an island $\mathcal{I}$, while another phase occurs at late times, and a non-trivial island $\mathcal I$ emerges to keep the entanglement entropy (\ref{eq_QESinRad}) from divergence after the Page time \cite{Almheiri:2019yqk}.

\begin{figure}
  \centering
 \subfigure[]{\label{fig_DDH_tv}
  \includegraphics[scale=0.4]{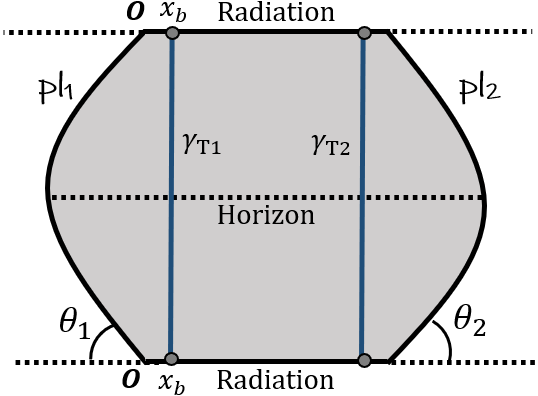}}
  \hspace{40pt}
  \subfigure[]{\label{fig_DDH_mix}
  \includegraphics[scale=0.4]{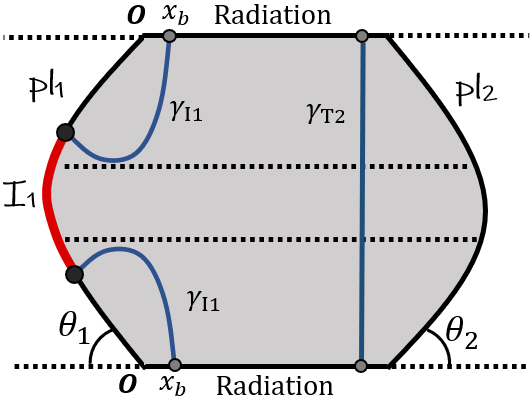}}
  \hspace{40pt}
  \subfigure[]{\label{fig_DDH_pl}
  \includegraphics[scale=0.4]{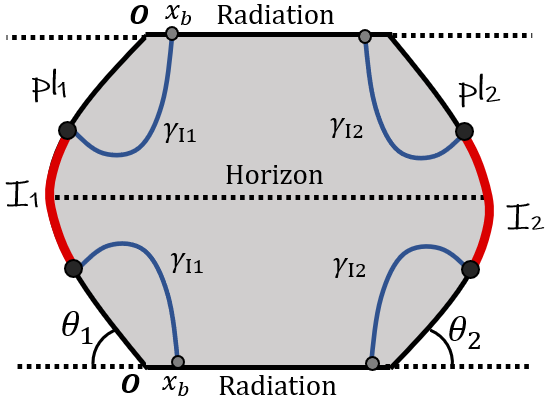}}
  \hspace{40pt}
  \subfigure[]{\label{fig_DDH_rd}
  \includegraphics[scale=0.4]{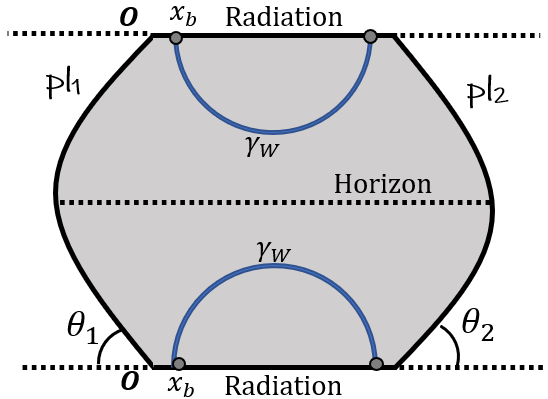}}

\caption{The configurations of HRT surfaces in four possible phases: (a): the trivial phase;
(b): the half-island phase; (c): the island phase; (d): the wormhole phase, with the HRT surfaces being plotted in blue.}\label{fig_DDH}
\end{figure}

For simplicity, we consider $\theta_1 = \theta_2$ and $\lambda_2\geq \lambda_1$ throughout this paper, and the involvement of a second Planck brane renders four possible phases for the configuration of HRT surfaces at most\footnote{Actually, without this assumption, the ambient geometry with two Planck branes will generally lead to five possible phases at most, where the extra phase is similar to Fig.~\ref{fig_DDH_mix}.} -- which are illustrated in Fig.~\ref{fig_DDH}. 

As proposed previously in \cite{Ling:2020laa,Ling:2021vxe}, we will still call the phases as shown in Fig.~\ref{fig_DDH_tv} and \ref{fig_DDH_pl} \textbf{the trivial phase} and \textbf{the island phase} respectively, since the configurations of HRT surfaces are similar to the cases with one Planck brane. 
In addition, it is interesting to notice that the presence of the second Planck brane brings two more possible phases during evolution. 
The first possible phase is called \textbf{the half-island phase} -- Fig.~\ref{fig_DDH_mix}, at which only one island emerges on the brane due to the fewer DOF on it. 
Another possible phase may occur at the saturation of the whole system, which we call it \textbf{the wormhole phase} -- Fig.~\ref{fig_DDH_rd}. 
In this phase, the entanglement wedge connecting $\mathcal{B}_1$ and $\mathcal{B}_2$ appears in the bulk, representing the entanglement between them. We will call it a ``wormhole'' connecting these two black holes, following the idea of $\textbf{ER}=\textbf{EPR}$ conjecture \cite{Maldacena:2013xja} (see also \cite{Dai:2020ffw} for some constraints on experiments). 
The dynamics of the wormhole formation will be analyzed in the next section.

Technically, the endpoints of HRT surfaces should locate near the branes. 
Consider these HRT surfaces anchored at $x=x_b$ and $x=1-x_b$ respectively on the conformal boundaries -- Fig.~\ref{fig_DDH}. They measure the entanglement between two subsystems: one consists of $\mathcal{B}_1$, $\mathcal{B}_2$ and part of baths within the region $ x \in [0, x_b) \cup (1-x_b,1]$, while another consists of the remaining bath with $x \in [x_b, 1-x_b]$. 
Conventionally, we still call the former the black hole subsystem, and the latter the radiation subsystem.

\subsection{The entropy and mutual information in different phases}
From now on, to avoid cumbersome statements we will call the trivial, half-island, island and wormhole phase simply as Ph-\textbf{T}, Ph-\textbf{H}, Ph-\textbf{I}, and Ph-\textbf{W} respectively. 
From (\ref{eq_SR}), the entropy density of the radiation subsystem $\mathcal{R}$ can be determined by
\begin{align}
s[\mathcal{R}]=&\frac{4G_N^{(d+1)} S[\mathcal{R}]}{L^{d-1} V}\nonumber\\  
=&\frac{1}{V}\min_{\mathcal I} \bigg\{\textbf{A}[\gamma_{T_1}]+\textbf{A}[\gamma_{T_2}],\textbf{A}[\gamma_{I_1}]+\textbf{A}[\gamma_{T_2}]+\textbf{A}[\partial\mathcal{I}_1],\nonumber\\
&\textbf{A}[\gamma_{I_1}] + \textbf{A}[\gamma_{I_2}]+\textbf{A}[\partial\mathcal{I}_1]+\textbf{A}[\partial\mathcal{I}_2],\textbf{A}[\gamma_{W}]\bigg\}.\label{eq_sr}
\end{align}
Here $V$ is the infinite volume of the relevant spatial directions. For instance, for $d=3$ we have $V=\int dw_1$. $\textbf{A}[G]$ is the area of the corresponding extremal surface $G$. Moreover, in the second line, we have set $L=1$.
While for each black hole subsystem $\mathcal{B}_i \,(i=1,2)$, the formula of entropy density is the same as that in the ordinary double holography with one Planck brane -- see Fig.~\ref{fig_SDH}, which is
\begin{align}
  s[\mathcal{B}_i]&=\frac{1}{V}\min_{\mathcal I'}\bigg\{ \textbf{A}[\gamma_{T_i}],\textbf{A}[\gamma_{I_i}]+\textbf{A}[\partial\mathcal{I'}_i]\bigg\},\quad i=1,2.\label{eq_sb}
  \end{align}
Moreover, the density of mutual information between any two subsystems $\mathcal{K}$ and $\mathcal{K}'$ can be expressed as
\begin{equation}\label{eq_MI}
  I[\mathcal{K}:\mathcal{K}']=s[\mathcal{K}]+s[\mathcal{K}']-s[\overline{\mathcal{K}\cup\mathcal{K}'}],
\end{equation}
with $\mathcal{K}\in\{\mathcal{B}_1,\mathcal{B}_2,\mathcal{R}\}$. For simplicity, all these quantities are dimensionless. Following the transformation (\ref{eq_dim}), their dimensions can be recovered as 
\begin{equation}\label{eq_dim2}
\left\{w_0,T_h,\mu,V,s[\mathcal{K}],I[\mathcal{K}:\mathcal{K}']\right\} \to \left\{w_0/z_h,T_h z_h,\mu z_h,V/ z_h,s[\mathcal{K}]z_h,I[\mathcal{K}:\mathcal{K}'] z_h\right\}.
\end{equation} 

\begin{figure}
  \centering
 \subfigure[]{\label{fig_PhaseT}
  \includegraphics[scale=0.33]{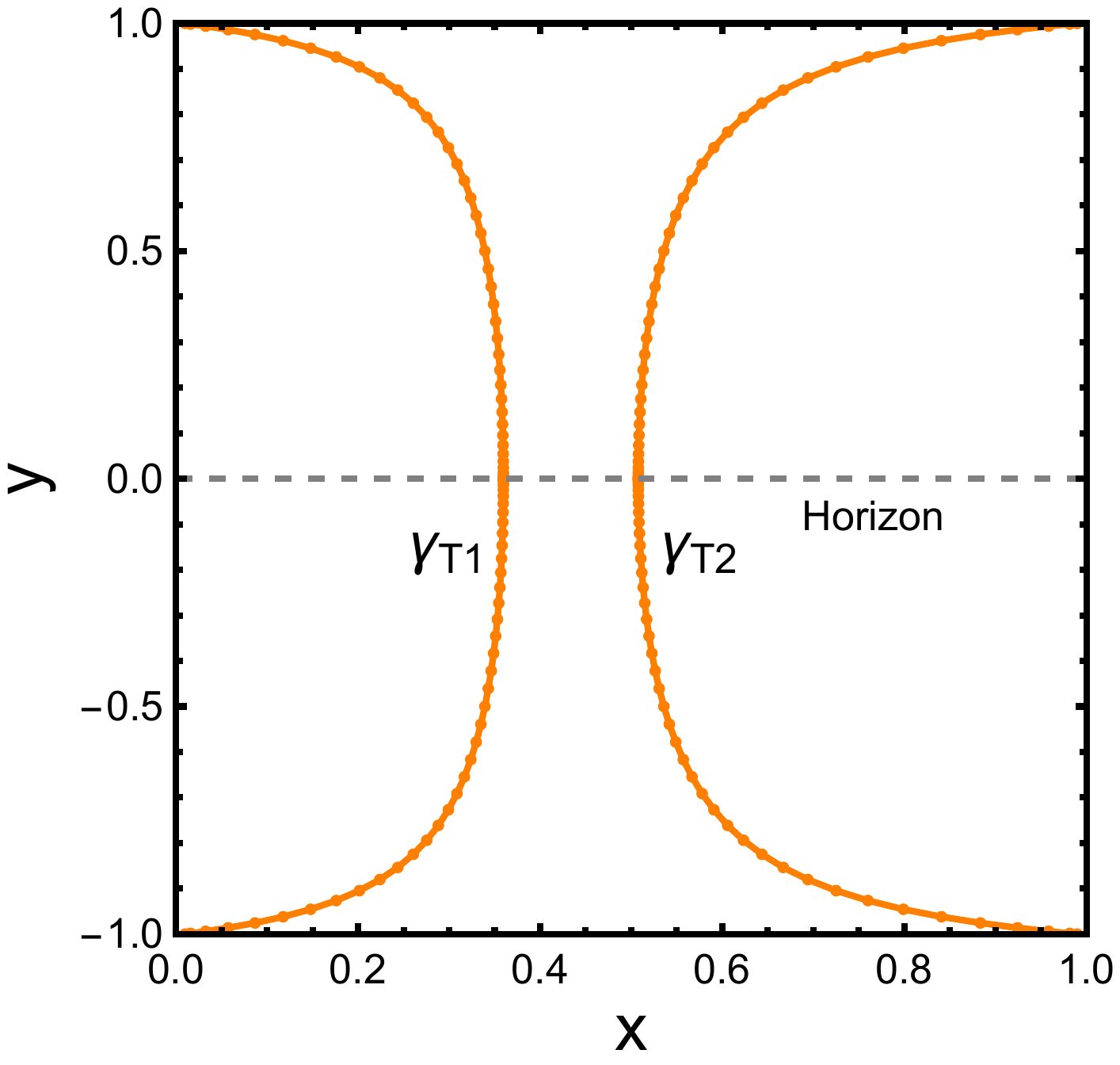}}
  \hspace{0pt}
  \subfigure[]{\label{fig_PhaseI}
  \includegraphics[scale=0.33]{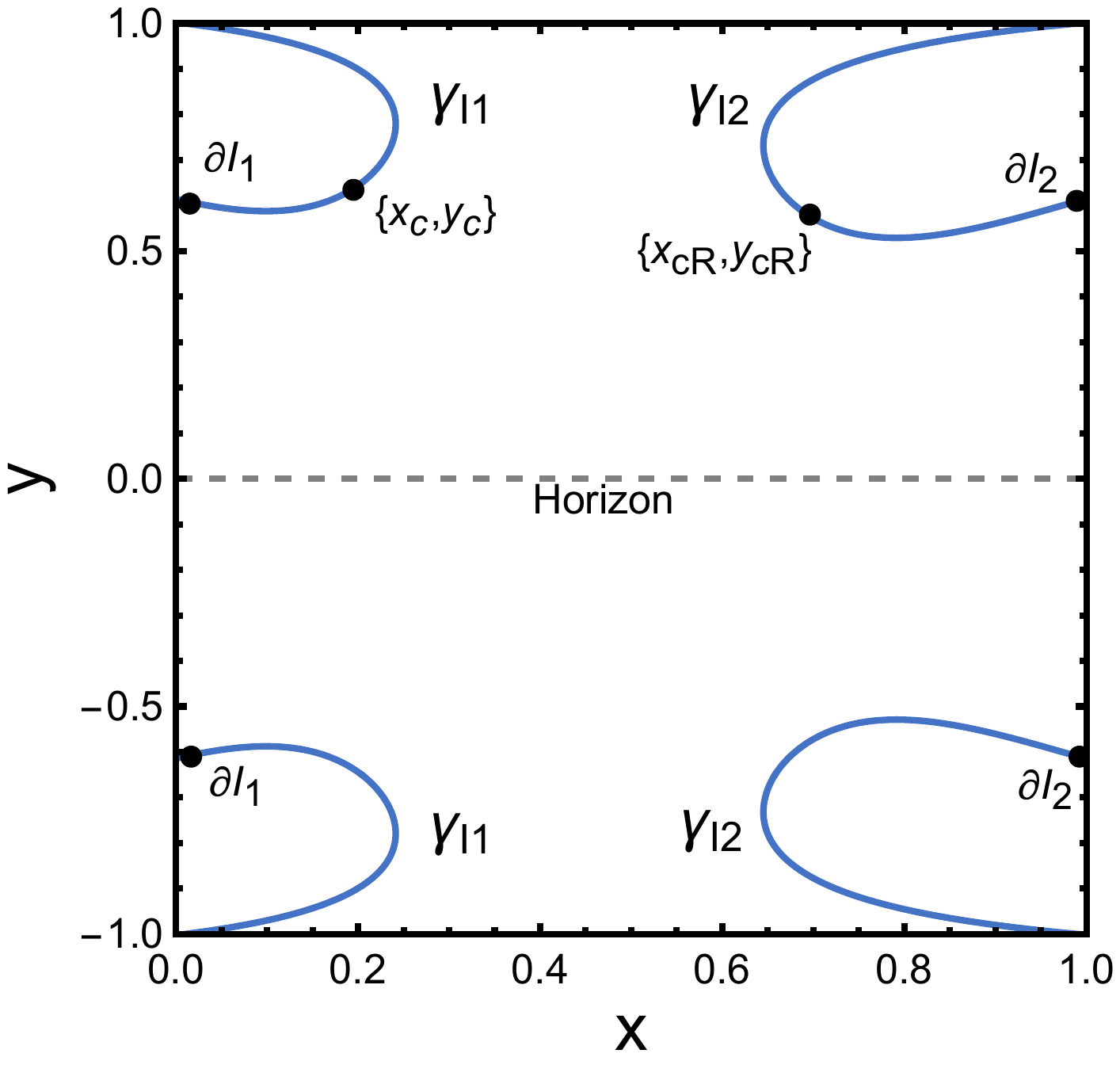}}
  \hspace{0pt}
  \subfigure[]{\label{fig_PhaseW}
  \includegraphics[scale=0.33]{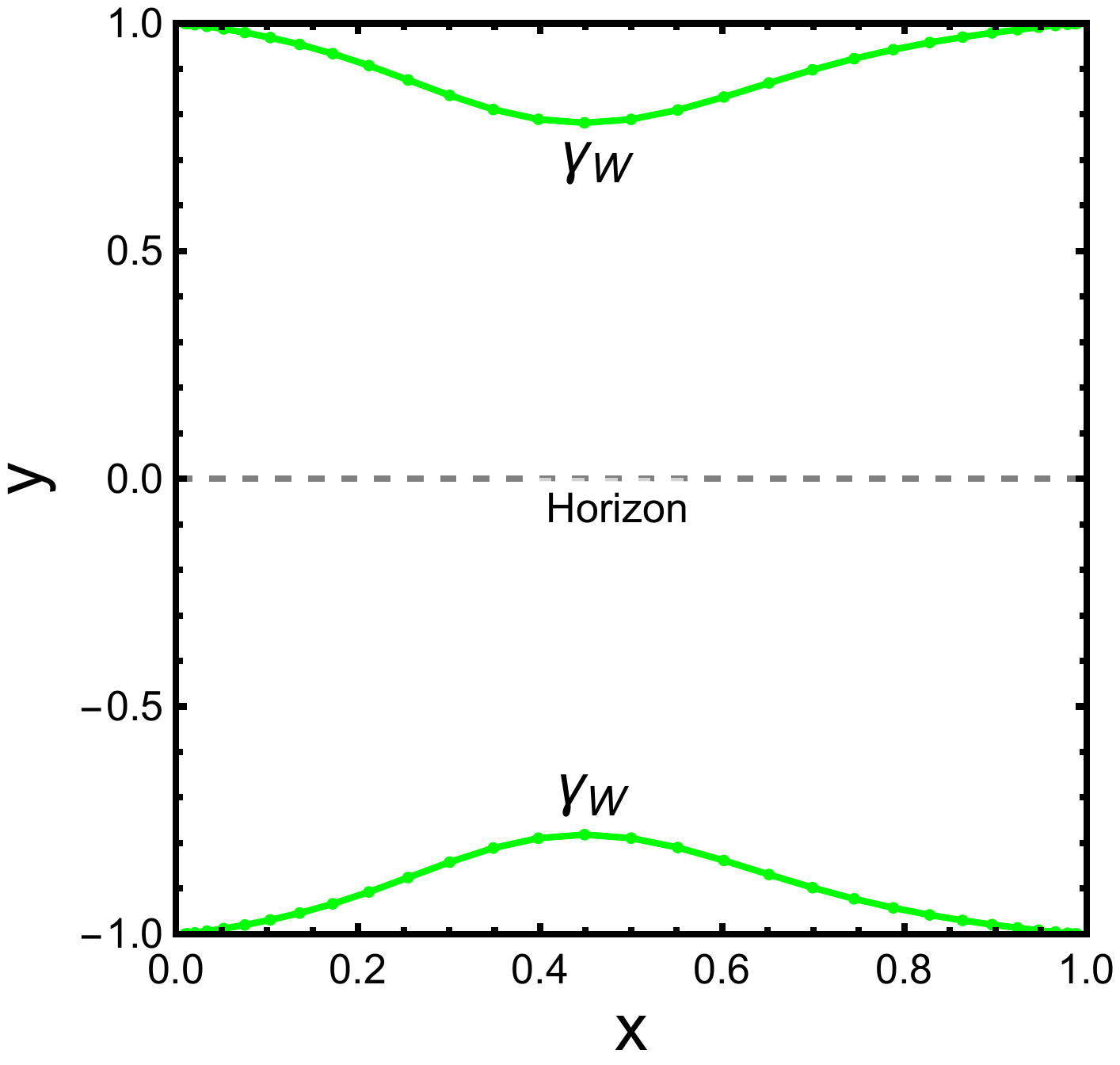}}

\caption{The extremal surface in (a): trivial phase, (b): island phase and (c): wormhole phase with $\{L,\theta_1,\theta_2,\lambda_1,\lambda_2,\mu,w_0\}=\{1,\pi/4,\pi/5,1,1,1,1/2\}$. Note that the minus sign in $y$-direction represents another asymptotic region, and is just convenient for drawing figures.}\label{fig_Phase}
\end{figure}

Next, we will derive the expressions for densities $\textbf{A}/V$ in each phase in parallel. 
First, for the extremal surface $\gamma_{T_{i}}$ at $t=0$, we introduce the parameterization $x=x_{i}(y)\, (i=1,2)$ -- see Fig.~\ref{fig_PhaseT} for a concrete example, which leads to the corresponding density
\begin{equation}\label{eq_Atv}
  \frac{\textbf{A}[\gamma_{T_{i}}]}{2 V}\Bigg|_{t=0}=  \int_0^1 \frac{dy}{(1-y^2)^2}\sqrt{Q_5\left(4\frac{Q_2}{P(y)}+Q^4(2y Q_3+x'_{i}(y))^2\right)}.
\end{equation}

 Then, for $\gamma_{I_{i}}$ together with $\partial\mathcal{I}_{i}\, (i=1,2)$, we introduce two different ways of parameterization in different intervals. 
 In the $(x,y)$ plane, for the curve in $y \in [y_c,1]$, we introduce $x=x_{i}(y)$ just as the parameterization of $\gamma_{T_i}$, while for the curve in $x \in [x_c,1]$, we introduce $y=y_{i}(x)$ instead, with $y'_{i}(x_c)=x'_{i}(y_c)^{-1}$ -- see Fig.~\ref{fig_PhaseI}. 
 Note that the anchoring point $y_{1}(0)=y_{I_{1}}$ on the brane can be read off from the integration procedure. Thus, the surface with the extremal area can be found by running the anchoring point all over the brane. 
 Subsequently, the corresponding density associated with that anchoring point can be expressed by
\begin{align}
  \frac{\textbf{A}[\gamma_{I_{1}}]}{2V} =&  \int_{y_c}^1 \frac{dy}{(1-y^2)^2}\sqrt{Q_5\left(4\frac{Q_2}{P(y)}+Q^4(2y Q_3+x'_{1}(y))^2\right)}
  \nonumber\\
   &+\int_{0}^{x_c} \frac{dx}{\left(y_{1}(x)^2-1\right)^2}\sqrt{Q_5 \left(\frac{4 Q_2 y'_{1}(x)^2}{P(y_{1}(x))}+Q_4 \left(1+2 y_{1}(x) Q_3 y'_{1}(x)\right)^2\right)}.\label{eq_Apl1}\\
   \frac{\textbf{A}[\partial\mathcal{I}_{1}]}{2V}=&
   \frac{\lambda_{1} \sqrt{Q_5}}{1-y_{I_1}^2}\Bigg|_{x=0}.\label{eq_Area1}
 \end{align}
 Similarly, for $\gamma_{I_{2}}$ anchored at $y_{2}(1)=y_{I_{2}}$, we have
 \begin{align}
  \frac{\textbf{A}[\gamma_{I_{2}}]}{2V} =&  \int_{y_{cR}}^1 \frac{dy}{(1-y^2)^2}\sqrt{Q_5\left(4\frac{Q_2}{P(y)}+Q^4(2y Q_3+x'_{2}(y))^2\right)}
  \nonumber\\
   &+\int_{x_{cR}}^{1} \frac{dx}{\left(y_{2}(x)^2-1\right)^2}\sqrt{Q_5 \left(\frac{4 Q_2 y'_{2}(x)^2}{P(y_{2}(x))}+Q_4 \left(1+2 y_{2}(x) Q_3 y'_{2}(x)\right)^2\right)}.\label{eq_Apl2}\\
   \frac{\textbf{A}[\partial\mathcal{I}_{2}]}{2V}=&
   \frac{\lambda_{2} \sqrt{Q_5}}{1-y_{I_2}^2}\Bigg|_{x=1}.\label{eq_Area2}
 \end{align}

 Finally, for the surface $\gamma_{W}$, we introduce the parameterization $y=y(x)$-- Fig.~\ref{fig_PhaseW}, which leads to the corresponding density
 \begin{equation}\label{eq_Ard}
   \frac{\textbf{A}[\gamma_{W}]}{2V}=  \int_{x_b}^{1-x_b} \frac{dx}{\left(y(x)^2-1\right)^2}\sqrt{Q_5 \left(\frac{4 Q_2 y'(x)^2}{P(y(x))}+Q_4 \left(1+2 y(x) Q_3 y'(x)\right)^2\right)}.
 \end{equation}

With these expressions at hand, we are now ready to investigate the entanglement properties of the system.
In the next section, we will elaborate on the description of the entanglement entropy and mutual information in the probe limit, which renders complicated phase structures. Further, we will take the backreaction of branes into account and study its effect on both the entanglement properties and phase structures.

\section{Page curves and entanglement phase structures}\label{sec_pc&eps}
Page curve plays a significant role in understanding the black hole information paradox.
In our case, there are also Page curves caused by exchanging hawking modes of separate black holes.
However, usually, it is difficult to obtain the Page curve in the doubly holographic model since the backreaction of the brane to the ambient geometry is very hard to solve such that one lacks the data inside the event horizon during evolution \cite{Ling:2020laa}. 

Fortunately, this difficulty can be circumvented by considering the special case with the probe limit \cite{Ling:2020laa}. 
In such cases, Planck branes $\bm{pl}_i$ apply no backreaction to the ambient geometry. 
This allows us to consider the entanglement properties with respect to time. 
The strategy to calculate the Page curves is as follows: first, we analytically calculate the growth rate of entropy densities, and then numerically find the saturation values of the Page curves by properly choosing the distance $w_0$ between $\mathcal{B}_1$ and $\mathcal{B}_2$. 
As a result, the full Page curves can be obtained by this hybrid method. 

The upshot is that the unitary evolution can be preserved not only by the emergence of islands but also by the formation of a wormhole. 
In the next subsection, we will first calculate the growth rates of quantum information measures such as entanglement entropy and mutual information.

\subsection{The growth of entropy and mutual information densities}
In the probe limit, i.e. $|\pi/2- \theta_i | \ll 1$ together with $\lambda_i \ll 1$,  we have $$g \to \bar{g},\qquad Q_6 \to  \mu,$$ and hence, the corresponding ambient geometry in any dimensions can be treated as the standard RN-AdS black holes as shown in (\ref{eq_RNBH}) with $0\leq w \leq w_0$. Now we obtain the data of the geometry inside the horizon, which allows us to keep track of the growth of $\gamma_{T_i}$ with time. 
To see this, we firstly express the density functional of $\gamma_{T_i}$ in the Eddington-Finkelstein coordinates $\{v,z,w,w_j\}$ as
\begin{equation}\label{eq_TBint}
  \frac{\textbf{A}[\gamma_{T_{i}}]}{V}(t)=\int \frac{ d\Xi_i}{z_i(\Xi_i)^{d-1}}\sqrt{-v_i'(\Xi_i ) \left[f(z_i(\Xi_i )) v_i'(\Xi_i )+2 z_i'(\Xi_i)\right]}, \qquad i=1,2,
\end{equation}
where $\Xi_i$ is the intrinsic parameter on $\gamma_{T_i}$, and the corresponding time on the boundary system is $$t=v+\int \frac{dz}{f(z)}.$$

Due to the cyclic coordinate $v$ in (\ref{eq_TBint}), the integral of motion can be obtained as
\begin{equation}\label{eq_C}
  C=\frac{f(z) v'+z'}{z^{d-1}\sqrt{-v' \left[f(z) v'+2 z'\right]}}.
\end{equation}
Furthermore, since the action in (\ref{eq_TBint}) is
invariant under reparametrizations, we can
 freely choose the integrand as
\begin{equation}\label{eq_C2}
  \sqrt{-v' \left[f(z) v'+2 z'\right]}=z^{d-1}.
\end{equation}
We subsequently substitute both (\ref{eq_C}) and (\ref{eq_C2}) into
(\ref{eq_TBint}), and then the result is 
\begin{align}
  \frac{d}{dt}\frac{\textbf{A}[\gamma_{T_{i}}]}{V}&= \frac{\sqrt{-f(z_{max})}}{z_{max}^{d-1}}, \qquad i=1,2,\label{eq_GroAre}\\
  t&=\int_0^{z_{max}} dz \frac{C z^{d-1}}{f(z)\sqrt{f(z)+C^2z^{2d-2}}}\label{eq_time}.
\end{align}
Here $z_{max}$ denotes the turning point of the trivial surface $\gamma_{T_i}$ and the derivation is presented in Appendix~\ref{app_GrowthRate}. 
Moreover, the relation between $z_{max}$ and the integral of motion $C$ is
given by
\begin{equation}\label{eq_ConQua}
  f(z_{max})+C^2 z_{max}^{2d-2}=0.
\end{equation}

At late times, the extremal surface $\gamma_{T_i}$ tends to surround a special slice, with $z_{max}=z_M$
\cite{Hartman:2013qma}. 
Define
\begin{equation}
  F(z):=\frac{\sqrt{-f(z)}}{z^{d-1}},
\end{equation}
it is easy to show that $C^2=F(z_{max})^2$ will keep growing until approaching
an extremum at $z_{max}=z_M$, where we have the following relation
  \begin{equation}\label{eq_Fp}
    F'(z_M) = (1-d) z_M^{-d}
    \sqrt{-f\left(z_M\right)}-\frac{z_M^{1-d
    } f'\left(z_M\right)}{2
    \sqrt{-f\left(z_M\right)}} = 0.
  \end{equation}
By substituting the solution of (\ref{eq_Fp}) into (\ref{eq_GroAre}), we finally get the growth rates at late times as
\begin{equation}\label{eq_GroRat}
  \lim_{t\rightarrow \infty} \frac{d}{dt}\frac{\textbf{A}[\gamma_{T_{i}}]}{V} =  F(z_M), \qquad i=1,2.
\end{equation}

As a result, the growth rate of the entropy density of $\mathcal{R}$ at late times is
\begin{equation}
  \lim_{t\rightarrow \infty} \frac{d}{dt}s[\mathcal{R}]=\left\{
\begin{array}{lcl}
2 F(z_M),       &      & { \text{Ph-\textbf{T}},}\\
F(z_M) ,        &      & {\text{Ph-\textbf{H}},}\\
0  ,            &      & {\text{Ph-\textbf{I} and Ph-\textbf{W}}.}
\end{array} \right.
  \end{equation}

Note that the first is always twice as much as the second since the entropy on one of the branes reaches saturation in Ph-\textbf{H}. Specifically, after recovering the dimension, for the neutral cases where $\mu \to 0$, we have
\begin{equation}\label{eq_ntc}
  \lim_{t\rightarrow \infty \atop \mu \to 0} \frac{d}{dt}s[\mathcal{R}] = a b_n T_h^{d-1},
\end{equation}
where $b_n=2^{2d+\frac{1}{d}-3}\pi^{d-1}d^{\frac{3}{2}-d}(-1+d)^{\frac{1-d}{d}}(-2+d)^{\frac{d-2}{2d}}$. 
Moreover, $a=2$ if the system is in Ph-\textbf{T}, $a=1$ if the system is in Ph-\textbf{H}, and $a=0$ if the system is in Ph-\textbf{I} or Ph-\textbf{W}. 
While for the extremal cases where $T_h \to 0$, we have
\begin{align}\label{eq_cgc}
  \lim_{t\rightarrow \infty \atop T_h \to 0} \frac{d}{dt}s[\mathcal{R}] = a b_c T_h \mu^{d-2}.
\end{align}
Here $b_c=2\pi \sqrt{\frac{1}{d(d-1)}} \left[\frac{d(d-1)}{(d-2)^2}\right]^{\frac{2-d}{2}}$ and the linear behavior with respect to $T_h$ indicates that the near horizon geometry of the near-extremal black hole is AdS$_2\times R^{d-1}$ spacetime. 
In this case, the evolution is nearly frozen due to the low temperature and the entropy density barely grows.

Further in Ph-\textbf{W}, the entropy density of $\mathcal{R}$ (\ref{eq_sr}) saturates, while the mutual information density (\ref{eq_MI}) between $\mathcal{B}_1$ and $\mathcal{B}_2$ might not. 
Similarly, the late-time growth of the mutual information density can be expressed as
\begin{equation}
  \lim_{t\rightarrow \infty} \frac{d}{dt}I[\mathcal{B}_1 : \mathcal{B}_2]=\left\{
\begin{array}{lcl}
2 F(z_M),       &      & { \text{both}\;s[\mathcal{B}_i] \, (i=1,2) \; \text{are growing,}}\\
F(z_M) ,    &      & {\text{only} \;s[\mathcal{B}_1] \;\text{is growing,}}\\
0  ,    &      & {\text{both}\;s[\mathcal{B}_i] \, (i=1,2) \; \text{are saturated.}}
\end{array} \right.
  \end{equation}
Hence, the linear growth of mutual information density is similar to that of entanglement entropy density as shown in (\ref{eq_ntc}) and (\ref{eq_cgc}).

The similar derivations for remaining $s[\mathcal{K}]$ and $I[\mathcal{K} : \mathcal{K'}]$ will not present here. 
These quantum information measures are crucial for analyzing the evolution process including the formation of wormholes. 

\subsection{Dynamical evolution and phase structures in the probe limit}\label{sec_pagecurves}

In this subsection, we will elaborate on the entanglement phase structures during evolution. 
First, we investigate the cases with equal-sized black holes, where $\mathcal{B}_1$ and $\mathcal{B}_2$ contain the same DOF, which is simple but enough to generate a wormhole.
Then, we study the cases with different-sized black holes, where the smaller (larger) black hole $\mathcal{B}_1$($\mathcal{B}_2$) contains fewer(more) DOF.

For numerical convenience, here we continue to work with the coordinate set $\{t,y,x,w_1\}$ in $4$-dimensional spacetime. In addition, the UV cut-off near the conformal boundary is fixed to be $\epsilon=1-1/100$. We also use scale-free parameters in the following discussion, such as $\{\tilde{w}_0,\tilde{t},\tilde{T}_h,\tilde{\mu},\tilde{s}[\mathcal{R}],\tilde{I}[\mathcal{B}_1 : \mathcal{B}_2]\}=\{w_0/w_b, t/w_b, T_h w_b, \mu w_b, s[\mathcal{R}] w_b, I[\mathcal{B}_1 : \mathcal{B}_2] w_b\}$, where $w_b=w_0 x_b$.

\begin{figure}
  \centering
  \subfigure[]{\label{fig_PD3dS}
  \includegraphics[width=0.4\linewidth]{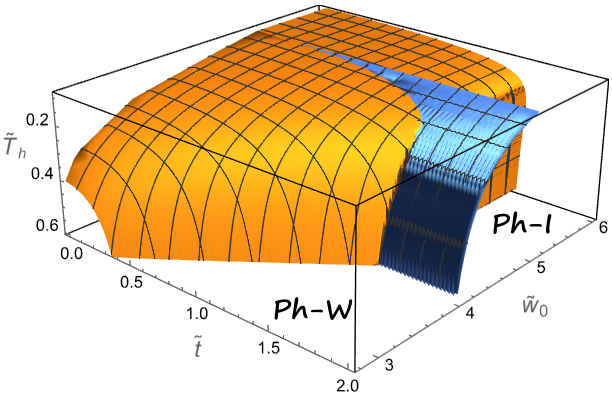}}\\
  \hspace{0pt}
  \subfigure[]{\label{fig_PD2ds2}
   \includegraphics[width=0.31\linewidth]{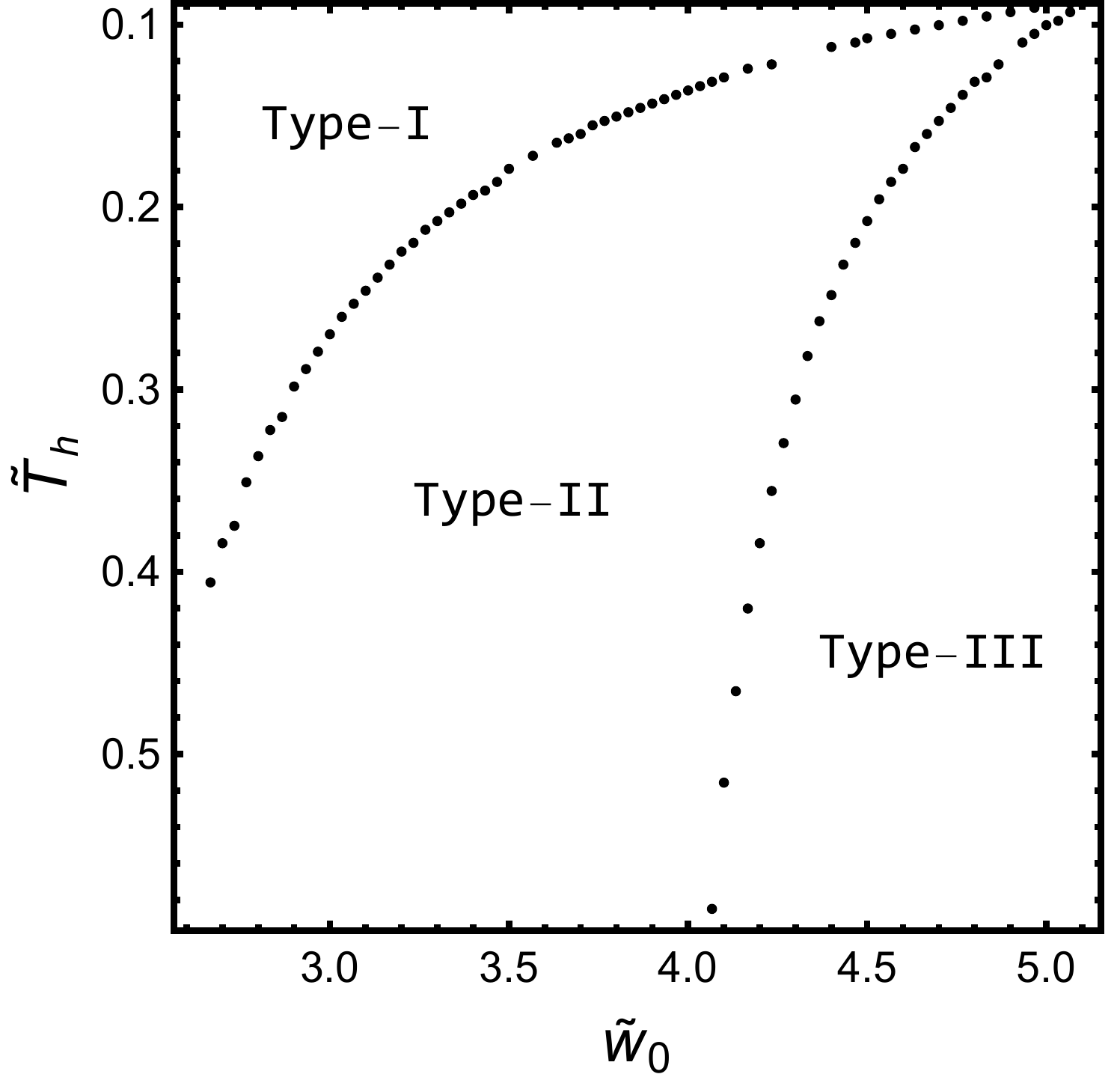}}
  \hspace{0pt}
  \subfigure[]{\label{fig_PD2dS1}
  \includegraphics[width=0.3\linewidth]{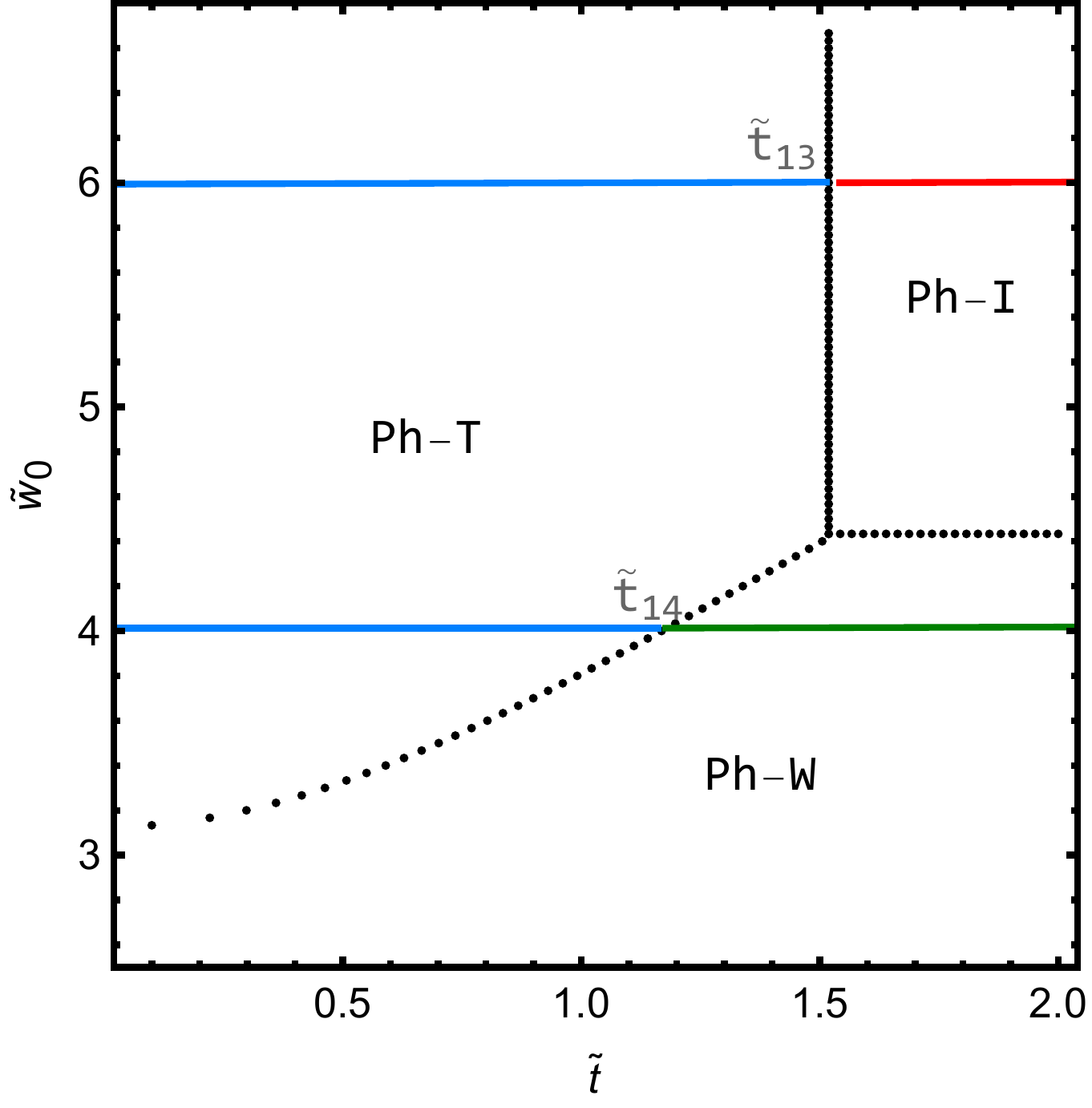}}
    \hspace{0pt}
  \subfigure[]{\label{fig_PCI1}
  \includegraphics[width=0.3\linewidth]{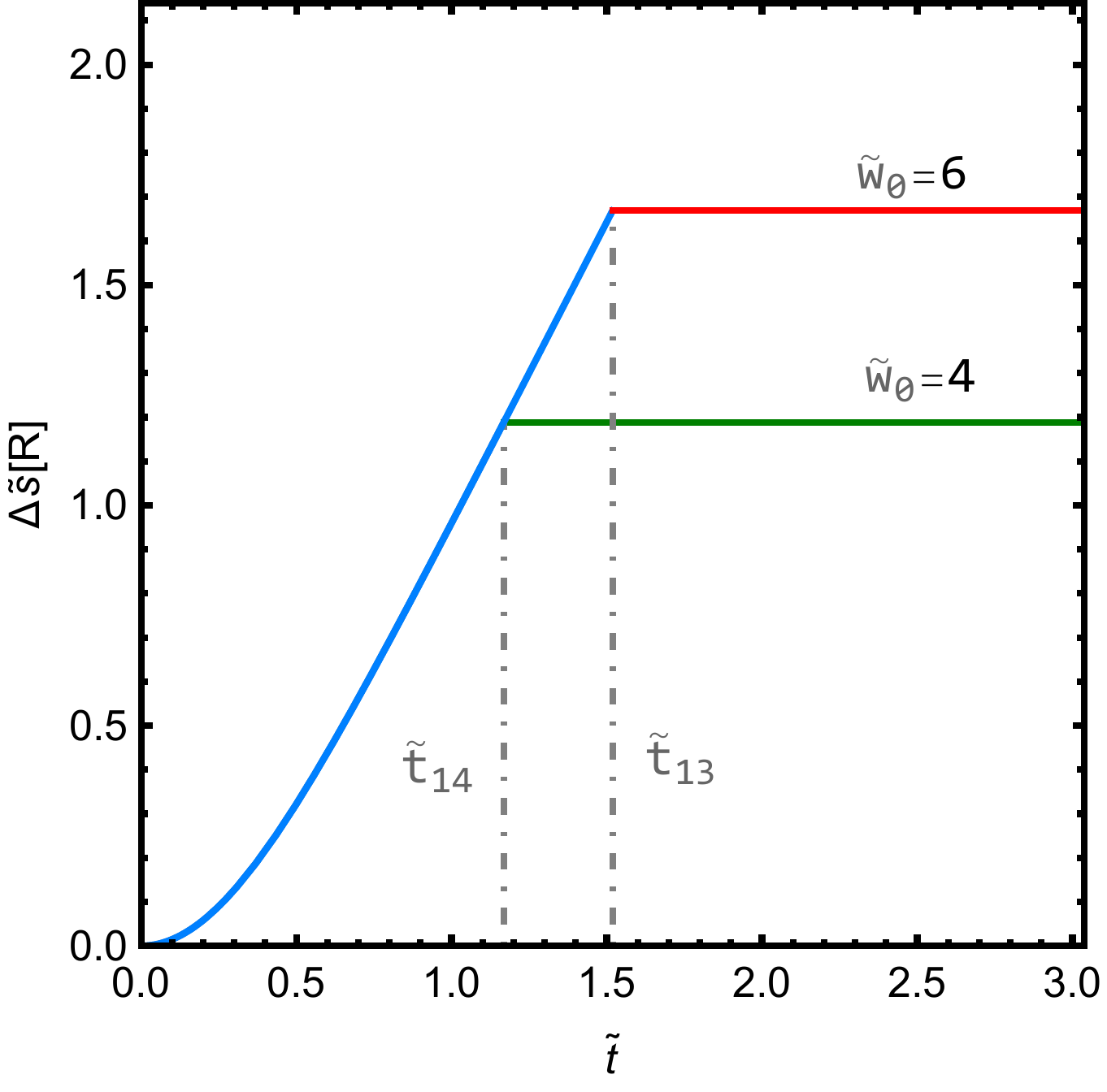}}
\caption{(a): The phase diagram in $\{\tilde{w}_0,\tilde{t},\tilde{T}_h\}$. The region inside the orange surface is in Ph-\textbf{T}. The region outside the orange surface is divided by the blue surface into Ph-\textbf{W} and Ph-\textbf{I}.
(b): The different types of evolution, with $\tilde{w}_{c1}(\tilde{T}_h)$ and $\tilde{w}_{c2}(\tilde{T}_h)$ being two critical distances. (c): For $\tilde{T}_h=6/8\pi$, the projected phase diagram in $\{\tilde{w}_0,\tilde{t}\}$. (d): Two Page curves for radiation $\mathcal{R}$, with Page time $\tilde{t}=\tilde{t}_{14}$ and $\tilde{t}=\tilde{t}_{13}$ respectively. Segments in blue, green and red, represent the combined system is in Ph-\textbf{T}, \textbf{W} and \textbf{I}, respectively.}\label{fig_TransitionPhaseS}
\end{figure}

\begin{figure}
  \centering
  \subfigure[]{\label{fig_GS0S}
  \includegraphics[width=0.3\linewidth]{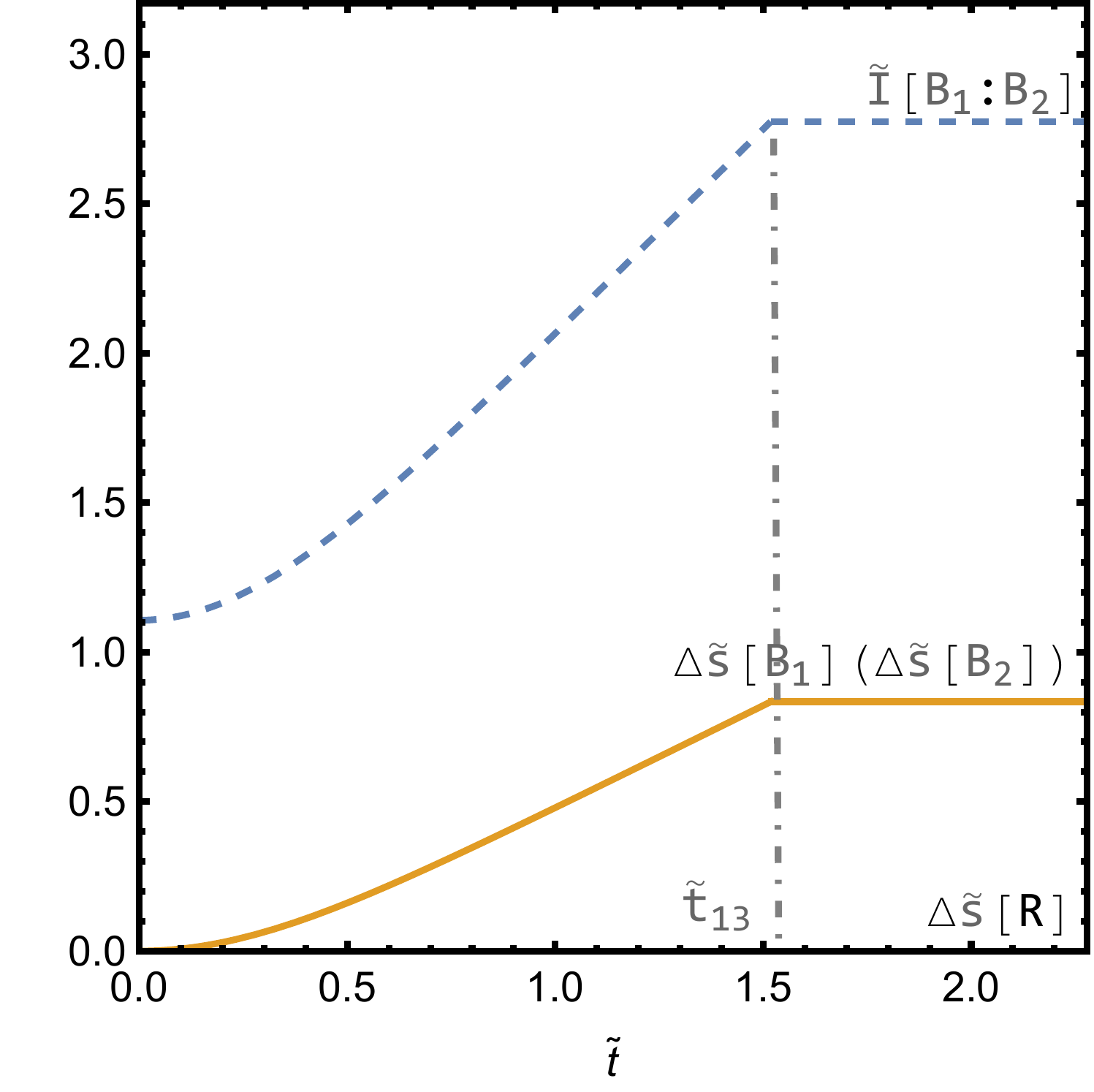}}
  \hspace{0pt}
 \subfigure[]{\label{fig_GS1S}
  \includegraphics[width=0.3\linewidth]{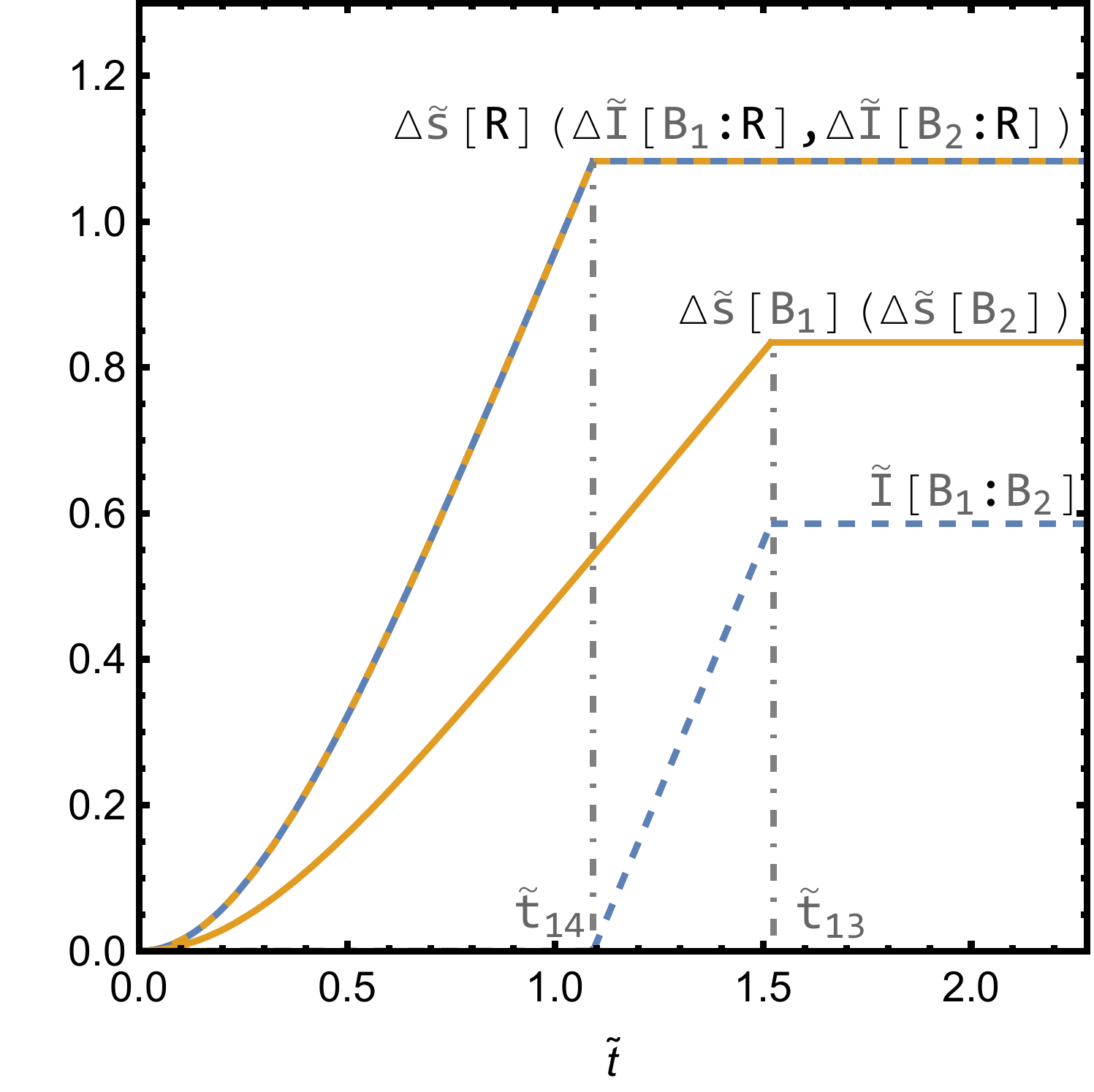}}
  \hspace{0pt}
  \subfigure[]{\label{fig_GS2S}
  \includegraphics[width=0.3\linewidth]{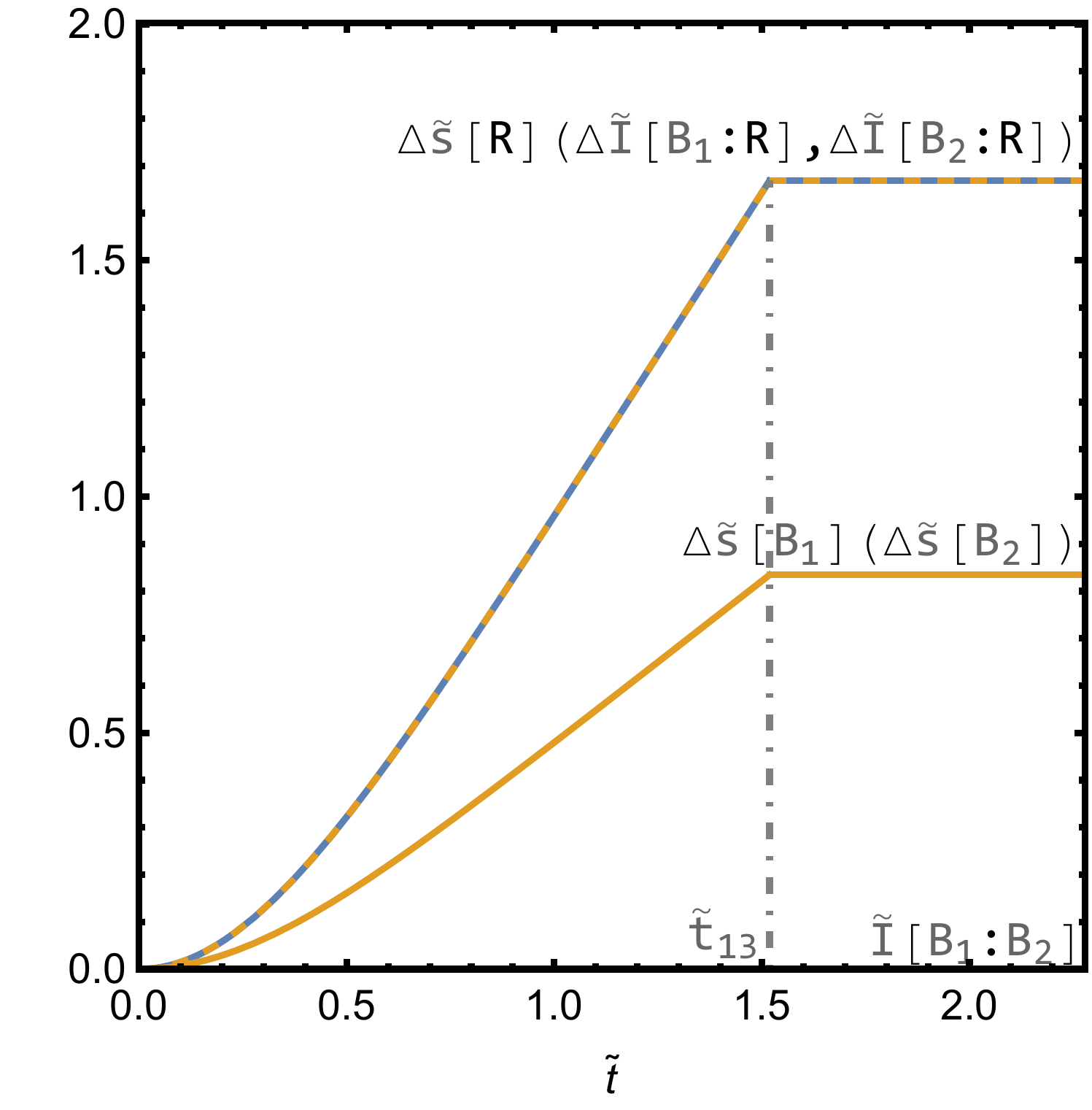}}
\caption{The evolution on entropy and mutual information density at different distances (a): $\tilde{w}_0=2.67$, (b): $\tilde{w}_0=3.91$, (c): $\tilde{w}_0=6.67$, while other parameters are specified to $\{L,\theta_1,\theta_2,\lambda_1,\lambda_2,\tilde{\mu},\tilde{T}_h\}=\{1,\pi/2,\pi/2,0.3,0.3,0,6/8\pi\}$. Note that $\Delta \tilde{\mathcal{Q}}:=\tilde{\mathcal{Q}}(t)-\tilde{\mathcal{Q}}(0)$, where $\mathcal{Q}= s,\; I$.}\label{fig_GrowthRateSS}
\end{figure}

\subsubsection{Black holes of equal size}
Different phases are undergone by the combined system during evolution -- Fig.~\ref{fig_TransitionPhaseS}, at different distances $\tilde{w}_0$ between $\mathcal{B}_1$ and $\mathcal{B}_2$ as well as temperature $\tilde{T}_h$. As shown in Fig.~\ref{fig_PD3dS}, there are two critical surfaces in orange and blue, which separate Ph-\textbf{T}, \textbf{I} and \textbf{W}. The reason for no Ph-\textbf{H} is that the black hole subsystems with the same DOF are saturated at the same time. 

For instance, at the beginning of evolution, the hot systems with subsystems $\mathcal{B}_1$ and $\mathcal{B}_2$ being far apart are always in Ph-\textbf{T}, which indicates that black hole subsystems are not entangled with each other in this phase. While the cold systems with adjacent subsystems are generally in Ph-\textbf{W}, which manifests that in these cases $\mathcal{B}_1$ and $\mathcal{B}_2$ are entangled with each other and hence, there is a wormhole connecting them. 

Based on these observations, we extract three characteristic types of evolution -- Fig.~\ref{fig_PD2ds2}. Intuitively, these types are distinguished by the formation of wormholes. First, for adjacent black holes where the distance $\tilde{w}_{0}$ is less than a critical scale $\tilde{w}_{c1}(\tilde{T}_h)$, wormholes have emerged from the beginning.
Second, for middle-ranged black holes where $ \tilde{w}_{c1}(\tilde{T}_h) < \tilde{w}_0 < \tilde{w}_{c2}(\tilde{T}_h)$, wormholes are generated in the intermediate stage. Finally, for distant black holes where $\tilde{w}_{0}$ is greater than another critical scale $\tilde{w}_{c2}(\tilde{T}_h)$, wormholes never appear. Again, to avoid cumbersome statements we will name these types of evolution simply as Type-\textbf{I}, Type-\textbf{II}, and Type-\textbf{III} respectively. 

Next, we will elaborate on these types of evolution by tracking the entanglement entropy density and mutual information density of each subsystem, and the detail is given as follows:

\begin{itemize}
  \item \textbf{Type I: the evolution of adjacent black holes,  $\bf{\tilde{w}_0<\tilde{w}_{c1}(\tilde{T}_h)}$}\\
  For adjacent black holes, the whole system is always in Ph-\textbf{W} during evolution -- Fig.~\ref{fig_GS0S}. 
  
  At the beginning, a wormhole has already been generated between the adjacent black holes $\mathcal{B}_{1}$ and $\mathcal{B}_{2}$, so they share nonzero mutual information density $\tilde{I}[\mathcal{B}_1 : \mathcal{B}_2]$ at $\tilde{t}=0$. Whereafter, the entanglement between $\mathcal{B}_{1}$ and $\mathcal{B}_{2}$ becomes stronger owing to the accumulation of exchanged Hawking modes between them. By the entanglement wedge reconstruction, the information on both Planck branes can only be encoded in $\bf{\mathcal{B}_{1}} \cup \bf{\mathcal{B}_{2}}$, but not in the radiation subsystem -- see Fig.~\ref{fig_DDH_rd}. 
  
  Until $\tilde{t}=\tilde{t}_{13}$, black hole subsystem $\mathcal{B}_1 \, (\mathcal{B}_2)$ is fully correlated with $\mathcal{R} \cup \mathcal{B}_2 \, (\mathcal{B}_1)$, and hence, the mutual information density $\tilde{I}[\mathcal{B}_1 : \mathcal{B}_2]$ approaches saturation. In this circumstance, in addition to the former reconstruction, we can also extract information on brane $\bm{pl}_1 (\bm{pl}_2)$ from the combined system $\mathcal{R} \cup \mathcal{B}_2 \, (\mathcal{B}_1)$.
  
  These phenomena are different from those in the standard double holography supporting one Planck brane \cite{Almheiri:2019hni,Almheiri:2019psf,Almheiri:2019qdq,Almheiri:2019psy,Almheiri:2019yqk,Ling:2020laa}, and are also direct consequences of the formation of the wormhole.

  \item \textbf{Type II: the evolution of middle-ranged black holes,  $\bf{\tilde{w}_{c1}(\tilde{T}_h)<\tilde{w}_0<\tilde{w}_{c2}(\tilde{T}_h)}$}\\
  For Middle-ranged black holes -- Fig.~\ref{fig_GS1S}, the system falls in Ph-\textbf{T} at the beginning. In this phase, $\mathcal{B}_1$ and $\mathcal{B}_2$ start to exchange Hawking modes with the nearby subsystem $\mathcal{R}$, so $\tilde{s}[\mathcal{R}]$ starts to grow. 

At $\tilde{t}=\tilde{t}_{14}$, the system undergoes a phase transition from Ph-\textbf{T} to Ph-\textbf{W}, in which the wormhole  occurs. The radiation subsystem $\mathcal{R}$ is fully entangled with  $\mathcal{B}_1 \cup \mathcal{B}_2$ ($\tilde{s}[\mathcal{R}]$ saturates), and simultaneously, $\mathcal{B}_1$ begins to interact with $\mathcal{B}_2$ ($\tilde{I}[\mathcal{B}_1 : \mathcal{B}_2]$ and $\tilde{s}[\mathcal{B}_i]$ starts to grow). Until $\tilde{t}=\tilde{t}_{13}$, both the entropy and mutual information
densities saturate.

As a result, we obtain a dynamical curve of $\tilde{s}[\mathcal{R}]$ -- Fig.~\ref{fig_GS1S}. Although it exhibits similar behaviors as in the standard double holography, in this circumstance, the unitary evolution is preserved by the formation of wormholes rather than the emergence of islands.

  \item \textbf{Type III: the evolution of distant black holes,  $\bf{\tilde{w}_0>\tilde{w}_{c2}(\tilde{T}_h)}$\\} 
  For black hole subsystems which are far apart -- Fig.~\ref{fig_GS2S}, the evolution also tends to begin with Ph-\textbf{T}. That is, $\bf{\mathcal{B}_{1}}\cup\bf{\mathcal{B}_{2}}$ starts to be entangled with  $\bf{\mathcal{R}}$ as the process goes on.
  
  After $\tilde{t}=\tilde{t}_{13}$, the radiation subsystem $\mathcal{R}$ is fully entangled with $\mathcal{B}_1 \cup \mathcal{B}_2$, where $\tilde{s}[\mathcal{R}]$ saturates, as well as $\tilde{s}[\mathcal{B}_i]$, but the growth of $\tilde{I}[\mathcal{B}_1 : \mathcal{B}_2]$ is suppressed by the long distance. 
  
  Note that we acquire a Page curve here without the formation of wormholes. Therefore, the unitarity during evolution is preserved by the emergence of islands. In this sense, the situation is much similar to those in literature.
\end{itemize}

\begin{figure}
  \centering
  \subfigure[]{\label{fig_TP0}
  \includegraphics[width=0.4\linewidth]{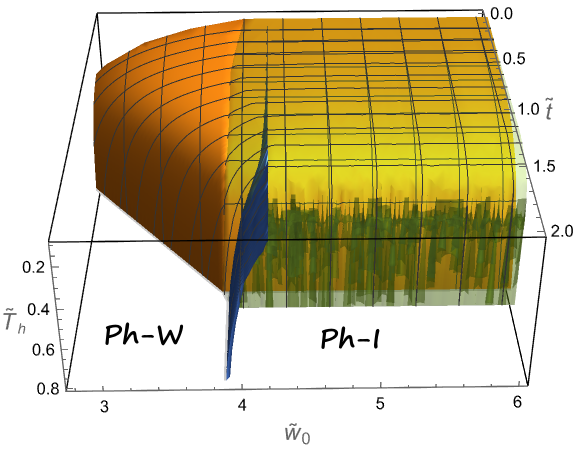}}\\
  \hspace{0pt}
  \subfigure[]{\label{fig_TP1}
   \includegraphics[width=0.31\linewidth]{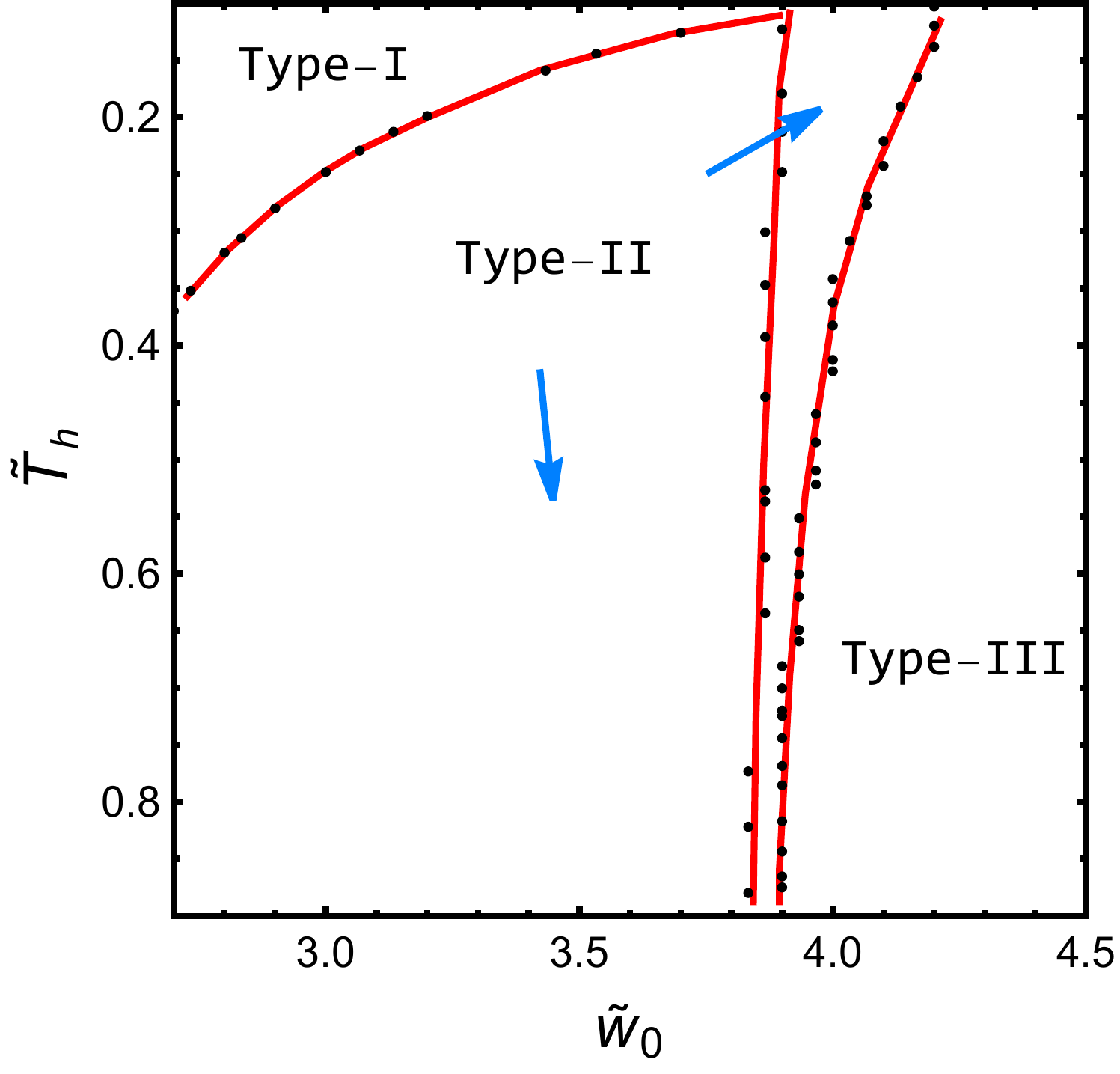}}
     \hspace{0pt}
  \subfigure[]{\label{fig_TP2}
  \includegraphics[width=0.3\linewidth]{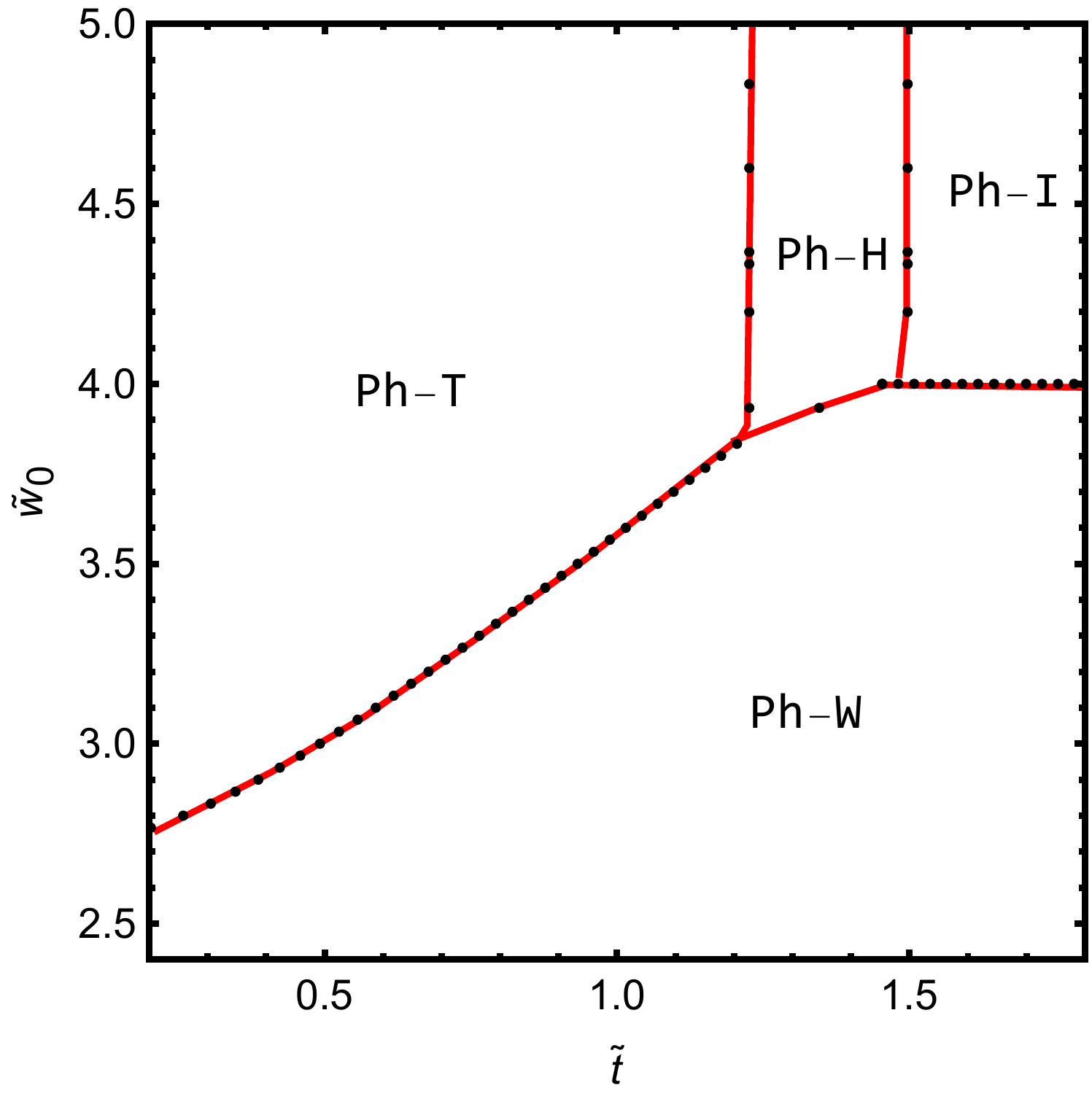}}
\caption{(a): Phase diagram in $\{\tilde{w}_0,\tilde{t},\tilde{T}_h\}$. The region inside the orange surface is in Ph-\textbf{T}. The region enclosed by the orange and the green surface is in Ph-\textbf{H}. While the region outside the green surface is divided by the blue surface into Ph-\textbf{W} and Ph-\textbf{I}.
(b): The different types of evolution. Here the left-most red curve $\tilde{w}_{c1}(\tilde{T}_h)$ and the right-most red curve $\tilde{w}_{c2}(\tilde{T}_h)$ are two critical distances. (c): For $\tilde{T}_h=0.36$, the projected phase diagram in $\{\tilde{w}_0,\tilde{t}\}$.}\label{fig_TransitionPhase}
\end{figure}

\subsubsection{Black holes of different sizes}
\begin{figure}
  \centering
  \subfigure[]{\label{fig_GS0}
  \includegraphics[width=0.3\linewidth]{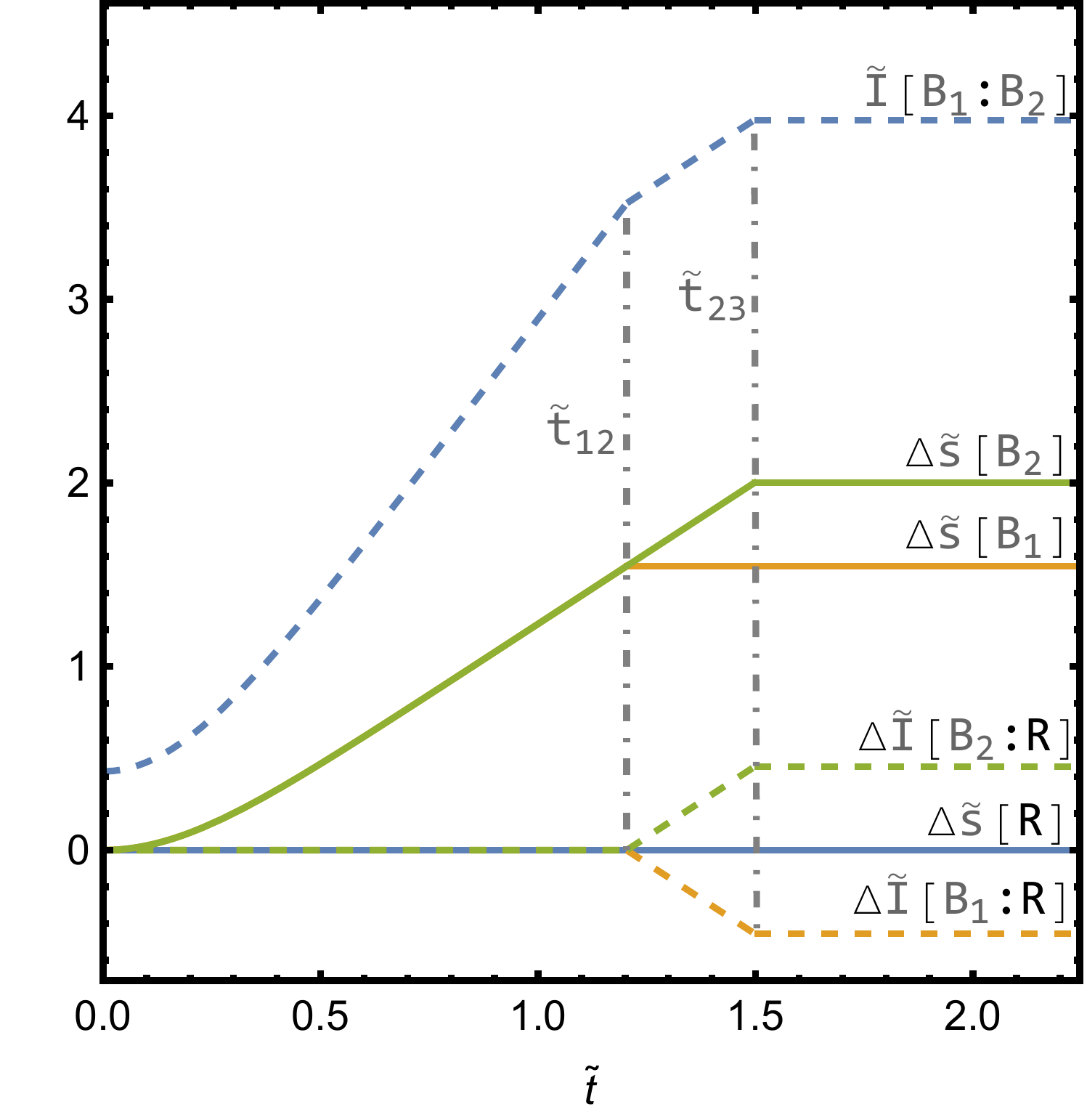}}
  \hspace{0pt}
 \subfigure[]{\label{fig_GS1}
  \includegraphics[width=0.31\linewidth]{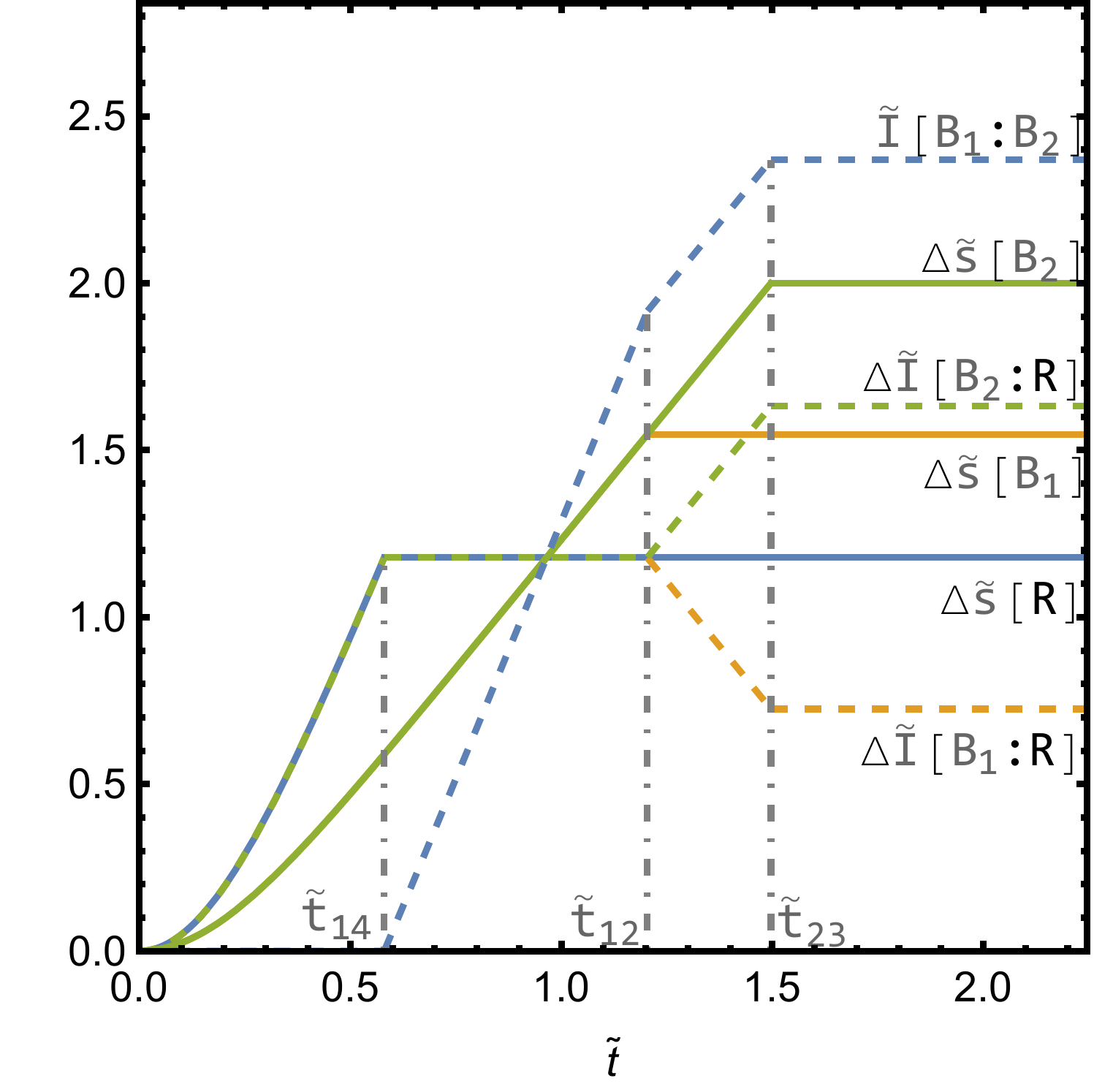}}\\
  \hspace{0pt}
  \subfigure[]{\label{fig_GS2}
  \includegraphics[width=0.3\linewidth]{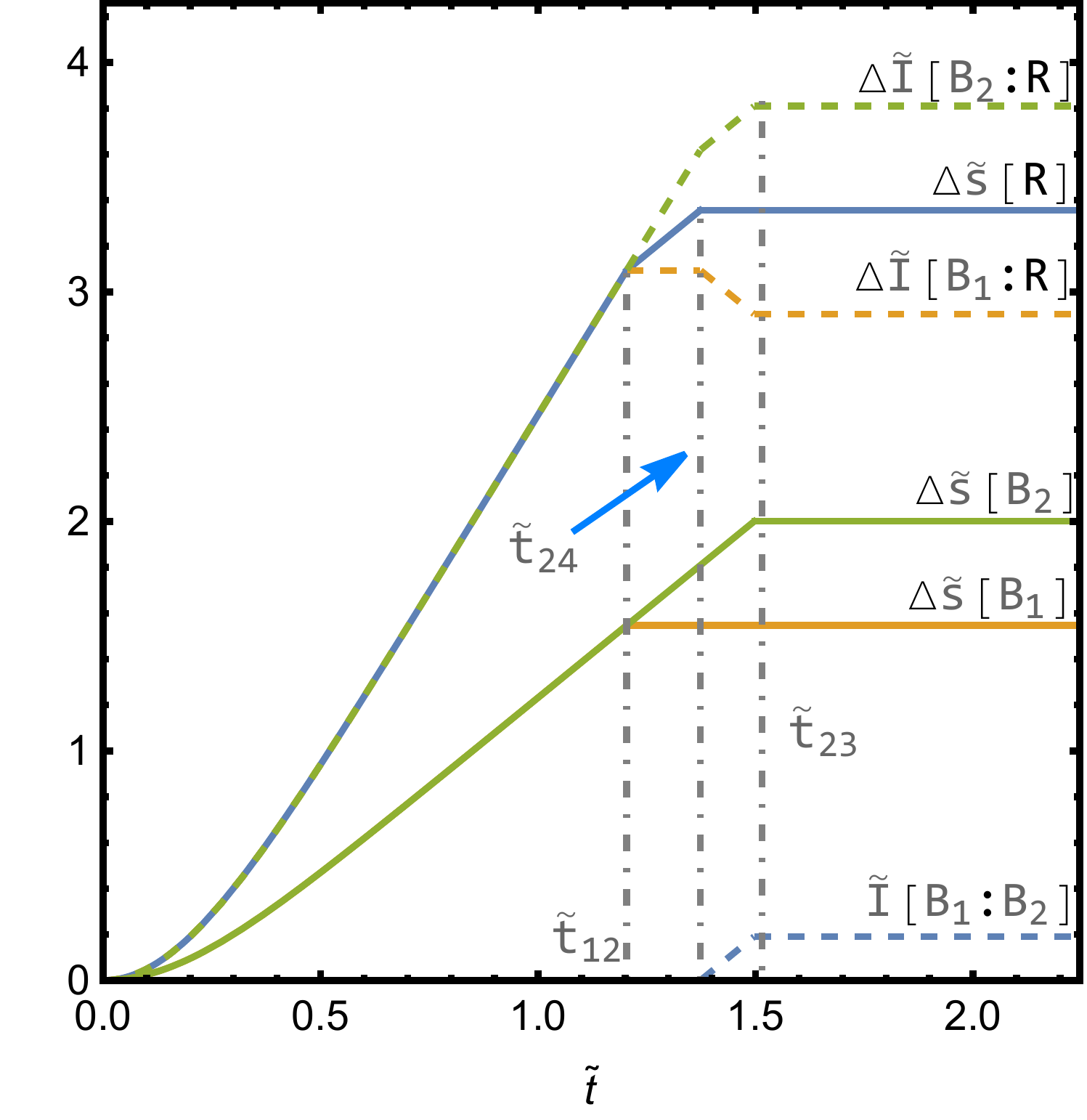}}
  \hspace{0pt}
  \subfigure[]{\label{fig_GS3}
  \includegraphics[width=0.3\linewidth]{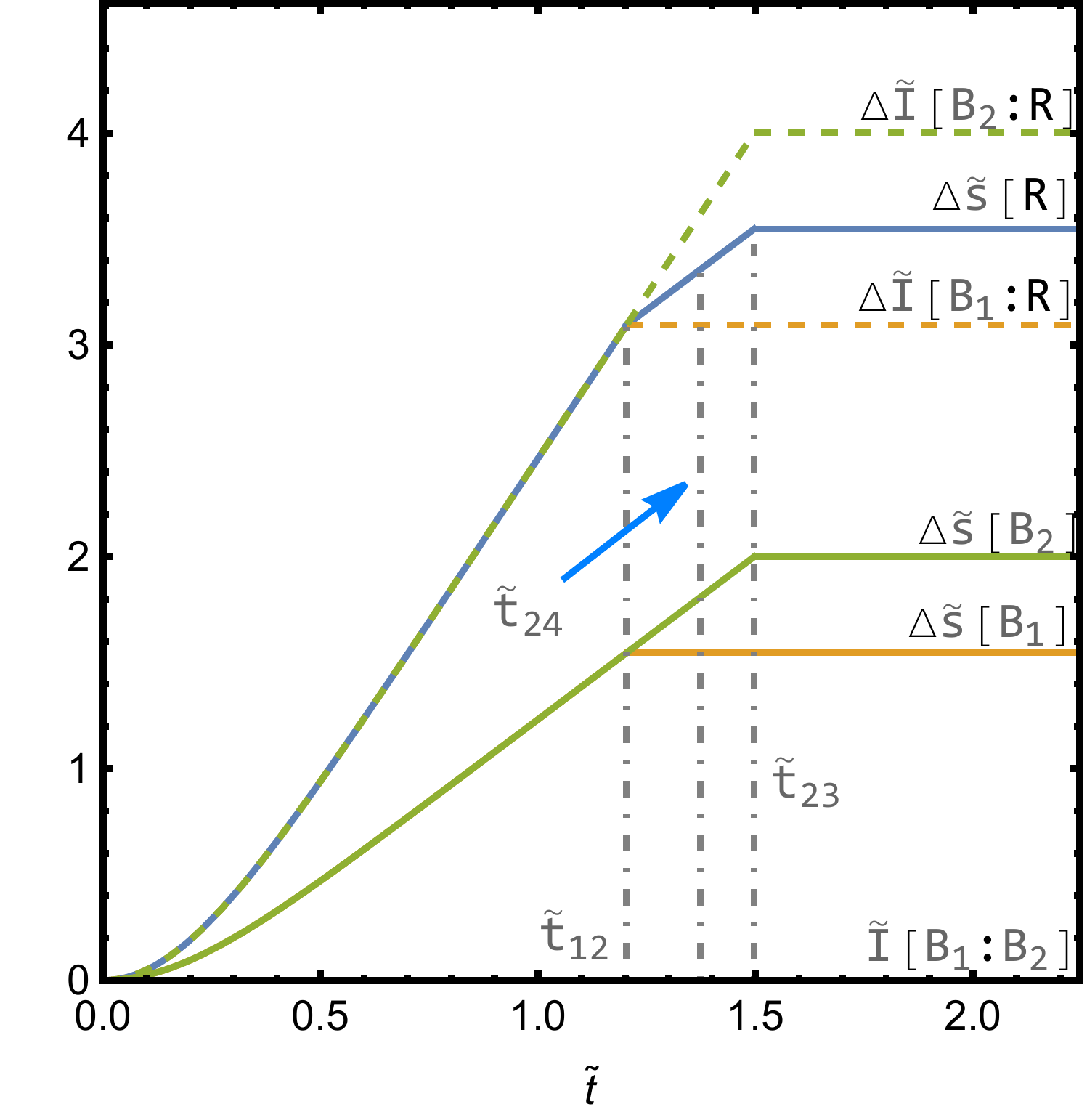}}
\caption{The evolution on entropy and mutual information density at different distances (a): $\tilde{w}_0=8/3$, (b): $\tilde{w}_0=47/15$, (c): $\tilde{w}_0=119/30$ and (d): $\tilde{w}_0=20/3$, while other parameters being specified to $\{L,\theta_1,\theta_2,\lambda_1,\lambda_2,\tilde{\mu},\tilde{T}_h\}=\{1,\pi/2,\pi/2,0,0.3,1,9/8\pi\}$. Note that $\Delta \tilde{s}[\mathcal{R}]:=\tilde{s}[\mathcal{R}](t)-\tilde{s}[\mathcal{R}](0)$.}\label{fig_GrowthRateS}
\end{figure}

In the following, we will generalize the discussion to the cases with different-sized black holes, where the smaller(larger) black hole $\mathcal{B}_1$($\mathcal{B}_2$) contains fewer(more) DOF. The phase diagram is demonstrated in Fig.~\ref{fig_TransitionPhase}, by varying both the distance $\tilde{w}_0$ and Hawking temperature $\tilde{T}_h$. We remark that one more possible phase (Ph-\textbf{H}) appears in this case, due to the different instants of the saturation of $\tilde{s}_{{B}_i}$. Nevertheless, we point out that the addition of Ph-\textbf{H} will not change the classification of evolution. Thus we still envisage the evolution with the following types:

\begin{itemize}
  \item \textbf{Type I: the evolution of adjacent black holes}\\
  For adjacent black holes, the full system still tends to stay in Ph-\textbf{W} -- Fig.~\ref{fig_GS0}. Nonzero mutual information density at $\tilde{t}=0$ indicates the existence of the wormhole connecting $\mathcal{B}_1$ and $\mathcal{B}_2$. As time passes by, the accumulation of Hawking modes directly enhances the entanglement between them. 
  
  At $\tilde{t}=\tilde{t}_{12}$, the entropy density $\tilde{s}[\mathcal{B}_1]$ of the smaller black hole subsystem $\mathcal{B}_1$ saturates first, while that of the larger one $\mathcal{B}_2$ does not. As a result, the entanglement between $\mathcal{B}_1$ and $\mathcal{R}$ will continue to ``pass'' into $\mathcal{B}_2$ (where $\tilde{I}[\mathcal{B}_1 : \mathcal{R}]$  decreases while $\tilde{I}[\mathcal{B}_2 : \mathcal{R}]$ increases). From the brane perspective, in this phase, we are allowed to reconstruct the information inside the black hole on brane $\bm{pl}_1$ from the boundary region $\mathcal{R}$. Finally, at $\tilde{t}=\tilde{t}_{23}$, all measures under consideration reach equilibrium.

  \item \textbf{Type II: the evolution of Middle-ranged black holes}\\
  For middle-ranged black holes, the system undergoes either Ph-\textbf{T,W} or Ph-\textbf{T,H,W}, and then reaches the saturation. The desired wormhole will be generated during the evolution in both routes. 

  Let us firstly discuss the evolution undergoing Ph-\textbf{T,W} phases-- Fig.~\ref{fig_GS1}. At early times, $\mathcal{B}_1$ and $\mathcal{B}_2$ start to exchange Hawking modes with $\mathcal{R}$. At $\tilde{t}=\tilde{t}_{14}$, $\mathcal{R}$ is fully entangled with the surroundings, and hence, $\mathcal{B}_1$ begins to be correlated with $\mathcal{B}_2$. Until $\tilde{t}=\tilde{t}_{12}$, the entropy density of $\mathcal{B}_1$ saturates, while that of $\mathcal{B}_1$ does not. Thus the entanglement between $\mathcal{B}_1$ and $\mathcal{R}$ passes into $\mathcal{B}_2$ all the way until the equilibrium is reached at $\tilde{t}=\tilde{t}_{23}$.
 
  As for the second case -- Fig.~\ref{fig_GS2}, the main difference from the former is that the entropy density of $\mathcal{B}_1$ saturates prior to that of $\mathcal{R}$. After the saturation of $\mathcal{B}_1$ at $\tilde{t}=\tilde{t}_{12}$, $\mathcal{B}_2$ continues to interact with $\mathcal{R}$. Until the saturation of $\mathcal{R}$ at $\tilde{t}=\tilde{t}_{24}$, the entanglement between $\mathcal{B}_1$ and $\mathcal{R}$ is reassigned into $\mathcal{B}_2$.
  We remark that in this case, the wormhole occurs later and the maximal entanglement is much lower, reflecting the intuitive fact that the farther the distance $\tilde{w}_0$, the harder it is for $\mathcal{B}_1$ and $\mathcal{B}_2$ to be correlated. 

We also point out that in both cases unitarity is preserved by the formation of a wormhole rather than an island.

  \item \textbf{Type III: the evolution of distant black holes}\\
  For distant black holes, the system is in Ph-\textbf{T} at the beginning and finally saturates in Ph-\textbf{I}.
  
  In this type of evolution, the entropy density of $\mathcal{B}_1$ saturates first at $\tilde{t}=\tilde{t}_{12}$, while those of $\mathcal{B}_2$ and $\mathcal{R}$ saturate later at $\tilde{t}=\tilde{t}_{23}$. As a result, the entanglement between $\mathcal{B}_1$ and $\mathcal{B}_2$ continues to grow until $\tilde{t}=\tilde{t}_{23}$.
  
  Similar to the Type-\textbf{III} of black holes with equal size, the wormhole is never generated if only considering the leading order of mutual information and unitarity is preserved by the emergence of islands as described in literature.
\end{itemize}

In this subsection, we have demonstrated that when the amount of DOF on the brane is small such that the backreaction of the brane can be ignored, the evolution of the system can be classified into three universal types according to the distance between the black holes. Then a natural issue arises and remains unanswered. That is, whether and how these types of evolution will be affected by the amount of DOF on the branes. Therefore, inspired by the work in \cite{Ling:2020laa}, we tend to take the backreaction of branes into account and investigate its effect on the existence of the wormhole in the next subsection.

\subsection{The wormhole phase in backreacted spacetime}\label{sec_num}
To illustrate the universality of the wormhole phase, or more precisely, the universality of the evolution characterized by three types, in this subsection, we take the backreaction into account, which is equivalent to joining DOF on the black hole subsystems $\mathcal{B}_1$ and $\mathcal{B}_2$. Moreover, in the probe limit, we need an artificial nonzero scale $w_b$ to produce a dynamical Page curve, but in backreacted cases, we can safely eliminate it. Therefore, we change the definition of symbol ``tilde'', which results in new scale-free parameters, such as $\{\tilde{w}_0,\tilde{t},\tilde{\mu},\tilde{s}[\mathcal{R}],\tilde{I}[\mathcal{B}_1 : \mathcal{B}_2]\}=\{w_0 T_h, t T_h, \mu / T_h, s[\mathcal{R}] / T_h, I[\mathcal{B}_1 : \mathcal{B}_2] / T_h\}$.

\begin{figure}
  \centering
  \subfigure[]{\label{fig_HRTBR}
  \includegraphics[width=0.3\linewidth]{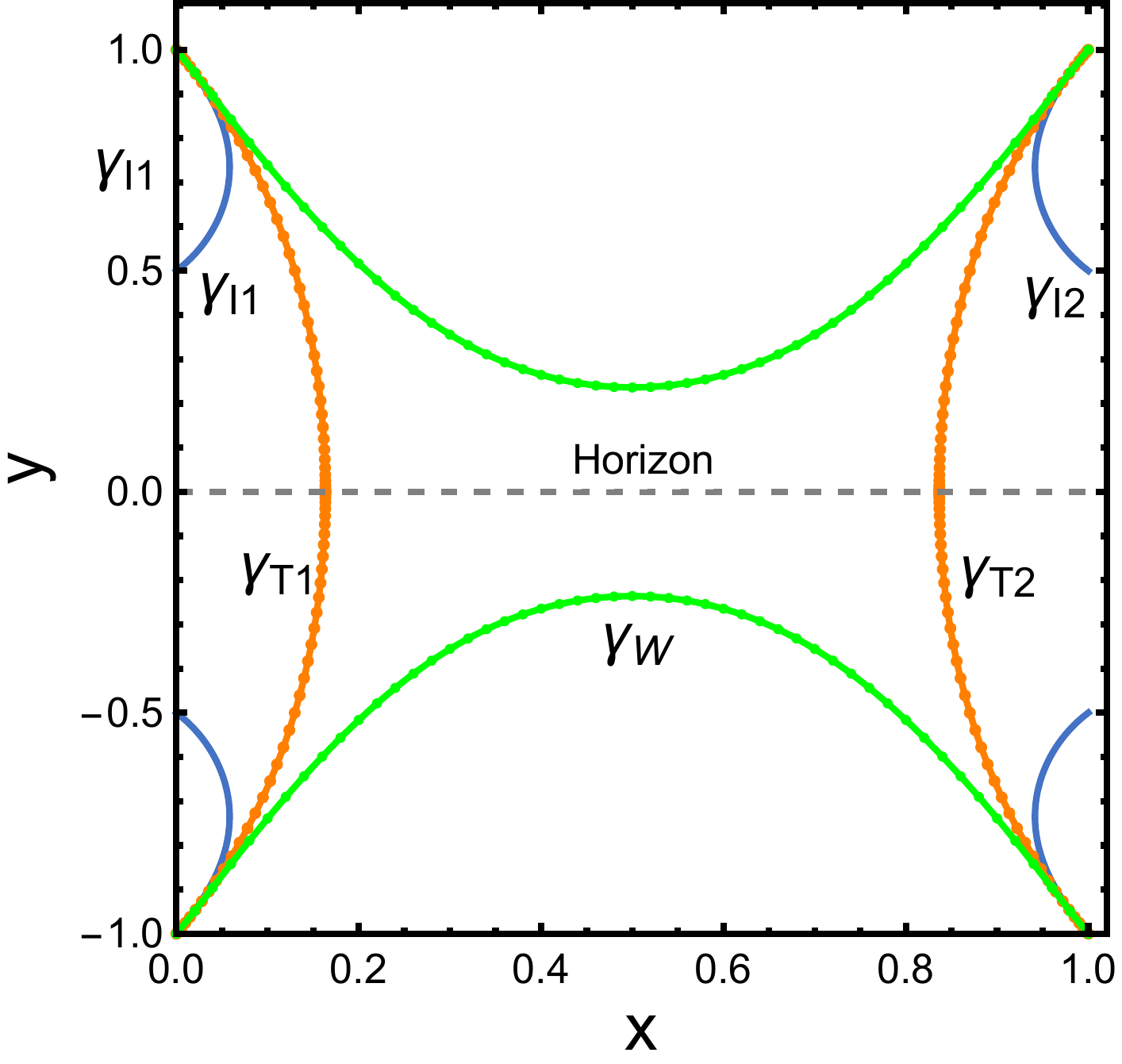}}
  \hspace{0pt}
 \subfigure[]{\label{fig_PatDBR}
  \includegraphics[width=0.3\linewidth]{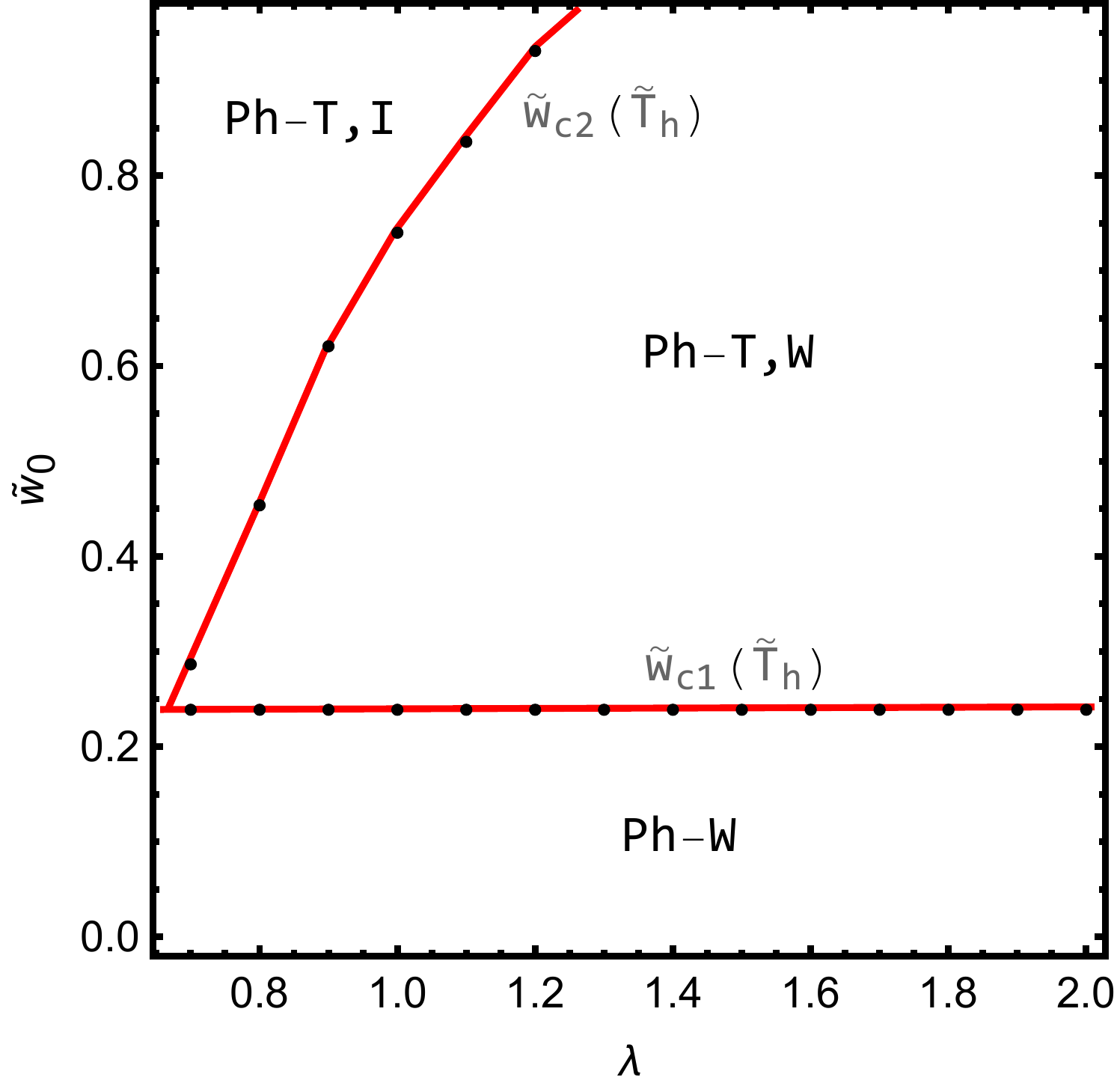}}
\caption{(a): Different configurations of HRT surfaces at $\tilde{t}=0$, with $\{L,\theta_1,\theta_2,\tilde{\mu},\lambda,\tilde{w}_0\}=\{1,3\pi/16,3\pi/16,0,2,3\pi/2\}$. (b): Different types during evolution, with $\{L,\theta_1,\theta_2,\tilde{\mu}\}=\{1,3\pi/16,3\pi/16,0\}$. Green dots represent that the system undergoes Ph-\textbf{T,I} during evolution. Orange dots represent the system undergoes Ph-\textbf{T,W}. While Blue dots represent the system always stays in Ph-\textbf{W}.}\label{fig_HRTandPatDinBR}
\end{figure}

For simplicity, here we explore neutral cases with the same DOF on branes. That is to say, we specify $\tilde{\mu}=0$ and $\theta_1=\theta_2=3\pi/16$, while vary $\lambda:=\lambda_1=\lambda_2$ and $\tilde{w}_0$. The corresponding HRT surfaces and types of evolution are shown in Fig.~\ref{fig_HRTBR} and \ref{fig_PatDBR}, respectively. It can be seen that there are also two critical scales $\tilde{w}_{c1}$ and $\tilde{w}_{c2}$ which classify the evolution into three types. For $\lambda > 0.7$, $\tilde{w}_{c2}$ grows with $\lambda$, while $\tilde{w}_{c1}$ is rarely changed with $\lambda$. This result indicates that with more DOF, $\mathcal{B}_1$ and $\mathcal{B}_2$ are more likely to be correlated, and thus the wormhole is more likely to be generated during evolution. 

\section{SYK clusters coupled to Majorana chains}\label{sec_SYK}
In this section, a similar setup in the context of SYK models is constructed by identifying $\mathcal{B}_1$ and $\mathcal{B}_2$ as distinct SYK systems, and radiation $\mathcal{R}$ as Majorana chains connecting them. This combined system is studied by exact diagonalization (ED). The entanglement entropy and mutual information of subsystems are investigated during the evolution, and the results are compared with those obtained in the previous section. 

\subsection{The setup}
\begin{figure}
  \centering
  \includegraphics[width=\linewidth]{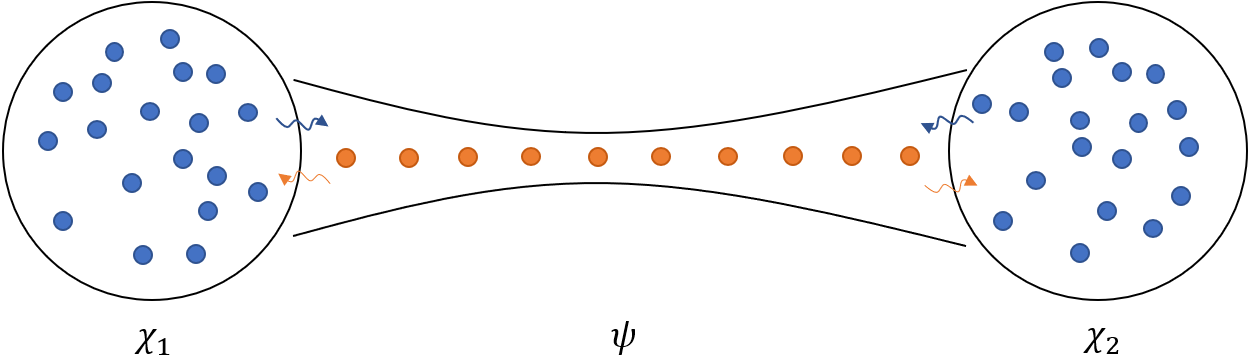}
\caption{The combined system, where the SYK subsystems are connected by the Majorana chain.}\label{fig_Hamiltonian}
\end{figure}

To be specific, the SYK systems labeled by $\chi_1$ and $\chi_2$ are composed of $N_{\chi_1}$ and $N_{\chi_2}$ Majorana fermions with random couplings among them, respectively, while the Majorana chains are described by $\psi(x)$ on $N_{\psi}$ separate sites with hopping $\Lambda/2$, and the fermions on the first and last sites are coupled with nearby SYK systems respectively. The combined system is sketched in Fig.~\ref{fig_Hamiltonian}. The corresponding Hamiltonian of the whole system is given by
\begin{align}
    H_L  =&H_{\chi_{1L}}+H_{\chi_{2L}}+H_{\psi_L}+H_{int} \nonumber\\
    =&\sum_{i<j<k<l}^{N_{\chi_1}/2} (J_{\chi_{1L}})_{ijkl} \chi_{1L,i} \chi_{1L,j} \chi_{1L,k} \chi_{1L,l}
	 +\sum_{i<j<k<l}^{N_{\chi_2}/2} (J_{\chi_{2L}})_{ijkl} \chi_{2L,i} \chi_{2L,j} \chi_{2L,k} \chi_{2L,l}\nonumber\\
	& +\frac{i\Lambda}{2} \sum_{x}^{N_{\psi}/2}  \psi_L(x) \psi_L(x+1)
	+\frac{i V_1 \sqrt\Lambda}{\sqrt{N_{\chi_1}/2}} \sum_{i}^{N_{\chi_1}/2}  \chi_{1L,i} \psi_L(1)
	 +\frac{i V_2 \sqrt\Lambda}{\sqrt{N_{\chi_2}/2}} \sum_{i}^{N_{\chi_1}/2}  \psi_L(N_{\psi}/2)\chi_{2L,i} 
\end{align}
where  $J_i\;(i=\chi_{1L},\chi_{2L})$ are in Gaussian distribution, which satisfy
\begin{equation}
    \overline{(J_i)_{ijkl}}=0, \qquad \overline{[(J_i)_{ijkl}]^2}=\frac{3! \mathcal{J}_i^2}{N_K^3}.
\end{equation}

Starting with a thermofield double (TFD) state, with six subsystems $\chi_{1L}, \chi_{2L}, \psi_L$ and $\chi_{1R}, \chi_{2R}, \psi_R$, the time evolution of the TFD state can be constructed as
\begin{equation}
    \ket{\text{TFD}(t)}=\frac{1}{\sqrt{Z(\beta)}} \exp^{- ( \beta + 4 i t) H_L /2 } \ket{I}_{\chi_{1L},\chi_{1R}}\ket{I}_{\psi_L,\psi_R}\ket{I}_{\chi_{2L},\chi_{2R}},
\end{equation}
where $\ket{I}_{K_L,K_R}$ is a maximally entangled state between $K_L$ and $K_R$ subsystems, where $K\in\{\chi_1,\chi_2,\psi\}$. On the SYK systems, the state is defined as
\begin{align}
    \left(\chi_{1L, j}+i \chi_{1R, j}\right)|I\rangle_{\chi_{1L}, \chi_{1R}}=0, \quad j=1,2, \ldots, N_{\chi_1}/2,\\
    \left(\chi_{2L, j}+i \chi_{2R, j}\right)|I\rangle_{\chi_{2L}, \chi_{2R}}=0, \quad j=1,2, \ldots, N_{\chi_2}/2.
\end{align}
On the Majorana chains, the state is defined as
\begin{equation}
    \left(\psi_{L}(x)+i \psi_{R}(x)\right)|I\rangle_{\psi_{L}, \psi_{R}}=0, \quad x= 1,2,\cdots,N_{\psi}/2.
\end{equation}

 Next, the density matrix of the subsystem $K$ can be obtained by tracing out the DOF of its complement $\bar{K}$, which is given by
\begin{align}
\rho_K(t)&=\text{Tr}_{\bar{K}}\left(\ket{\text{TFD}(t)}\bra{\text{TFD}(t)}\right).
\end{align}
For instance, when $K=\chi_1$, we have $\bar{K}=\chi_2 \cup \psi$. Then the entanglement entropy of a system $K$ is defined by
\begin{equation}
    S[K]=-\text{Tr}_{\bar{K}}(\rho_K \log \rho_k),
\end{equation}
and the mutual information between $K$ and $K'$ is defined to be
\begin{equation}
    I[K:K']=S[K]+S[K']-S[\overline{K\cup K'}].
\end{equation}

\subsection{The entanglement properties of subsystems}
In this subsection, we investigate different measures in quantum information to keep track of the entanglement properties during evolution.
Precisely, $\chi_1$ and $\chi_2$ serve as the black hole subsystems $\mathcal{B}_1$ and $\mathcal{B}_2$ respectively, while $\psi$ is regarded as the radiation subsystem $\mathcal{R}$. The amount of DOF in $\mathcal{B}_1$ and $\mathcal{B}_2$ are referred to as the number of fermions in $\chi_1$ and $\chi_2$, namely $N_{\chi_1}$ and $N_{\chi_2}$, and the distance between $\mathcal{B}_1$ and $\mathcal{B}_2$ is captured by the number of sites $N_{\psi}$ in Majorana chains. 

\begin{figure}
  \centering
  \subfigure[]{\label{fig_EqualVNear}
  \includegraphics[width=0.31\linewidth]{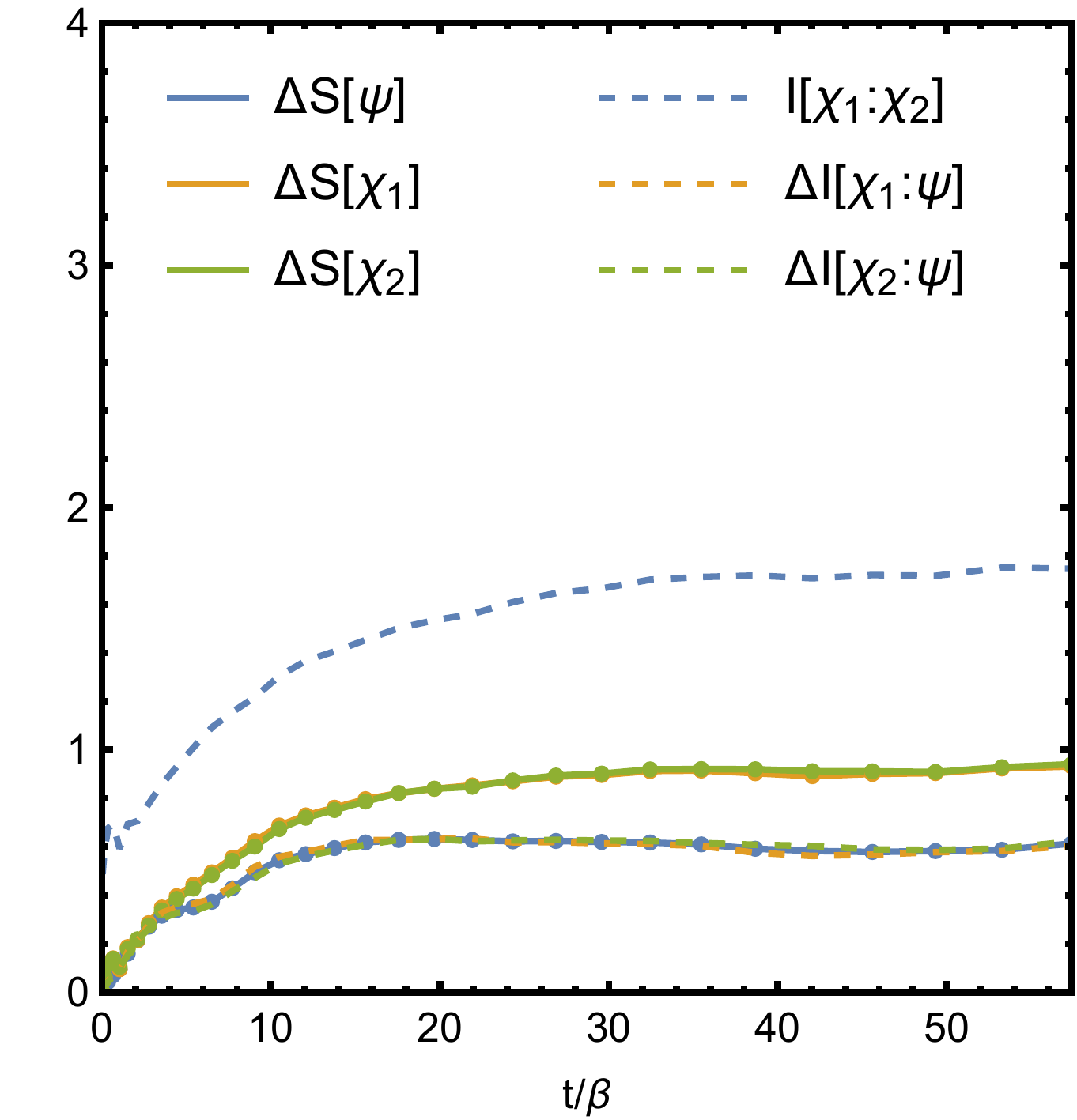}}
  \hspace{0pt}
 \subfigure[]{\label{fig_EqualFar}
  \includegraphics[width=0.31\linewidth]{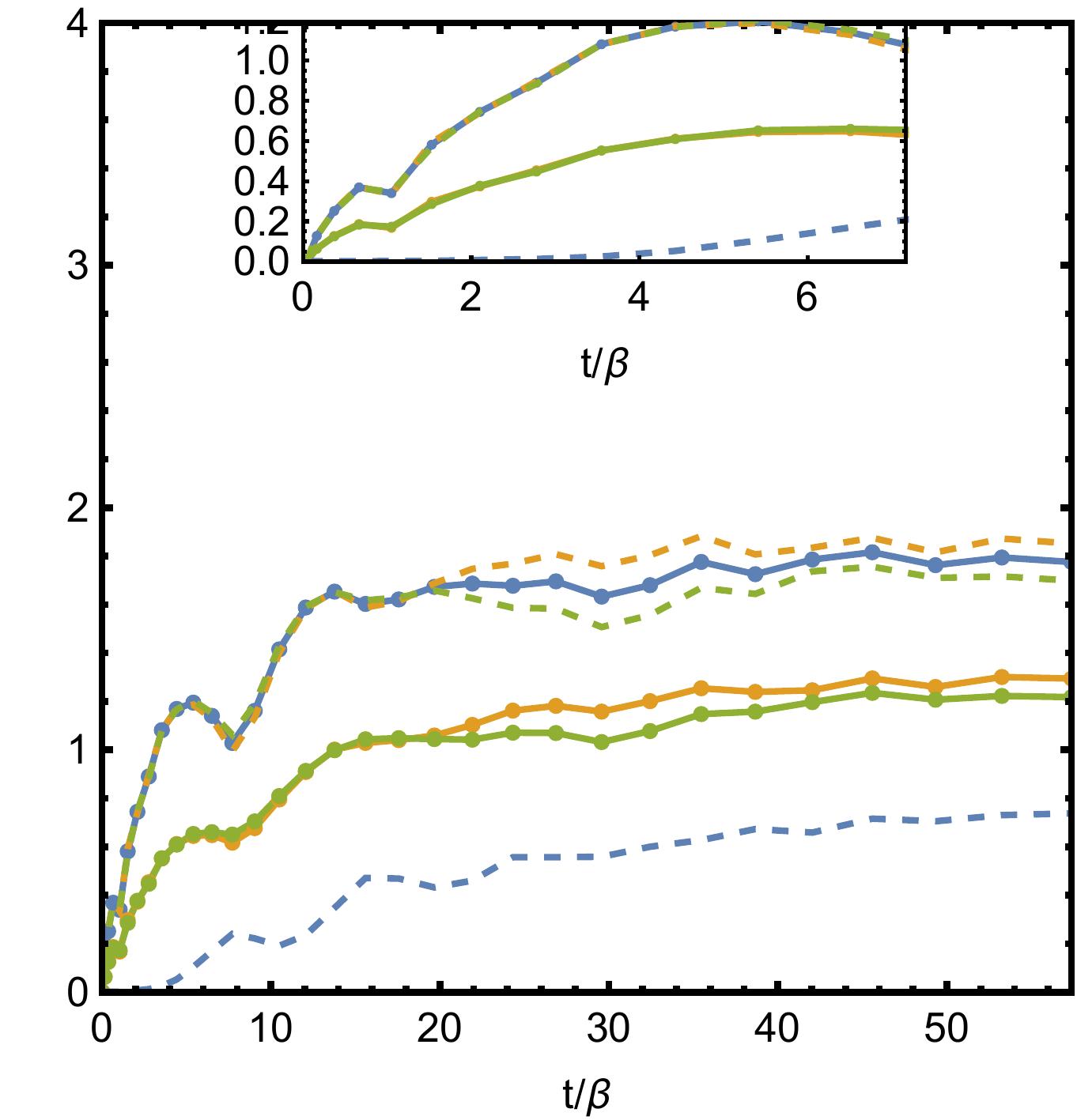}}
  \hspace{0pt}
  \subfigure[]{\label{fig_EqualVFar}
  \includegraphics[width=0.31\linewidth]{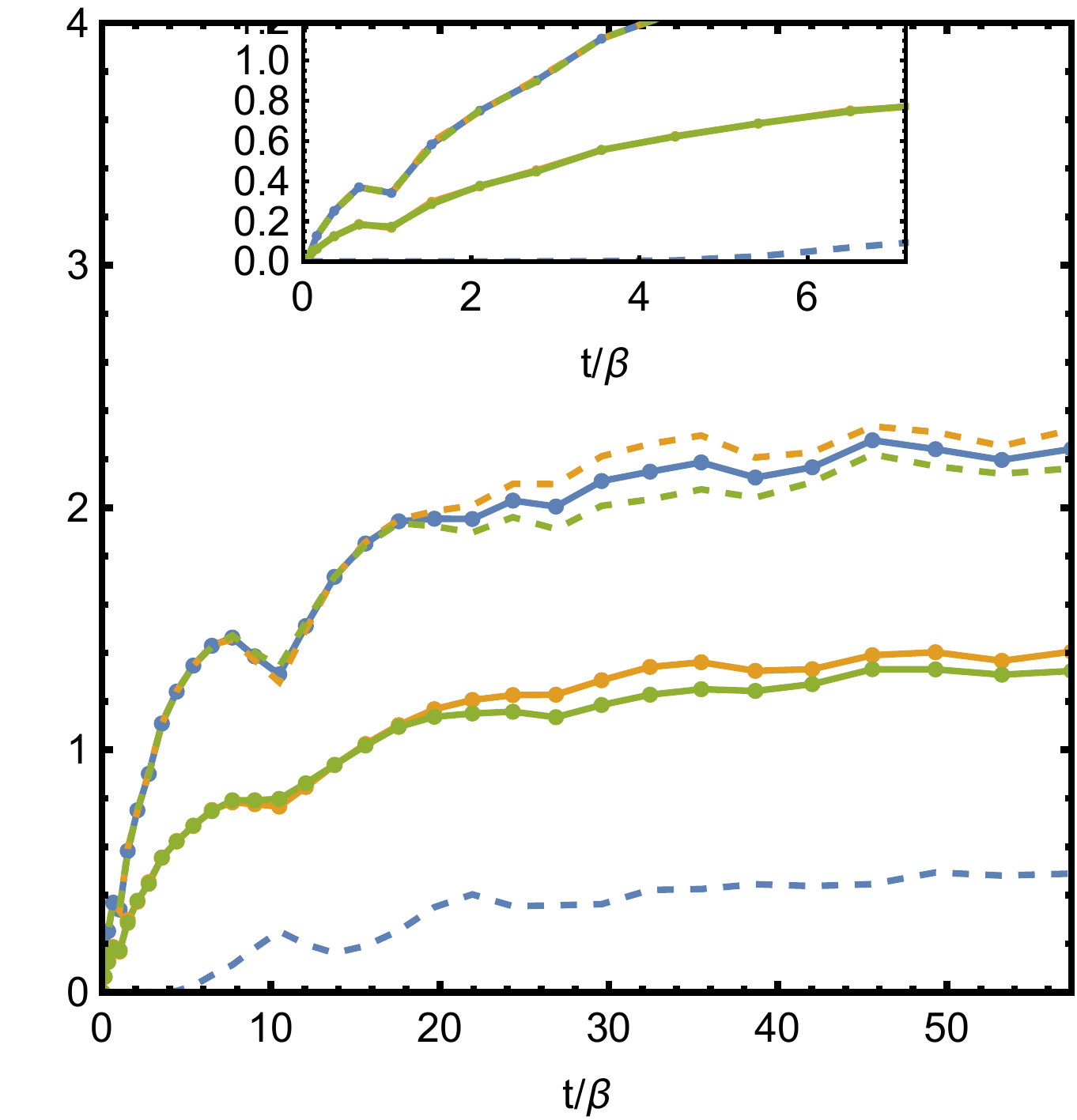}}\\
  \hspace{0pt}
  \subfigure[]{\label{fig_EqualVNear2}
  \includegraphics[width=0.31\linewidth]{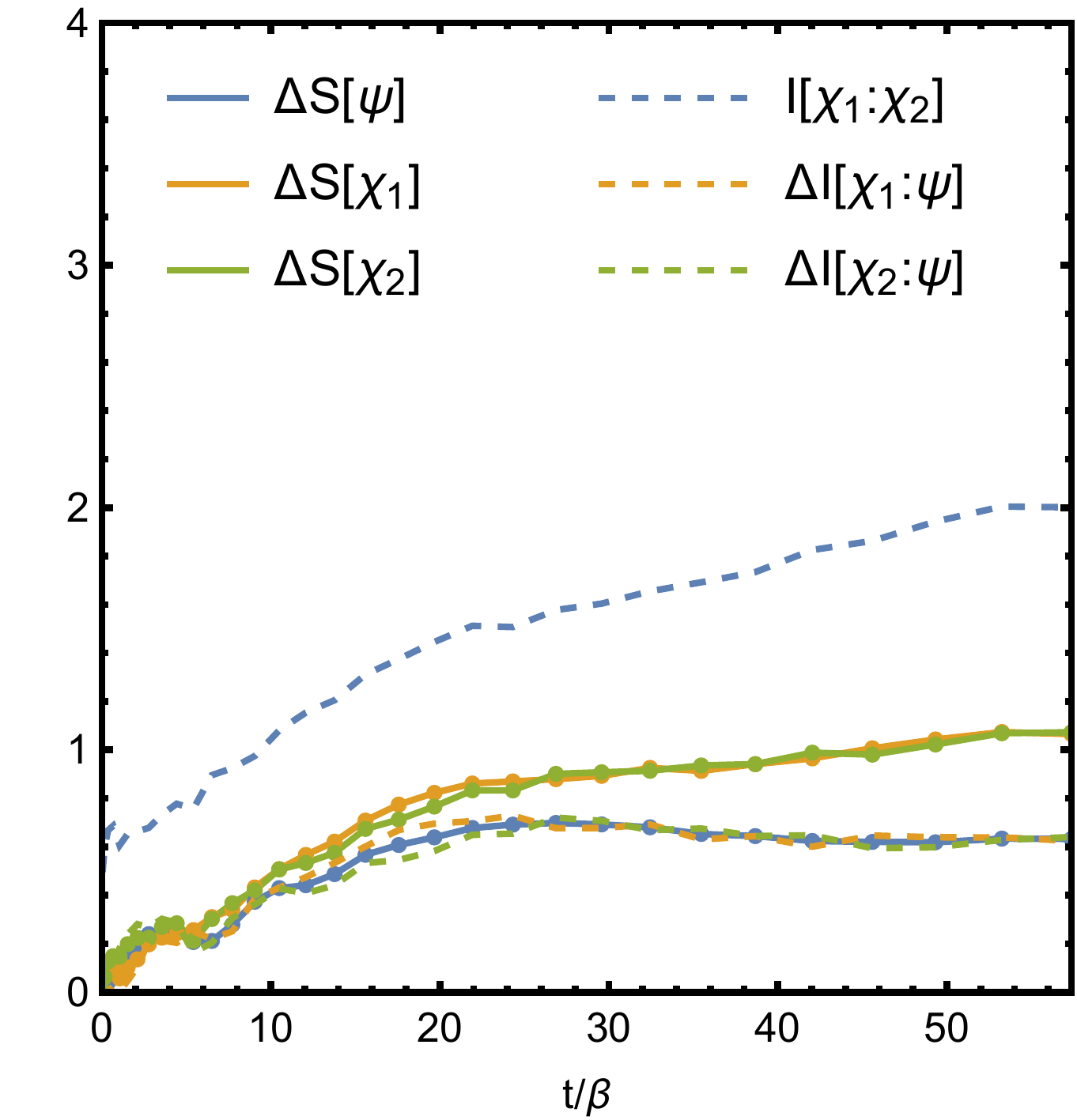}}
  \hspace{0pt}
 \subfigure[]{\label{fig_EqualFar2}
  \includegraphics[width=0.31\linewidth]{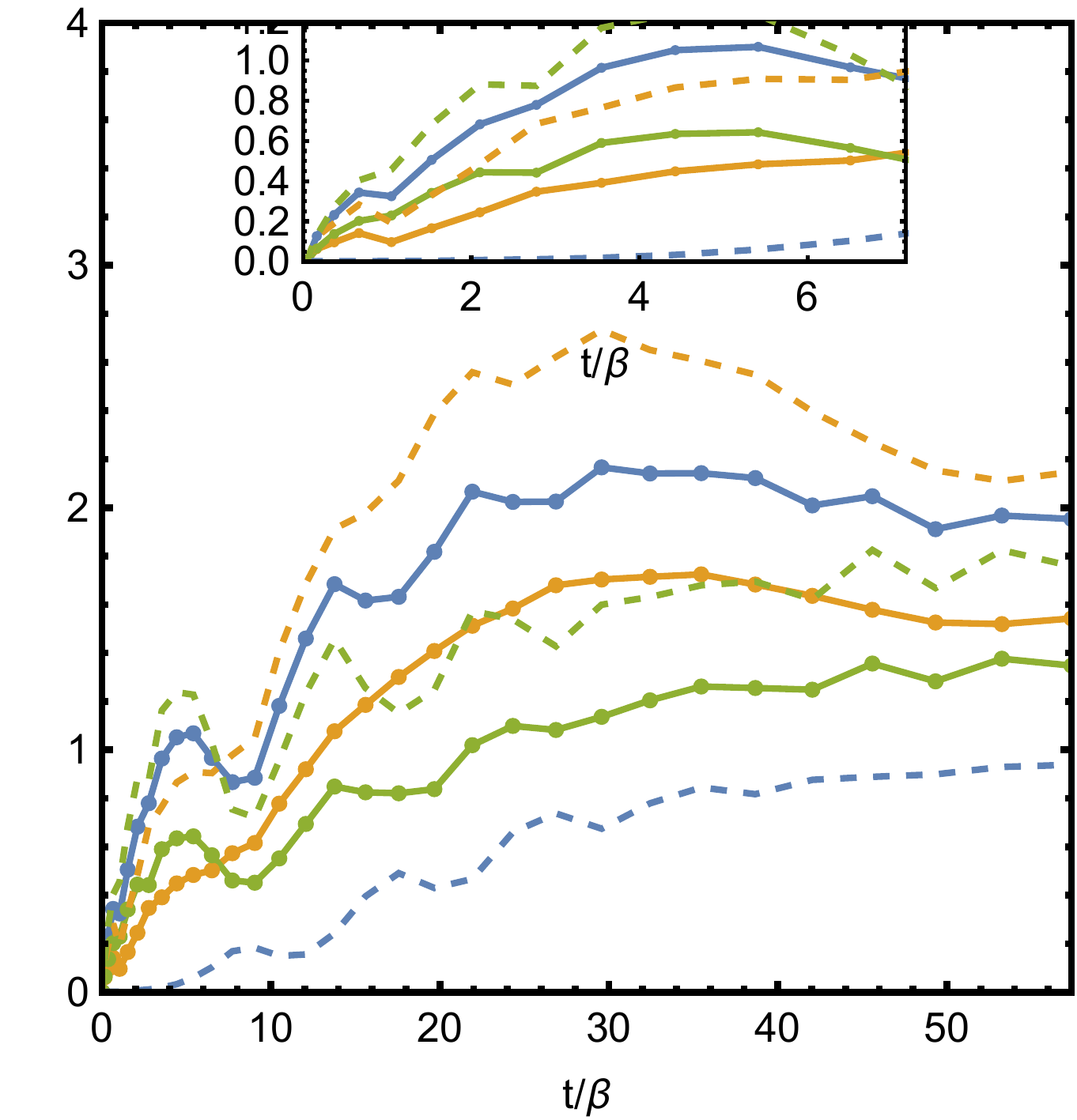}}
  \hspace{0pt}
  \subfigure[]{\label{fig_EqualVFar2}
  \includegraphics[width=0.31\linewidth]{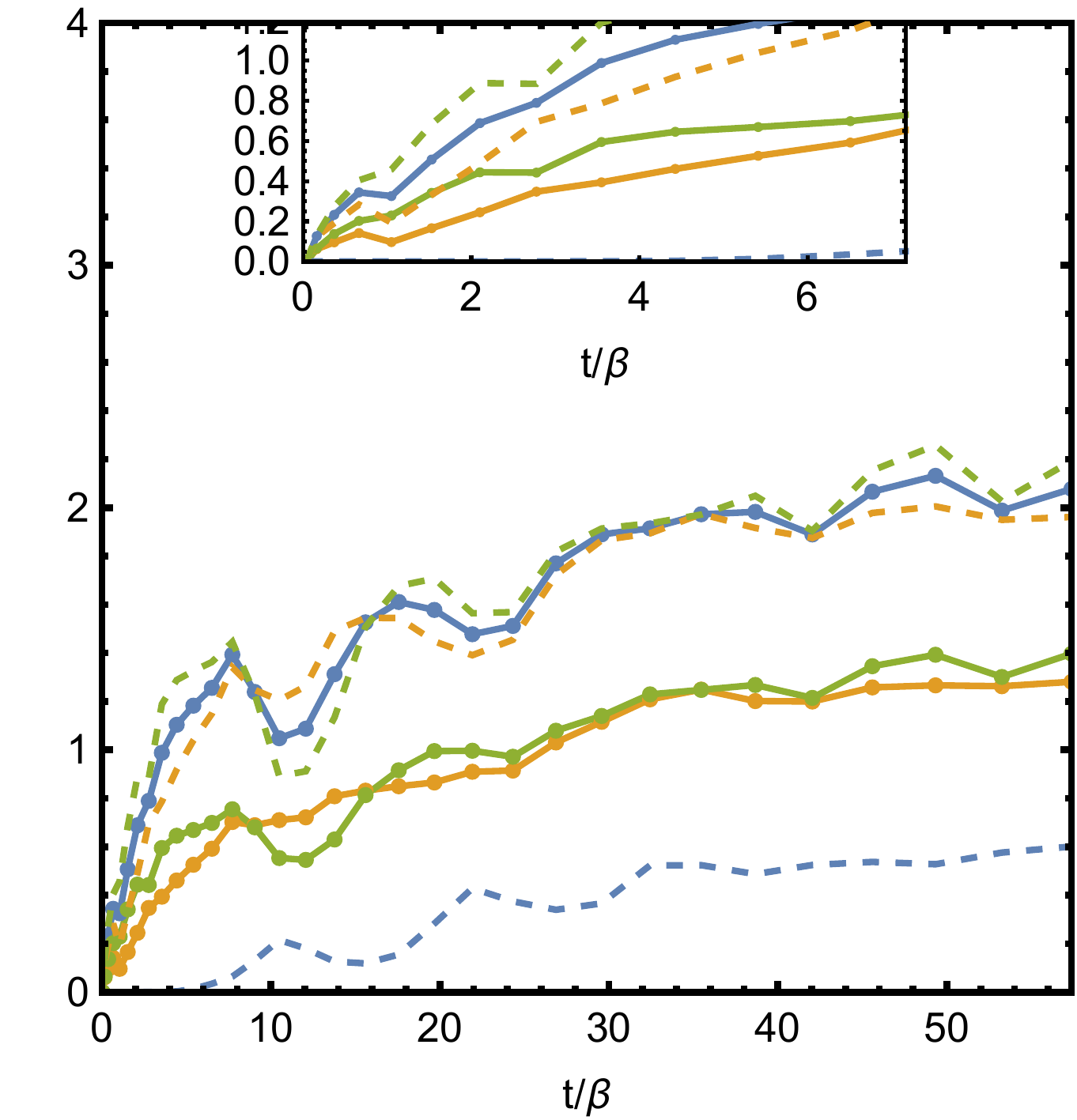}}
\caption{The evolution on entropy and mutual information densities, with $N_{\chi_1}=N_{\chi_2}=12$, and $\{J_1 \beta, J_2 \beta, \Lambda\beta, V_1 \sqrt{\Lambda} \beta, V_2 \sqrt{\Lambda} \beta\}=\{5,5,1,3,3\}$. For the first line, an average of 21 samples is plotted in each panel, while for the second line, a single sample is plotted in each panel. In both lines, $N_{\psi}=4,12,16$ from left to right.}\label{fig_EqMIS}
\end{figure}

\begin{figure}
  \centering
  \subfigure[]{\label{fig_NEqualNear}
  \includegraphics[height=0.33\linewidth]{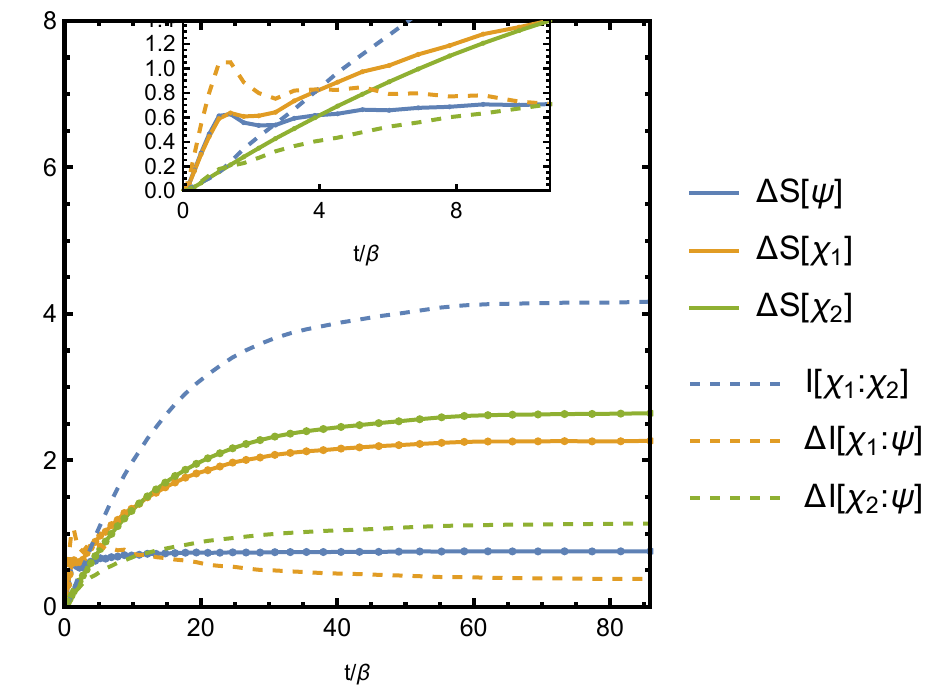}}
    \hspace{0pt}
  \subfigure[]{\label{fig_NEqualNear2}
  \includegraphics[height=0.33\linewidth]{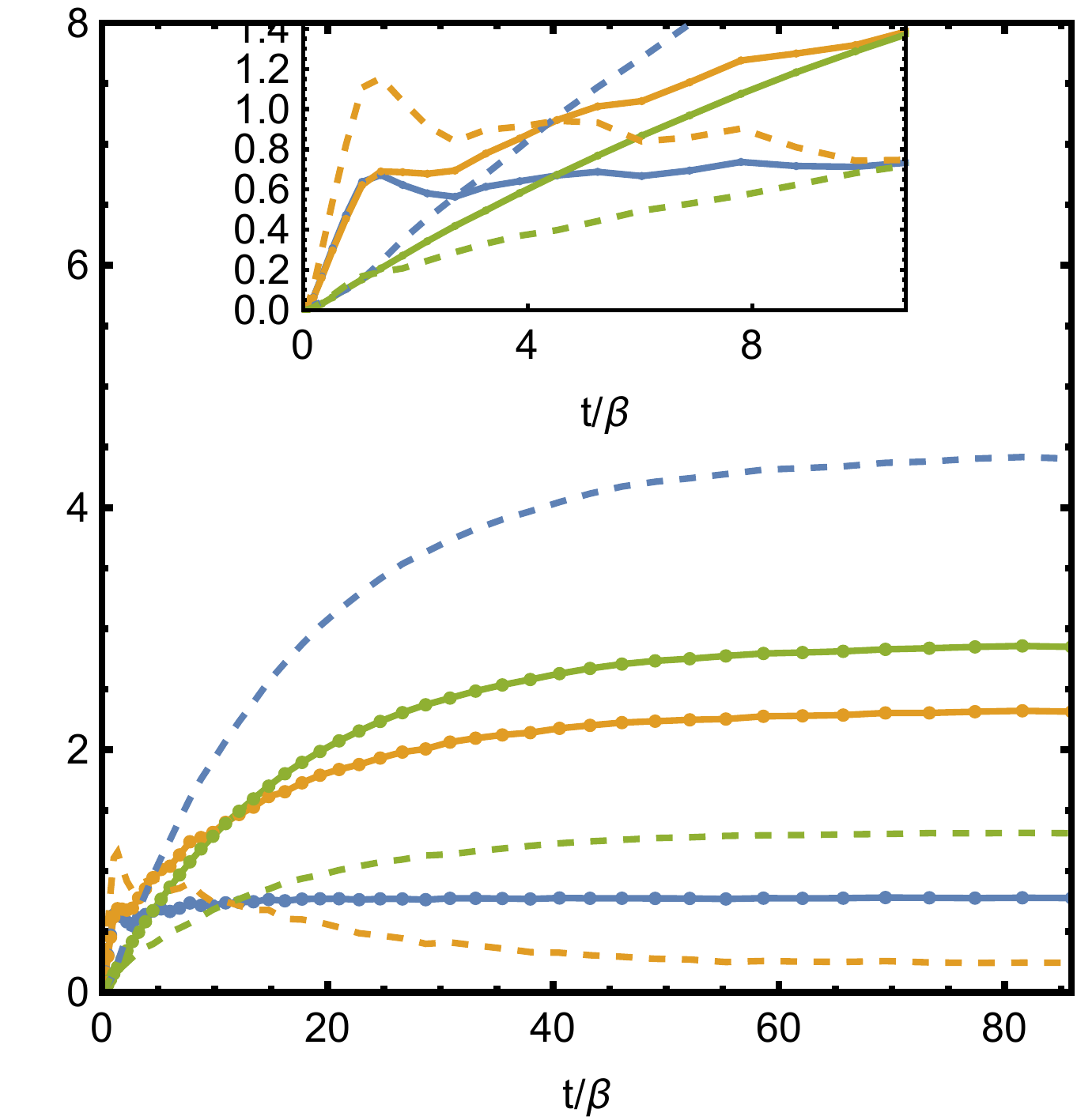}}
\caption{The evolution on entropy and mutual information densities, with $N_{\chi_1}\neq N_{\chi_2}$. An average over 40 samples with $\{N_{\chi_1},N_{\psi},N_{\chi_2}\}=\{12,4,16\}$ is plotted in the left panel, while another average over 10 samples with $\{N_{\chi_1},N_{\psi},N_{\chi_2}\}=\{12,4,20\}$ is plotted in the right panel. Other parameters are fixed to be $\{J_1 \beta, J_2 \beta, \Lambda \beta, V_1 \sqrt{\Lambda} \beta, V_2 \sqrt{\Lambda} \beta\}=\{3,3,3,1,0.3\}$.}\label{fig_NeqMIS}
\end{figure}

In principle, we hope the combined SYK system can not only resemble a classical gravity system, but also reach local thermalization quickly for each subsystem, which specifies the hierarchies between the parameters in the above Hamiltonian. The former requires $1 \ll  J_i \beta \ll N_{\chi_i}$ and the latter requires $ J_i \beta \gg V_i \sqrt{\Lambda} \beta$. In practice, although it is difficult to realize all the large hierarchies in ED, we still maintain the proper gaps between the parameters. The numerical computations of the entanglement entropy and mutual information are illustrated in Figs.~\ref{fig_EqMIS} and \ref{fig_NeqMIS}. Their evolution still shares many qualitative features with the entanglement evolution in the gravity system, as discussed below, which suggests the universal evolution of the entanglement of subsystems. At the end of this subsection, we will also discuss some distinctions between the present setup and the gravitational one.

\paragraph*{Equal-sized cases with $N_{\chi_1}=N_{\chi_2}$ -- Fig.~\ref{fig_EqMIS}}
    \begin{itemize}
    \item For adjacent SYK subsystems with small $N_\psi$, the mutual information $I[\chi_1:\chi_2]$ is positive and grows immediately at the early stage -- Fig.~\ref{fig_EqualVNear} and \ref{fig_EqualVNear2}. This indicates that the SYK systems have already been correlated at $t/\beta=0$, which agrees with Type-\textbf{I} evolution -- Fig.~\ref{fig_GS0S}, where the wormholes have been generated for adjacent black holes. 
    \item With the growth of the number of sites $N_{\psi}$ in Majorana chains, the growth of $I[\chi_1:\chi_2]$ is delayed, and $I[\chi_1:\chi_2]$ saturates at a lower level.
     The first agrees with the result that the appearance of Ph-\textbf{W} is delayed with the increment of the distance of black holes in Type-\textbf{II} evolution -- Fig.~\ref{fig_GS1S}, where the unitarity is preserved via this phase transition. While the second implies that the evolution approaches Type-\textbf{III} -- Fig.~\ref{fig_GS2S}, with the growth of distance.
    \item For all of $N_\psi$,  $S[\chi_1]$ are always close to $S[\chi_2]$, since $\chi_1$ and $\chi_2$ have the same number of Majorana fermions and their couplings obey the same Gaussian distribution. In the gravity system, we always find $\tilde{s}_{\mathcal{B}_1}=\tilde{s}_{\mathcal{B}_2}$, when $\mathcal{B}_1$ and $\mathcal{B}_2$ contain the same amount of DOF.
\end{itemize}

\paragraph*{Different-sized cases with $N_{\chi_1}<N_{\chi_2}$ -- Fig.~\ref{fig_NeqMIS}}
\begin{itemize}
    \item The bifurcation among $I[\chi_1:\psi]$, $I[\chi_2:\psi]$ and $S[\psi]$ occurs at the scale that $S[\chi_1]$ nearly saturates, where $S[\psi]$ has reached saturation and the entanglement between $\chi_1$ and $\psi$ starts to ``pass'' into $\chi_2$ . This phenomenon is similar to that in the gravity system, where the entanglement transmits into $\mathcal{B}_2$ at late times, as shown in Fig.~\ref{fig_GS1}.
    \item The growth rate of $I[\chi_1:\chi_2]$ has fallen nearly by half at the scale of the bifurcation -- Fig.~\ref{fig_NEqualNear2}, which also agrees with the result in Fig.~\ref{fig_GS1}.
    \item $S[\chi_1]$ always saturates earlier and at a level lower than $S[\chi_2]$, due to fewer DOF in $\chi_1$ compared with those in $\chi_2$. In the gravity system, this result is also manifest.
\end{itemize}

At the end of this section, we remark that aside from these similarities, there are also some distinctions in this quantum mechanical setup compared to the gravity system, which mainly C{stems} from the deviation from both the large $N$ limit and strongly coupled limit, especially in the Majorana chain.
First, in the combined SYK model, $\chi_1$ and $\chi_2$ will always be entangled within the range of parameters, and it is reasonable to expect the existence of entanglement in any choice of parameters. While in the gravity system, since we only observe the results in the leading order, the entanglement between $\mathcal{B}_1$ and $\mathcal{B}_2$ always vanishes as long as the distance is far enough. Second, the saturation of entropies is always smooth in the combined SYK model, which is distinct from the first-order phase transition of entropies in the gravity system. Third, some decaying oscillations occur in the growth of the entanglement entropy and mutual information, due to the lack of thermalization in the Majorana chain.

\begin{figure}
  \centering
  \includegraphics[width=\linewidth]{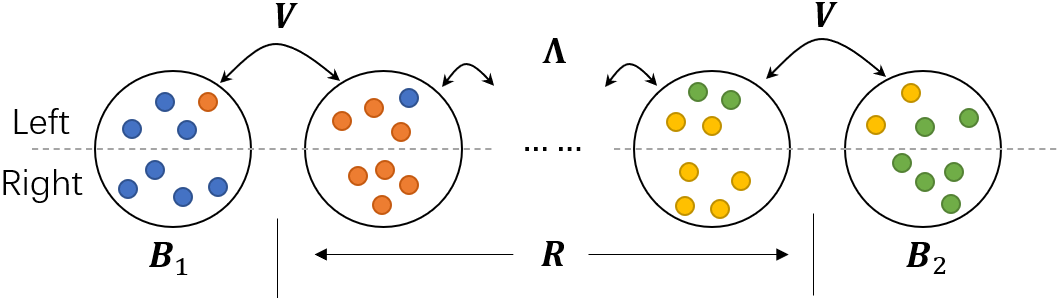}
\caption{A chain model consisting of EPR clusters at early times, where one \textbf{right} qubit has only been entangled with one \textbf{left} qubit in the same color. Without loss generality, we simply require the \textbf{right} qubits remain still, while only those on the left will exchange.}\label{fig_EPR Clusters}
\end{figure}

\section{A toy model for the evolution of entanglement: EPR clusters}\label{sec_tm}

We propose a toy model to mimic the evolution of two-body entanglement in the previous systems.
Considering several clusters of EPR pairs, each pair consists of one \textbf{left} and one \textbf{right} qubit, which together build a maximally entangled state. Collecting all the \textbf{left} and \textbf{right} qubits respectively, the resulting huge maximally entangled state resembles a TFD state, which is the model we are mainly interested in in this section.

Specifically, we denote $M$ clusters of EPR pairs as $\{G_i|i=1,2,..,M\}$, each of them containing $N_i$ EPR pairs, and denote the total number of EPR pairs as $N=\sum_{i=1}^M N_i$. Then, consider the time evolution only on the \textbf{left} qubits under a simple interaction --- exchanging one \textbf{left} qubit in $G_i$ with one \textbf{left} qubit in $G_j$ at the frequency of $J_{ij}$, which plays the role of a coupling constant. Note that the \textbf{right} qubits remain static during evolution. We define $N_{ij}$ as the number of \textbf{left} qubits in cluster $G_j$ entangled with the \textbf{right} qubits in cluster $G_i$, which satisfies the constraints $N_i=\sum_{j=1}^M N_{ij}=\sum_{j=1}^M N_{ji}$. The evolution equation for $N_{ij}$ in mean field theory is 
\begin{align}
    \frac{d}{dt} N_{ij}
    =\sum_{k=1}^M J_{jk}\left[-\frac{N_{ij}}{N_j} \left(1-\frac{N_{ik}}{N_k}\right)
    +\left(1-\frac{N_{ij}}{N_j}\right)\frac{N_{ik}}{N_k}\right].
\end{align}
In the unit of $2\ln2$, the entanglement entropy of $G_i$ is $S_i=N_i-N_{ii}$, and the mutual information between $G_i$ and $G_j$ is $I_{i;j}=N_{ij}+N_{ji}$.

Now we divide all clusters into three subsystems, namely $ \mathcal{B}_1=G_1,\  \mathcal{B}_2=G_M$, and $\mathcal{R}=\cup_{i=2}^{M-1}G_i$, which respectively serve as two black hole subsystems and radiation subsystems in the previous holographic model. Then, we specify the coupling constants $J_{ij}$ as $J_{12}=J_{(M-1)M}=V,\ J_{i(i+1)}=\Lambda,  \ i\in\{2,3,...,M-2\}$, with others being $0$, and $N_\alpha=N_{\alpha+1}=n_{\mathcal{R}}$, with $\alpha \in \{ 2,\cdots, M-2\}$. As a result, these clusters together construct a chain model, which resembles the previous holographic setup -- Fig.~\ref{fig_EPR Clusters}. In this combined system, we have the following identifications:
\begin{itemize}
    \item $N_1,\ N_M, n_\mathcal{R}$ are regarded as the DOF in the black hole and radiation subsystems.
    \item $M$ is referred to as the distance of distinct black hole subsystems.
    \item $V$ characterizes the strength of interactions between $ \mathcal{B}_{1,2}$ and $\mathcal{R}$.
    \item $\Lambda$ characterizes the speed of propagation of qubits in $\mathcal{R}$.
\end{itemize}
In this context, the entanglement entropies and mutual informations can be expressed as
\begin{align}
    &S[ \mathcal{B}_1]=S_1,&
    &S[ \mathcal{R}]=\sum_{i=2}^{M-1}(I_{i:1}+I_{i:M})/2,& 
    &S[ \mathcal{B}_2]=S_M,& \\
    &I[ \mathcal{B}_1: \mathcal{R}]=\sum_{i=2}^{M-1} I_{1:i},& &I[ \mathcal{R}: \mathcal{B}_2]=\sum_{i=2}^{M-1} I_{M:i},& 
    &I[ \mathcal{B}_2: \mathcal{B}_1]=I_{M:1}.&
\end{align}

\begin{figure}
  \centering
  \includegraphics[height=0.3\linewidth]{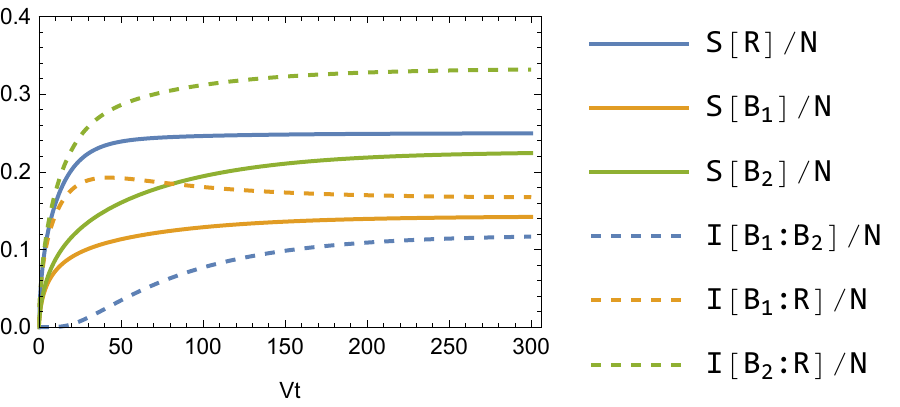}
\caption{The evolution on entropy and mutual information in the EPR clusters, where $M=16,N_1=5,n_R=1,N_M=10$ and $V=\Lambda=1$.}\label{fig_EPRASym}
\end{figure}

The evolution of entanglement for typical parameters of $M\gg1$ and $n_R<N_1<N_M$ is illustrated in Fig.~\ref{fig_EPRASym}. The results are also similar to those obtained in the previous holographic model:
\begin{itemize}
    \item Overall, entanglement entropy and mutual information grow at early times, because of the exchanging process, and finally reach equilibrium.
    \item The growth of $I[\mathcal{B}_1,\mathcal{B}_2]$ exhibits a time delay related to $M$, because the qubits that are exchanged between $\mathcal{B}_1$ and $\mathcal{B}_2$ have to traverse $\mathcal{R}$.
    \item $I[\mathcal{B}_1,\mathcal{R}]$ grows at first and then shrinks because both $\mathcal{B}_1$ and $\mathcal{R}$ have to be entangled with the largest subsystem $\mathcal{B}_2$ and lose some entanglement between each other according to the monogamy of entanglement.
\end{itemize}

\section{Conclusions and discussions}\label{sec_c&d}
In this paper, we have constructed three different kinds of models to demonstrate the dynamic process that two black holes can be entangled by emitting radiation, which exhibits a generic phenomenon in a quantum system with black holes. 

First of all, from the bulk gravity perspective, the doubly holographic model has been constructed numerically, where the holographic CFT matter sectors are dual to a higher dimensional bulk and the black holes $\mathcal{B}_1$ and $\mathcal{B}_2$ are replaced by two Planck branes in the bulk. 
The bulk geometry is solved by the Einstein-Deturck equations with Neumann boundary conditions on the branes. 
The DOF on branes $\bm{pl}_i$ are specified both by the angle $\theta_i$ and the DGP coupling $\lambda_i$. 
Restricting $\mathcal{B}_1$ to be a smaller black hole with fewer DOF, we have found four possible phases of QES during evolution, including a novel wormhole phase, which corresponds to the formation of wormholes owing to the exchange of Hawking modes between black holes. 
In the probe limit, we have elaborated on the entanglement phase diagrams and dynamical evolution for black holes of both the same and different sizes.
In general, the entanglement entropy of each black hole subsystem  increases at early times due to the exchange of Hawking modes with the environment, while saturates at late times. 
Furthermore, three universal types of evolution can be distilled: first, for adjacent black hole subsystems, a wormhole has been generated from the beginning, and no dynamical Page curve of radiation is found. 
Second, for middle-ranged black holes, the instant of wormhole formation is delayed by the distance, and unitarity of the evolution is preserved. 
Third, for distant black hole subsystems, the entanglement between black holes vanishes up to the leading order of quantum corrections in the bulk, because of the long distance. Consequently, unitarity in this type will only be preserved by the emergence of islands. 
Moreover, for the cases with $\mathcal{B}_1$ being a smaller black hole, the entanglement between the smaller black hole and radiation will leak into the larger one at late times.

The second model we have proposed is a quantum mechanical system that is composed of two separate SYK clusters $\chi_1$ and $\chi_2$ connected by a Majorana chain $\psi$. 
In this setup, SYK clusters are regarded as black hole subsystems and Majorana chains serve as radiation. 
Concretely, the DOF both in SYK clusters and the Majorana chains are specified by the number of fermions they contain, namely, $N_{\chi_1},\ N_{\chi_2}$ and $N_\psi$. 
To resemble the former gravity setup and consider the limitation of the numerical calculation, we further require the weak hierarchies $N_{\chi_i}> \beta J_i>1$ and $\beta J_i> V_i \sqrt{\Lambda}\beta$, where $\beta J_i$ and $V_i \sqrt{\Lambda}\beta$ are the dimensionless couplings inside the SYK system $\chi_i$, and between the SYK system $\chi_i$ and the Majorana chain $\psi$. 
With the finite number of fermions, the system is solved by exact diagonalization, and we have numerically computed the same entropy measures in this coupled SYK model.
We have investigated both the equal-sized SYK clusters with $N_{\chi_1}=N_{\chi_2}$ and the different-sized SYK clusters with $N_{\chi_1}<N_{\chi_2}$.
In summary, with short Majorana chains, the SYK clusters will be entangled from the beginning. 
With the increment of the chain length, the instant of the entanglement between the SYK clusters is delayed and the maximum of the entanglement is suppressed. 
In addition, in the cases of the SYK clusters with different sizes, the entanglement between the ``smaller'' SYK cluster $\chi_1$ and the Majorana chain $\psi$ will pass into the ``larger'' SYK cluster $\chi_2$.

The third model consists of several EPR clusters $\{G_i|i=1,2,\cdots,M\}$ forming a chain structure including two outermost clusters $G_1,\ G_M$ referred to as black holes $\mathcal{B}_1,\ \mathcal{B}_2$ and the inner clusters serving as radiation $\mathcal{R}$. 
Concretely, the DOF in each subsystem is specified by the number of EPR pairs it contains, and the distance of $\mathcal{B}_1$ and $\mathcal{B}_2$ is referred to as the number of clusters $M$. 
By exchanging particles with nearby clusters, the entanglement properties of the subsystems can be explored. 
Focusing on the case with $N_1<N_M$, and in the large distance limit $M\gg1$, we find the growth of entanglement between $\mathcal{B}_1$ and $\mathcal{B}_M$ exhibits a time delay. 
Also, $\mathcal{B}_1$ will lose some entanglement with $\mathcal{R}$ at late times due to the monogamy of entanglement.

Nevertheless, the deviation from the large $N$ and strong coupling limit in our quantum mechanical setup will generally lead to some discrepancy in the entropy and mutual information in comparison with the results in a gravity system. 
So we are looking forward to studying the entanglement by solving the Schwinger-Dyson equations of the coupled SYK model in the future.

It is also interesting to compare our model to \cite{Maldacena:2018lmt}, where an eternal traversable wormhole was constructed in both nearly-$AdS_2$ gravity and the SYK model by coupling two entangled subsystems. These two subsystems share the same Hamiltonian and are coupled by a double-trace interaction which matches their entanglement structure of the ground state, a nearly-TFD state. The ground state is eternally traversable, where a signal injected from one subsystem can travel into the holographic bulk and appears in the other subsystem \cite{Gao:2016bin}. Turning on this interaction does not affect the entanglement entropy between these two sides \cite{Chen:2019qqe}. 
While in our construction, the subsystems $\mathcal B_{1,2}$ or SYK$_{1,2}$ are not directly coupled, and their entanglement increases gradually as Hawking modes travel through the bath. 
In other words, the clusters SYK$_{1,2}$ in our system get entangled and reach the wormhole phase in a different manner.

In addition to the above discrepancies, our models also share some similarities with \cite{Maldacena:2018lmt}. On one side, our phase structures at late times are similar to those in \cite{Maldacena:2018lmt} at finite temperature. From the bulk gravity perspective, the entanglement wedge of $\mathcal B_{1} \cup \mathcal B_{2}$ is connected (disconnected) in the wormhole phase (island phase). This feature is similar to the connected (disconnected) bulk in the eternal-traversable-wormhole phase (two-black-hole phase) in \cite{Maldacena:2018lmt}. On the other side, the transitions between these two phases are also similar. To see this, we trace out the bath and consider the purity of the remaining two subsystems \footnote{Usually $\mathcal B_{1} \cup \mathcal B_{2}$ is not in a Gibbs state so the temperature is not well defined.}, which is negatively correlated to the second Renyi entropy of the bath. When the bath is small (huge), the purity is high (low), then the two subsystems tend to end in the wormhole phase (island phase). In this sense, this transition is similar to the Hawking-Page transition in \cite{Maldacena:2018lmt}, where at low (high) temperature or high (low) purity, the eternal-traversable-wormhole phase (two-black-hole phase) dominates.

Another interesting direction is to consider the dynamics in the full backreacted spacetime, which cannot be covered by our setup in this paper, due to the lack of metric data inside the event horizon. 
Nonetheless, by applying the ingoing Eddington coordinates in the Einstein-DeTurck formulation \cite{Figueras:2012rb}, this problem may be conquered. 
It must be pointed out that, the Einstein equations  become a mixed elliptic-hyperbolic system, and hence the local uniqueness of solutions is not guaranteed in this setup \cite{Dias:2015nua}. 
Beyond the stationary cases, it is also interesting to further generalize the numeric setup into the nonequilibrium cases \cite{Chen:2019uhq}.

Furthermore, observing various entropy computed in the coupled SYK model indicates that the result of each individual realization seems to have some noise around the averaged result, although our computation is limited by the number of fermions. 
This could be related to the recent ``half-wormhole'' interpretation of the noise of the spectral form factors~\cite{Saad:2018bqo,Saad:2021uzi,Saad:2021rcu,Mukhametzhanov:2021nea,Mukhametzhanov:2021hdi}. 
It is interesting to see in our case that the half-wormhole also contributes to the entanglement entropy. 
This is in fact a natural result from the holographic point of view; holographically the entropy is computed by the bulk gravitational path integrals that involve bulk geometries compatible with the replica trick in the boundary field theory. 
Then it is very likely that (half-)wormhole-like off-shell bulk geometries could also contribute to this computation, which could be the origin of the fluctuations of the entropy for each realization of the random coupling. 
It is interesting to show how various ``half-wormhole'' contributions enter the holographic computation of various entropies explicitly.
In particular, the current case involves two asymptotic boundaries connected by a pair of branes, which could be translated to a boundary with two pairs of matter trajectories and the half-wormhole analysis in~\cite{Peng:2021vhs}, see also the follow-up discussions in~\cite{Peng:2022pfa} could be applied directly. 
It is interesting to further optimize our numerical code, probably borrowing some numerical techniques from~\cite{Alet:2020ehp}, and verify this behavior in the future. 

\subsubsection*{Note added}
While this work was being completed Ref.~\cite{Afrasiar:2022ebi} appeared, where the authors considered a similar doubly holographic setup but in $AdS_3/BCFT_2$ duality and found a similar dynamic phase structure of entanglement. Our discussion applies to general dimensions, and also to the case where the black holes are different in size, realized via different tensions and DGP terms. 
Furthermore, we also construct two microscopic many-body models to study the evolution of entanglement. 
Finally, we discuss the time delay of entanglement due to the propagation of radiation and the loss of entanglement at late times due to the different sizes of the black holes.

\section*{Acknowledgments}
We are grateful to Yu-Chen Ding, Qing-Hua Zhu, Peng Liu, Chao Niu, Cheng-Yong Zhang, Yu Tian, Hong-Bao Zhang, Qian Chen, and Shao-Kai Jian for helpful discussions. Liu Yuxuan special thanks to his wife for supporting his work.
This work is supported in part by the National Natural Science Foundation of China under Grant No.~11875053, 12035016, and 12075298, and by China Postdoctoral Science Foundation, under the National Postdoctoral Program for Innovative Talents BX2021303, and also by Beijing Natural Science Foundation under Grant No. 1222031. CP is supported by NSFC NO. 12175237, the Fundamental Research Funds for the Central Universities, and by funds from the University of Chinese Academy of Sciences. ZYX also acknowledges support from the Deutsche Forschungsgemeinschaft (DFG, German Research Foundation) under Germany’s Excellence Strategy through the W\"urzburg-Dresden Cluster of Excellence on Complexity and Topology in Quantum Matter ct.qmat (EXC 2147, project id 390858490).

\bibliographystyle{unsrt}

\bibliography{ref}

\appendix
\section{Black hole geometry with Planck branes}\label{app_bhewpb}
As mentioned in (\ref{eq_RNBH}), the metric of the standard RN-AdS$_{d+1}$ black hole is
\begin{align}
  ds^2&=\frac{L^2}{z^2}\left[-f(z)dt^2+\frac{dz^2}{f(z)}+dw^2+\sum_{i=1}^{d-2}dw_j^2\right],\\
  A=&\mu\left(1-z^{d-2}\right)dt, \\
  f(z)=&1-
  \left(1+\frac{d-2}{d-1} \mu^2\right) z^d + \frac{d-2}{d-1}\mu^2  z^{2d-2},
\end{align}
where the outer horizon has been scaled to be $z_h=1$. However, for the geometry with branes, it is convenient to transform the coordinate into
\begin{equation}\label{eq_app_trans1}
  w = W + F(W) z,
\end{equation}
where $F(W)$ is some smooth function of $W$.

A trivial example is the black hole geometry with one Planck brane at $ w + z \cot \theta = 0$ \cite{Ling:2020laa}. One can set $F(W)$ to be a constant, which is $F(W) = - \cot \theta$ and hence, (\ref{eq_app_trans1}) becomes 
\begin{equation}
  w = W - \cot \theta z.
\end{equation}
In this context, the location of the brane becomes $W=0$ and we only focus on the region with $W\geq 0$. 

A slightly more complicated example is the black hole geometry with two Planck branes at $ w + z \cot \theta_1 = 0$ and $ w-w_0 - z \cot \theta_2 = 0$ respectively, as mentioned in the main context. In this case, we may  consider a function satisfying the following boundary conditions
\begin{align}
  F(0)= - \cot\theta_1 \quad \text{and} \quad F(w_0)= \cot\theta_2
\end{align}
at the same time. Therefore, a simple choice  is
\begin{equation}
  F(W)= \frac{(\cot \theta_2 +\cot \theta_1)}{w_0} W - \cot \theta_1
\end{equation}
and we keep our attention only on the region between the branes, which is $0\leq W \leq w_0$.  Then, for numerical convenience, the coordinates can further be  transformed to
\begin{equation}
  x=\frac{W}{w_0} \quad \text{and} \quad y=\sqrt{1-z},
\end{equation}
which lead to the form as shown in (\ref{eq_CoordinateTransform}).

\section{Derivation of the free parameter $\alpha_i$}\label{app_fixalpha} 
Near the boundary, the geometry is asymptotic to $AdS_{4}$ which in Poincaré coordinates is described by
  \begin{equation}\label{eq_app_mm}
  ds^2=\frac{L^2}{z^2}\left(-dt^2+dz^2+dw^2+ dw_1^2\right),
  \end{equation}
  For the Planck brane at $w + \cot \theta_1 z = 0$, the outward normal co-vector is 
  \begin{equation}\label{eq_app_icv}
    n= n_\mu dx^\mu = \frac{L}{z} \frac{-1}{\sqrt{1+ \cot^2 \theta_1}} \left(\cot \theta_1 dz +dw\right).
  \end{equation}
  Substituting (\ref{eq_app_mm}) and (\ref{eq_app_icv}) into the induced metric $(h_1)_{\mu\nu}=\eta_{\mu\nu} - n_\mu n_\nu$, we find the extrinsic and intrinsic curvature are
  \begin{equation}\label{eq_app_ec}
    K_1 = \frac{ 3 \cos \theta_1}{L} \quad \text{and} \quad R_{h_1}=-\frac{6 \sin^2\theta_1}{L^2}.
  \end{equation}
  Further, substituting (\ref{eq_app_ec}) into (\ref{eq_BCSonBrane}), the free parameter is fixed to be 
  \begin{equation}
    \alpha_1=\frac{2 \cos \theta_1 - \lambda_1 \sin^2 \theta_1}{L}.
  \end{equation}
  Similarly, for the brane at $(w-w_0) - \cot \theta_2 z =0 $, the outward normal co-vector is 
  \begin{equation}\label{eq_app_icv2}
    m= m_\mu dx^\mu = \frac{L}{z} \frac{1}{\sqrt{1+ \cot^2 \theta_2}} \left(-\cot \theta_2 dz +dw\right),
  \end{equation}
  and the corresponding free parameter is obtained as
  \begin{equation}
    \alpha_2=\frac{2 \cos \theta_2 - \lambda_2 \sin^2 \theta_2}{L}.
  \end{equation}

  \section{Monitoring the norm of DeTurck vector $\xi^2$}\label{app_converge} 
  In this appendix, we examine the numerical convergence of $\xi^2$ to distinguish the standard solution of Einstein equations (\ref{eq_eineq}) and the corresponding DeTurck equations (\ref{eq_EDE}).

\begin{figure}
    \centering
   \subfigure[]{\label{fig_N10}
    \includegraphics[scale=0.3]{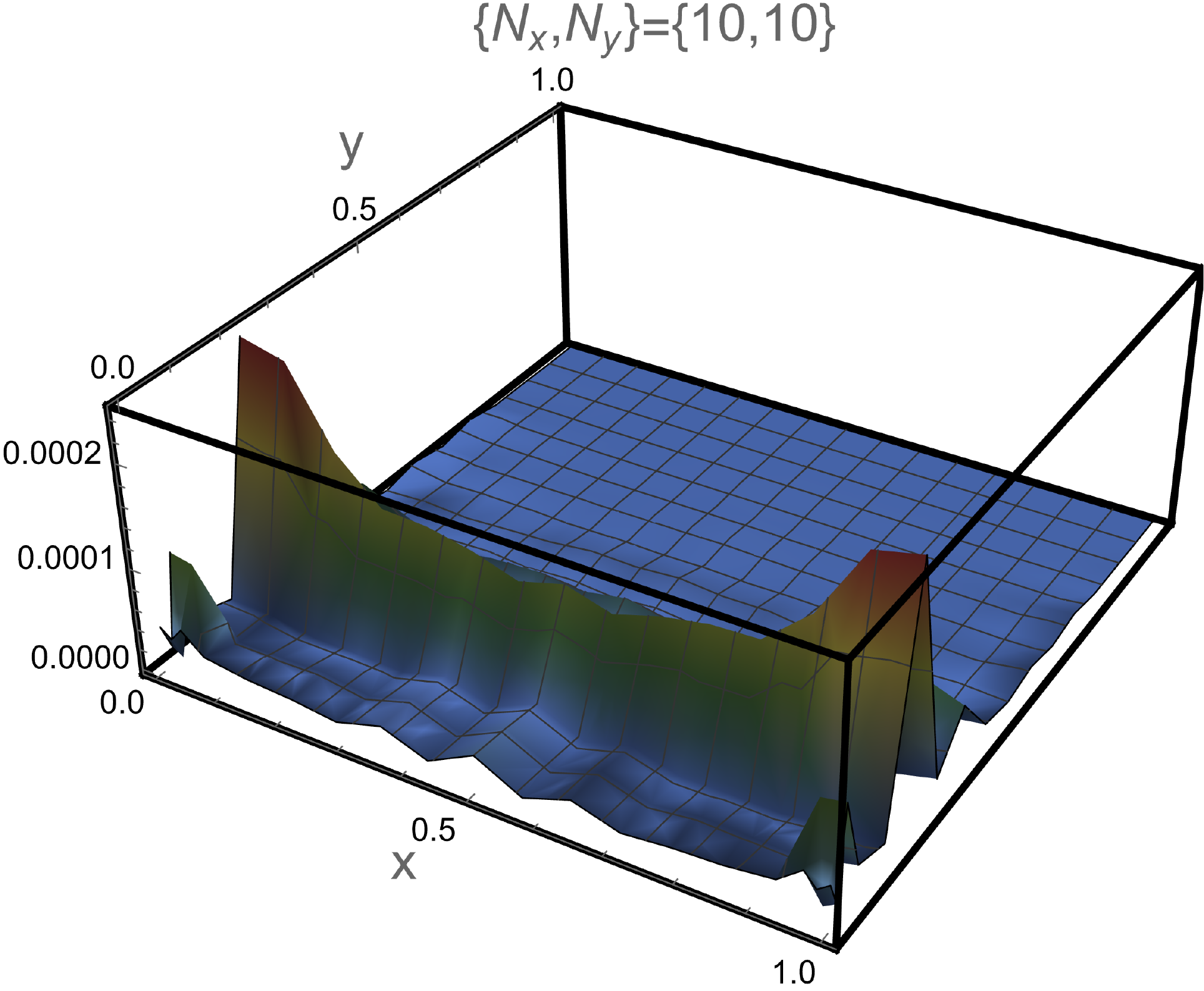}}
    \hspace{0pt}
    \subfigure[]{\label{fig_NumCon}
    \includegraphics[scale=0.4]{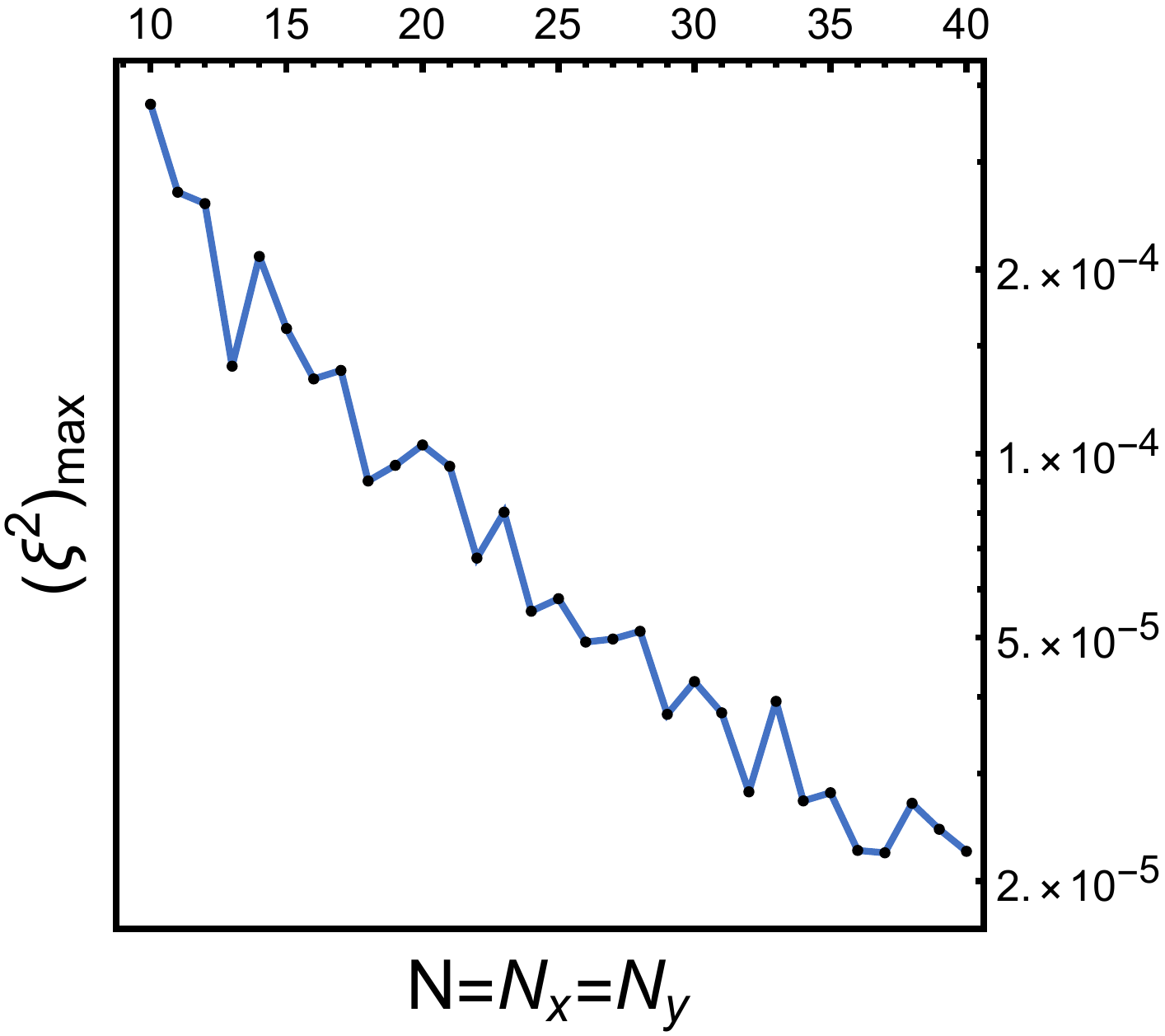}}
  \caption{Numerical convergence test with fixed parameters $\{L,\theta_1,\theta_2,\mu,w_0\} = \{1, 19\pi/40, 19\pi/40, 0, 2\}$. (a): Norm of the DeTurck vector $\xi^2(x,y)$ with gridpoints $\{N_x,N_y\}=\{10,10\}$. (b): Maximum of the norm as a function of gridpoints $N(=N_x=N_y)$.}
  \end{figure}

  By adjusting the number of Chebyshev grid points both along $x$ and $y$ directions, which we denote as $N_x$ and $N_y$ respectively, we monitor the norm of DeTurck vector $\xi^2 := \xi_\mu \xi^\mu $ -- see Fig.~\ref{fig_N10}. Specifically, for fixed grid points $\{N_x,N_y\}$ and other parameters such as $\{L,\theta_1,\theta_2,\mu,w_0\}$, we take the maximum of $\xi^2$ within domain $\{ (x,y) | 0\leq x \leq 1 \cap 0< y <1\}$, which we denote as $\xi^2_{max}$. Then varying $\{N_x,N_y\}$, we finally obtain $\xi^2_{max}$ as a function of grid points, as illustrated in Fig.~\ref{fig_NumCon}. The key point is that even though there is no proof of the non-existence of the DeTurck soliton with Neumann boundaries, with sufficient Chebyshev grid points, we are still capable to obtain a solution that is arbitrarily close to that of Einstein equations.

  \section{Derivation of the growth rate of entropy density}\label{app_GrowthRate}
Recall that the entropy density of Planck branes penetrating the horizon is written as
\begin{equation}\label{eq_app_Atr}
  s[\mathcal{R}_i]=\frac{\textbf{A}(\gamma_{T_{i}})}{2 V}(t)=\int \frac{ d\xi}{z^{d-1}}\sqrt{-v' \left[f(z) v'+2 z'\right]},
\end{equation}
where $i=1,2$ denote the different Planck branes.

In addition, the conserved momentum conjugated to $v$ and the invariance of the reparametrization can be expressed respectively as
\begin{equation}\label{eq_app_CM}
  C=\frac{f(z) v'+z'}{z^{d-1}\sqrt{-v' \left[f(z) v'+2 z'\right]}},
\end{equation}
and
\begin{equation}\label{eq_app_Rep}
  \sqrt{-v' \left[f(z) v'+2 z'\right]}=z^{d-1}.
\end{equation}
From (\ref{eq_app_CM}) and (\ref{eq_app_Rep}), we further obtained
\begin{align}
  z'(\xi)=&-z^{(d-1)}\sqrt{f(z)+C^{2}z^{2 (d-1)} },\\
  v'(\xi)=&\frac{C+\sqrt{\left(C^{2}+f(z)z^{-2  (d-1)}\right)}}{f(z)},
\end{align}
and also
\begin{equation}
  v'(z)=-\frac{1+\frac{Cz^{(d-1)}}{\sqrt{f(z)+C^{2} z^{2 (d-1)}}}}{f(z)},
\end{equation}
where we have set $L=1$.

Therefore, the expressions of time and area can be expressed respectively as
\begin{align}
  t=&\int dv +\int \frac{dz}{f(z)}=\int\left(v^{\prime}(z)+\frac{1}{f(z)}\right) d z\nonumber\\
  =&\int_0^{z_{max}}d z \frac{C z^{(d-1)}}{f(z) \sqrt{f(z)+C^{2}z^{2 (d-1)}}}\\
  \frac{\textbf{A}}{2 V}=&\int  d\xi = \int_0^{z_{max}}\frac{dz}{z'(\xi)}
  =\int_0^{z_{max}}\frac{dz}{z^{(d-1)}\sqrt{f(z) +C^{2}z)^{2 (d-1)} }}\nonumber\\
  =&\int_0^{z_{max}}\frac{\sqrt{z^{2 d-2} C^{2}+f(z)}}{f(z) z^{d-1}} d z-C \int_0^{z_{max}}\frac{C z^{d-1}}{f(z) \sqrt{f(z)+C^{2} z^{2 d-2}}} dz\nonumber\\
  =&\int_0^{z_{max}}\frac{\sqrt{z^{2 d-2} C^{2}+f(z)}}{f(z) z^{d-1}} d z - C t,
\end{align}
where $z_{max}$ is the turning point of $\gamma_{T_i}$ and in the following, we will use the fact that $z_{max}$ is evolving in time. Combining these two equations, the growth rate of the area can be obtained as
\begin{align}
  \frac{d}{dt}\frac{\textbf{A}}{2 V}=&\left(\frac{\sqrt{z_{max}^{2 d-2} C^{2}+f(z_{max})}}{f(z_{max}) z_{max}^{d-1}}\right) \frac{d z_{max}}{d t}-\frac{d C}{d t}t \nonumber\\
  &+\frac{d C}{d t} \int_{0}^{z_{max}}\frac{C z^{-1+d}}{f(z) \sqrt{C^{2} z^{-2+2 d}+f(z)}} d z-C\nonumber\\
  =&\frac{\sqrt{-f(z_{max})}}{z_{max}^{d-1}}.
\end{align}
Note that in the last step, we have used the result as shown in (\ref{eq_ConQua}).

\end{document}